\pdfoutput=1
\documentclass[11pt,a4paper]{article}

\usepackage{MySty}
\usepackage{amsmath,amsfonts,amssymb,graphics}
\usepackage[normalem]{ulem}
\usepackage{graphicx}
\usepackage{mathrsfs}
\usepackage{bbm}
\usepackage{pifont}
\usepackage{braket}
\usepackage{footmisc}
\usepackage[dvipsnames]{xcolor}
\usepackage{mathtools}
\usepackage{xspace}
\usepackage{booktabs}
\usepackage{siunitx}
\usepackage{slashed}
\usepackage{tikz}
\usetikzlibrary{positioning,shapes,arrows,calc}
\usetikzlibrary{decorations.pathmorphing}
\usetikzlibrary{decorations.markings}
\usepackage{standalone}
\usepackage{csquotes}
 
\graphicspath{{figs/}}

\title{\boldmath \Huge \bf{Diraxiogenesis}}
\author{Maximilian Berbig}
\affiliation{Bethe Center for Theoretical Physics \& Physikalisches 
Institut der Universit\"at Bonn,\\
Nußallee 12, Bonn, Germany}
\emailAdd{berbig@physik.uni-bonn.de}

\abstract{The family of Dirac Seesaw models offers an intriguing alternative explanation for the smallness of neutrino masses without necessarily requiring microscopic lepton number violation, when compared to the more familiar class of Majorana Seesaws.
A global $\text{U}(1)_\text{D}$ symmetry, that is explicitly broken by a higher dimensional scalar operator, ensures that the right handed neutrino does not couple directly to the Standard Model like Higgs and an exact gauged or residual lepton number symmetry  prohibits all Majorana masses. We demonstrate that all three Dirac Seesaws possess a Pseudo-Nambu-Goldstone boson associated with the $\text{U}(1)_\text{D}$ symmetry, that we call the Diraxion, whose cosmological dynamics have so far been left unexplored. Furthermore we illustrate that a Dirac-Leptogenesis version of the recently proposed Lepto-Axiogenesis scenario can be realized in this class of models, leading to a unified origin of the observed baryon asymmetry and dark matter relic abundance. Explaining only the baryon asymmetry can lead to potentially observable amounts of right handed neutrino dark radiation with  $\Delta N_\text{eff.}\lesssim 0.028$. On the other hand, if we only fix  the dark matter abundance via the kinetic misalignment mechanism, this set-up could lead to detectable signatures in proposed cosmic neutrino background experiments via decays of eV-scale Diraxions to neutrinos.  Here there is no domain wall problem, since topological defects  decay to a subleading fraction of relic Diraxions. A key ingredient of all Axiogenesis scenarios is the dynamics of relatively light scalar called the Saxion, that in our case has a mass at the GeV-scale and which  might reveal itself in heavy meson decays or collider searches.  Our setup predicts   isocurvature perturbations in baryons, dark matter and dark radiation sourced by fluctuations of the Saxion.}

\begin{document}
\maketitle

\section{Introduction}
Condensates of scalar fields, either elementary or composite, play an important role in our understanding of fermion and gauge boson mass generation. Apart from the celebrated Higgs mechanism these condensates also allow for the possibility of explaining either the observed dark matter relic abundance or the matter anti-matter asymmetry. Some of the earliest proposals for baryogenesis focused on the dynamics of oscillating scalar bosons \cite{Dimopoulos:1978kv}. With the advent of inflationary cosmology \cite{Guth:1980zm} it was realized that the vacuum expectation value (vev) of such a field can undergo a large excursion during the quasi de-Sitter phase of  the early universe \cite{Starobinsky:1994bd}. The requirement for this is the existence of a very flat or almost vanishing scalar potential, which is naturally realized in supersymmetric field theories such as the MSSM.
About 300 of these flat directions, that only receive their masses from supersymmetry breaking and higher dimensional effective operators or radiative corrections, are known \cite{Dine:1995kz,Gherghetta:1995dv}. This prompted the development of the Affleck-Dine mechanism \cite{Affleck:1984fy}, where the large effective vev generates the baryon asymmetry from a (potentially Planck scale) suppressed baryon number violating scalar interaction. To transfer this asymmetry from the scalar   sector to the baryons typically involves  decays of the condensate to thermal bath particles, which can be automatically included if the flat direction is responsible for inflation and the subsequent era of reheating (see e.g. \cite{Cline:2019fxx} for a recent example).\\
\\
While the aforementioned scenario adheres to well known   Sakharov-conditions \cite{Sakharov:1967dj} relying on CPT-conservation, there exist a class of scenarios in which CPT is spontaneously broken in the plasma of the early universe \cite{Cohen:1987vi,Cohen:1988kt}: The Pseudo-Nambu-Goldstone boson (PNGB) $\theta$ of an abelian symmetry, such as e.g. baryon number, oscillates in a cosine-potential. The initial field value   of its oscillations  can be at most $2\pi$ times the decay constant $f$ of the field. One typically requires a rather large $f$ to ensure that the underlying symmetry was broken before inflation and to suppress isocurvature perturbations, see e.g. \cite{Kusenko:2014uta}. Since the baryon asymmetry in these models is proportional to the oscillation frequency of the PNGB $\dot{\theta}$ , which is related to its mass, one typically also needs masses around \cite{Cohen:1987vi,Cohen:1988kt} or far above the electroweak scale \cite{Kusenko:2014uta} making these proposals hard to test in laboratory experiments.
Consult also reference \cite{DeSimone:2016ofp} for a review and  \cite{Domcke:2020kcp} for a systematic treatment of this scenario. The charge from the oscillating PNGB is transferred to the fermions $\psi$ in the plasma via a (schematic) derivative coupling $(\partial_\mu \theta) \overline{\psi}\gamma^\mu \psi $, which reduces to $\dot{\theta}\; \overline{\psi}\gamma^0 \psi$ for a homogeneous and isotropic PNGB field. Following from the fact that this only time-dependent term breaks Lorentz symmetry spontaneously in the early universe plasma, one can deduce that CPT is also spontaneously violated, hence the name of this scenario.\\
\\
A new approach pioneered by \cite{Co:2019wyp} under the name of \enquote{Axiogenesis} essentially combines the two aforementioned proposals with the phenomenologically attractive QCD axion \cite{Peccei:1977hh,Weinberg:1977ma,Wilczek:1977pj,Kim:1979if,Shifman:1979if,Zhitnitsky:1980tq,Dine:1981rt}. Here one can have large decay constants $f_a$ and an observable PNGB whose mass is inversely proportional to $f_a$, making the scenario potentially testable. A large field excursion of the radial mode of the complex scalar that houses the PNGB angle, known as the Saxion \footnote{This name is inspired by supersymmetric axion models and here we assume no Supersymmetry.}, can be used to convert its oscillatory motion in the early universe into a coherent rotation with $\dot{\theta}\neq 0$. The rotation in the axion direction couples to the QCD anomaly and induces chiral asymmetries for the quarks via the QCD sphaleron, which gets reprocessed into a B+L asymmetry via the electroweak sphaleron process. On top of that one can use the axion field velocity for a new scenario of dark matter known as kinetic misalignment \cite{Co:2019jts,Co:2020dya} (see also \cite{Barman:2021rdr}), unlike the conventional misalignment which works under the assumption of negligible velocity \cite{Preskill:1982cy,Abbott:1982af,Dine:1982ah}.
However this comes with the drawback of overproducing dark matter if one fixes the baryon asymmetry to its observed value \cite{Co:2019wyp}. This conclusion can be avoided if the electroweak phase transition is modified  \cite{Co:2019wyp}, one introduces additional sphalerons from e.g. a gauged $\text{SU}(2)_\text{R}$ \cite{Harigaya:2021txz}, one takes the chiral plasma instability into account \cite{Co:2022kul} or the axion possesses very large couplings to the weak anomaly \cite{Kim:2004rp}. The basic scenario can also work for more generic axion-like particles \cite{Co:2020xlh}, very heavy QCD axions \cite{Co:2022aav} or with multiple scalars \cite{Domcke:2022wpb}.  Additionally there can be rich gravitational wave signatures \cite{Co:2021lkc,Co:2021rhi,Gouttenoire:2021wzu,Gouttenoire:2021jhk,Madge:2021abk,Harigaya:2023ecg}. Another way to make Axiogenesis viable is to directly produce a B-L asymmetry instead of B+L via additional processes such as lepton number violating scatterings mediated by heavy Majorana neutrinos, known under the name of "Lepto-axiogenesis" \cite{Co:2020jtv,Kawamura:2021xpu,Barnes:2022ren}, or R-parity violating (RPV) supersymmetry \cite{Co:2021qgl} (see reference \cite{Dreiner:1997uz} for a review on RPV).\\
\\
In this work we follow the Lepto-axiogenesis route and try to explain the origin of the PNBG and the effective B-L violation from the same source. To this end we abandon the traditional QCD axion and focus on the PNGBs associated with neutrino mass generation. While the corresponding particle for Majorana neutrinos with the fitting name Majoron \cite{Gelmini:1980re} has been widely studied \cite{Chikashige:1980ui}, we choose to focus on the case of Dirac neutrinos. Here there are typically two symmetries \cite{Roncadelli:1983ty}: We assume global $\text{U}(1)_\text{D}$ to prevent the coupling between the Standard Model Higgs doublet and the right handed neutrinos together with a gauged or residual lepton number (or B-L) symmetry  ensuring the absence of all Majorana masses. Non perturbative quantum gravitational effects manifesting as Planck-scale suppressed higher dimensional effective operators explicitly break the $\text{U}(1)_\text{D}$ symmetry and we call the associated PNGB  Diraxion. The idea is that we produce equal and opposite asymmetries in the Standard Model fermions  and the right handed singlet neutrinos due to the conservation of the total B-L, which never get equilibrated \cite{Dick:1999je}. The electroweak sphaleron process is only sensitive to the 
the lepton doublet so effectively $\text{(B-L)}_\textbf{SM}$ is violated in the plasma and can be converted into a baryon asymmetry. 
A similar idea albeit with a QCD axion has been pursued in \cite{Chakraborty:2021fkp} for the case of composite right handed neutrinos. However due to the unspecified  UV nature of the non-perturbative mechanism behind the right handed neutrino formation the authors can only use the temperature from when on $\text{(B-L)}_\textbf{SM}$ is conserved as a free parameter.
Our scenario relies on calculable, perturbative models realizing parametrically small Dirac neutrino masses from threshold corrections in the Seesaw spirit. We discuss the three possible cases for these Dirac Seesaws and show that the singlet scalar that appears in all those constructions can house the Saxion and Diraxion. Similar to the original spontaneous baryogenesis proposal \cite{Cohen:1988kt}, we will assume that 
the underlying global symmetry is only broken by a higher dimensional operator 
responsible for both the Diraxion mass and converting the Saxion oscillation into a Diraxion rotation. Our scenario is inspired by models combining scalar field Leptogenesis with neutrino mass generation such as spontaneous Leptogenesis  from a very heavy oscillating Majoron \cite{Ibe:2015nfa}, which was recently reanalyzed and extended to rotating Majorons in \cite{Chun:2023eqc},  or inflationary Affleck-Dine Leptogenesis from the Type II Seesaw \cite{Barrie:2021mwi,Barrie:2022cub}. Affleck-Dine Dirac Leptogenesis from supersymmetric Dirac neutrinos involves two complex scalar fields, where the smallness of the $F$-term potential is connected to the tiny neutrino mass scale, and this scenario was investigated in \cite{Abel:2006hr,Senami:2007up}.
Unlike these models we predict sub-eV PNGBs and are able to explain the dark matter relic abundance via kinetic misalignment \cite{Co:2019jts}  or parametric resonance \cite{Harigaya:2015hha,Co:2017mop}. 
If our setup is only responsible for baryogenesis, dark radiation in the $\nu_R$ component can be produced with an amount of $\Delta N_\text{eff.}\lesssim 0.028$. Alternatively if we reproduce (a fraction of) the dark matter relic abundance, the Diraxion can be heavy and metastable enough for decays to neutrinos, leading to a discovery potential in next generation experiments investigating the cosmic neutrino background. Requiring the cogenesis of the baryon asymmetry together with dark matter on the other hand leads to unobservably small  $\Delta N_\text{eff.}$ and too light or long-lived Diraxions. While isocurvature fluctuations in baryons and dark matter are a generic prediction of this kind of scenario, we find that our setup also produces dark radiation isocurvature modes and all three kinds of perturbations are induced by Saxion fluctutations during inflation. Since the Diraxion only has suppressed couplings to photons we do not expect any signal in axion haloscopes or helioscopes. In our scenario we find slightly heavier Saxions than for Lepto-Axiogenesis \cite{Co:2020jtv} with masses around the GeV scale, which can be tested in collider experiments and rare decays of heavier mesons.\\
\\
The relevant parameter space of our analysis is spanned by the Diraxion mass $m_a$, the Saxion mass $m_S$, the Diraxion decay constant today $f_a$, the initial Saxion field value $S_i$ as well as  the domain wall number $N$, which coincides with the mass dimension of the non-renormalizable operator responsible for the Diraxion mass. We can eliminate $f_a$ by requiring that the interaction of the Saxion with the thermal bath is slow, which is equivalent to fixing the amount of right handed neutrino dark radiation produced via Freeze-In. Successful Saxion thermalization  from the Higgs portal interaction before an era of Saxion domination is viable for $S_i \lesssim 0.1 M_\text{Pl.}$. In order to keep the Diraxion light enough we fix $N=6$. This allows us to predict the masses $m_a$ and $m_S$ for successful cogenesis of the baryon asymmetry and dark matter relic abundance.\\
\\
Section \ref{sec:Models} gives an overview over the family of Dirac Seesaws together with the details about the Saxion and Diraxion. We illustrate the cosmological evolution leading to Saxion oscillations and ultimately a coherent rotation in the Diraxion direction in section \ref{sec:Dynamics}. Dirac-Lepto-Axiogenesis is the subject of section \ref{sec:Dirac}. Dark Matter and Dark Radiation are the topics of sections \ref{sec:Matter} and \ref{sec:Radiation} respectively. Estimates of the Dark Matter and Dark Radiation isocurvature perturbations can be found in \ref{sec:iso}. Section \ref{sec:therm} describes the required Saxion thermalization to avoid overclosure and excessive amounts of Dark Radiation. We discuss our findings for the regions of parameter space that realize both Leptogenesis and dark matter in \ref{sec:Diss} before we conclude and summarize in \ref{sec:Conclusion}.

\section{Models}\label{sec:Models}
\begin{table}[t]
\centering
 \begin{tabular}{|c|c||c|c|c|c||c|} 
 \hline
 model & field&   $\text{SU(3)}_\text{C}$ & $\text{SU}(2)_\text{L}$ & $\text{U}(1)_\text{Y}$ & $\text{U}(1)_\text{D}$ & \text{generations}\\
 \hline
  all &  $\nu_R$ & 1 & 1 & 0 & -1&3\\
  all&  $\sigma$ & 1 & 1 & 0 & 1 & 1  \\
\hline
\hline
 Type I &   $N_L$ & 1 & 1 & 0& 0 & 3\\   
 Type I&   $N_R$ & 1 & 1 & 0& 0 & 3\\   
\hline
\hline
  Type II &  $\eta$ & 1 & 2 & $1/2$ & 1& 1\\
\hline
\hline   
 Type III-a &   $T_L$ & 1 & 3 & 0 & 0 &3\\
 Type III-a &   $T_R$ & 1 & 3 & 0 & 0 &3\\
 \hline       
 Type III-b &    $D_L$& 1 & 2 & $-1/2$ & -1 &3\\
 Type III-b &   $D_R$ & 1 & 2 & $-1/2$ & -1 &3\\
\hline
 Type III &   $\Delta$ & 1 & 3 & 0 & 1 & 1 \\
\hline    
\end{tabular}
\caption{Charges and Representations under the SM gauge group and $\text{U}(1)_\text{D}$.}
\label{tab:charges-repsD}
\end{table}

\subsection{Dirac Weinberg-operator}
The famous dimension five  Weinberg-operator $(\overline{L} \tilde{H})(L^c i\sigma_2 H)$ with $\tilde{H} \equiv i \sigma_2 H^\dagger$ is the only gauge invariant combination of Standard Model (SM) fields that generates a Majorana mass for the left chiral neutrinos \cite{PhysRevLett.43.1566}. However it has long been known that by adding gauge singlet right chiral neutrinos $\nu_R$ together with a singlet scalar $\sigma$ one can realize a analogue of the Weinberg-operator for Dirac neutrinos
\begin{align}\label{eq:WeinbergDirac}
    \mathcal{L}_5 = \frac{c_\nu}{\Lambda_\text{UV}}\; \overline{L}\tilde{H}\; \sigma \nu_R + \text{h.c.},
\end{align}
where  $c_\nu$ is a dimensionless Wilson-coefficient that appears together with a UV cut-off-scale $\Lambda_\text{UV} \gg v_H$ and $v_H\equiv \SI{246}{\giga\electronvolt}$ is the vacuum expectation value (vev) of the neutral component of the SM Higgs field. After $\sigma$ condenses with the vev $v_\sigma$ the above operator leads to neutrino masses of 
\begin{align}
    m_\nu = c_\nu \frac{v_\sigma}{\Lambda_\text{UV}}v_H,
\end{align}
and the required suppression of the neutrino mass scale needs $\Lambda_\text{UV}\gg v_H, v_\sigma$ for order one $c_\nu$. The coupling of $\nu_R$ to the SM like Higgs via the operator 
\begin{align}
    Y_D \overline{L} \tilde{H} \nu_R + \text{h.c.}
\end{align}
can be forbidden by invoking a global $\text{U(1)}_\text{D}$, under which the SM is uncharged and no mixed anomalies appear. To form the Weinberg-Operator then requires that the charges of  $\nu_R$ and $\sigma$ satisfy
\begin{align}
    Q_\text{D}[\nu_R] + Q_\text{D}[\sigma]=0.
\end{align}
To ensure the absence of any perturbatively (in field theory) or non-perturbatively (from quantum gravity) generated \cite{Barbieri:1979hc,deGouvea:2000jp} Majorana mass terms for the SM neutrinos or heavy messenger fermions, we assume either an unbroken global symmetry like lepton number as a residual symmetry from a larger gauge symmetry, or a gauged symmetry like $\text{U(1)}_\text{B-L}$, which can give rise to Dirac masses depending on the choice of charges and the breaking pattern \cite{Ma:2014qra}.
In the following sections we discuss the tree-level UV-completions of the operator in \eqref{eq:WeinbergDirac}. We will not concern ourselves too much with the details of the aforementioned symmetry for the absence of Majorana masses such as anomaly cancellation.
Instead we focus on the cosmological evolution of the global $\text{U(1)}_\text{D}$ breaking. The explicit breaking via quantum gravitational effects might induce the following higher dimensional operator contributing to the Dirac neutrino masses
\begin{align}
    \mathcal{L}_\slashed{D} = c_n\; \overline{L}\tilde{H}\; \left(\frac{\sigma}{M_\text{Pl.}}\right)^n \nu_R + \text{h.c.}.
\end{align}
The largest contribution  for $v_\sigma \ll M_\text{Pl.}$ comes from $n=1$ and it is smaller than the contribution from the Dirac Weinberg-operator in \eqref{eq:WeinbergDirac} as long as
\begin{align}
    c_1 < 2.5\cdot \left(\frac{m_\nu}{\SI{0.05}{\electronvolt}}\right)\cdot \left(\frac{10^6\;\text{GeV}}{v_\sigma}\right).
\end{align}

\subsection{The three Dirac Seesaws}
Here we introduce the three tree-level UV-completions to the Weinberg-operator in \eqref{eq:WeinbergDirac} and focus on the heavy new messenger fields whose  threshold corrections lead to the tiny observed active neutrino masses. A systematic study of tree- and one-loop-level UV completions can be found in \cite{Ma:2016mwh,CentellesChulia:2018gwr,Yao:2018ekp}.

\subsubsection{Type I}

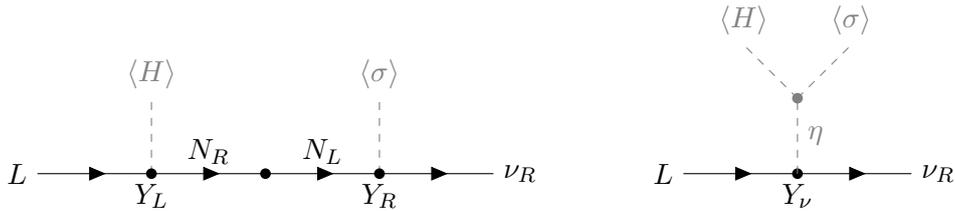
\begin{figure}[t]
 \centering
  \tikzset{
  blackline/.style={thin, draw=black, postaction={decorate},
    decoration={markings, mark=at position 0.6 with {\arrow[black]{triangle 45}}}},
     grayline/.style={thin, draw=gray, postaction={decorate},
    decoration={markings, mark=at position 0.6 with {\arrow[gray]{triangle 45}}}},
    blueline/.style={thin, draw=blue, postaction={decorate},
    decoration={markings, mark=at position 0.6 with {\arrow[blue]{triangle 45}}}},
    redline/.style={thin, draw=red, postaction={decorate},
    decoration={markings, mark=at position 0.6 with {\arrow[red]{triangle 45}}}},
    greenline/.style={thin, draw=green, postaction={decorate},
    decoration={markings, mark=at position 0.6 with {\arrow[green]{triangle 45}}}},    
    graydashed/.style={dashed, draw=gray, postaction={decorate},
    decoration={markings}},
   yellowdashed/.style={dashed, draw=orange, postaction={decorate},
    decoration={markings}},
    photon/.style={decorate, draw=red,
    decoration={coil,amplitude=12pt, aspect=0}},
  gluon/.style={dashed, decorate, draw=black,
    decoration={coil, segment length=5pt, amplitude=8pt}}
  line/.style={thick, draw=black, postaction={decorate},
    decoration={markings}}
}

\begin{tikzpicture}[node distance=1cm and 1cm]

%particles 
\coordinate[label = left: $L$] (start1);
\coordinate[right=6cm of start1,label=right: $ \nu_R$] (end);

%vertices
\coordinate[right=1.5cm of start1, label=below: $Y_L$] (H1);
\coordinate[right=1.5cm of H1] (sigma1);
\coordinate[right=1.5cm of sigma1, label=below: $Y_R$] (H2);

%vev insertions 
\coordinate[above=1cm of H1,label=above: $\color{gray} \braket{H}$] (vevH1);
\coordinate[above=1cm of H2,label=above: $\color{gray} \braket{\sigma}$] (vevH2);
\coordinate[below=1cm of sigma1,label=below: ] (vevS1);

%axis
\draw[blackline] (start1)   -- (H1);
\draw[blackline] (H1)   -- (sigma1);
\draw[blackline] (sigma1)   -- (H2);
\draw[blackline] (H2)   -- (end);

%vev insertions
\draw[graydashed] (H1)   -- (vevH1);
\draw[graydashed] (H2)   -- (vevH2);

%particle labels
\coordinate[right=0.75cm of H1, label=above: $N_R$];
\coordinate[left=0.75cm of H2, label=above: $N_L$];

%dots
\fill (H1) circle (2pt);
\fill (H2) circle (2pt);
\fill  (sigma1) circle (2pt);

%Second diagram
\coordinate[right=2.5 cm of end,label = left: $L$] (start2);
\coordinate[right=1.5cm of start2, label = below: $Y_\nu$] (vertex2);
\coordinate[right=1.5cm of vertex2, label = right : $\nu_R$] (end2);
\coordinate[above=1 cm of vertex2 ] (vertex3);
\coordinate[above=0.5 cm of vertex2, label = right: $\color{gray} \eta$ ] (test2);
\coordinate[above left =1cm of vertex3, label =  above: $\color{gray} \braket{H}$ ] (vev2);
\coordinate[above right =1cm of vertex3, label = above: $\color{gray} \braket{\sigma}$ ] (vev3);

\fill (vertex2) circle (2pt);
\fill[gray] (vertex3) circle (2pt);

\draw[blackline] (start2)   -- (vertex2);
\draw[blackline] (vertex2)   -- (end2);
\draw[graydashed] (vertex3) -- (vev2);
\draw[graydashed] (vertex3) -- (vev3);
\draw[graydashed] (vertex2) -- (vertex3);

\end{tikzpicture}
  \caption{Diagrammatic representation of the dimension 5 operators for the Type I  Dirac Seesaw $\textit{(left)}$ and Type II Dirac Seesaw $\textit{(right)}$  giving rise to Dirac masses for the active neutrinos. For the Type II case one could also consider an additional insertion of $\sigma$ not depicted here.}
  \label{fig:DiracSeesawI-II}
\end{figure}

The most basic Dirac Seesaw \cite{Roncadelli:1983ty,Roy:1983be} was discovered shortly after the more well known Majorana-Seesaw \cite{Minkowski:1977sc,Yanagida:1979as,Gell-Mann:1979vob,Glashow:1979nm,  PhysRevLett.44.912} and its phenomenology was 
explored in \cite{Cerdeno:2006ha}. The mechanism requires nothing more than at least two generations of super-heavy, electrically neutral vector-like fermions $N_{L,R}$ and the field content and charges can be found in table \ref{tab:charges-repsD}.
\begin{align}
    \mathcal{L}^{\text{(I)}} = Y_L\; \overline{L} \tilde{H} N_R + Y_R\; \overline{N_L}\sigma \nu_R +M_N\; \overline{N_L} N_R.
\end{align}
We assume that their masses $M_N$ are not connected to $\text{U(1)}_\text{D}$ and a  global or gauged $\text{U(1)}_\text{B-L}$ to be responsible for the absence of any Majorana masses. A diagrammatic representation of the mass generation mechanism was depicted on the left of figure \ref{fig:DiracSeesawI-II}. 
After integrating $N_{L,R}$ out, the light neutrino masses  in the single flavor approximation read
\begin{align}\label{eq:TypeISeesaw}
    m_\nu^\text{(I)} \simeq Y_L Y_R \; \frac{v_\sigma}{2 M_N} v_H,
\end{align}
where $Y_L v_H, \;Y_R v_\sigma \ll M_N$ was assumed and we can estimate
\begin{align}
    m_\nu^\text{(I)} \simeq \SI{0.05}{\electronvolt} \cdot Y_L Y_R \cdot \left(\frac{v_\sigma}{\SI{4}{\tera\electronvolt}}\right)\cdot \left(\frac{10^{16}\;\text{GeV}}{M_N}\right),
\end{align}
where we chose $M_N$ around the grand unification scale for illustration. Keeping $M_N$ fixed we can accommodate larger $v_\sigma$ by making the Yukawa couplings $Y_L Y_R$ smaller.
The one-loop correction to the active neutrino masses comes from closing the Higgs-singlet loop via an insertion of the mixed quartic coupling $\lambda_{\sigma H}$ \cite{Roncadelli:1983ty}
\begin{align}
     \delta  m^\text{(I)}_{\nu} = \frac{\lambda_{\sigma H}}{4\pi} \cdot   m_\nu^\text{(I)}
\end{align}
and it is subleading for perturbative values of $\lambda_{\sigma H}$. This variant of the Seesaw can be embedded in the grand unified theories based on $\text{SU(5)}$ \cite{Roncadelli:1983ty} or $\text{SO(10)}$ \cite{Peinado:2019mrn}. For example in the $\text{SO(10)}$ case this comes with the drawback of spoiling matter unification: $N_R$ fills up the 16-dimensional spinorial representation together with the rest of the SM fermions and $\nu_R, N_L$ have to be added as additional gauge singlets.
A simpler UV-completion based on $\text{U(1)}_\text{B-L}$ consists of giving all leptons $L,\nu_R, N_L, N_R$ vector-like charges normalized to one implying that B-L can be gauged and then breaking it via the vev of a scalar $\varphi$ with charge $|Q_\text{B-L}[\varphi]| > 2$  \cite{Ma:2014qra}, which has no direct couplings to fermions. This automatically forbids all renormalizable and effective operators leading to Majorana masses.

\subsubsection{Type II}\label{ec:typeII}
Instead of fermionic messengers fields that mix with the active neutrinos, for a Type II Seesaw one instead tilts a scalar potential slightly, to generate a tiny vev for a much heavier scalar field \cite{Lazarides:1980nt,Schechter:1980gr,Mohapatra:1980yp,PhysRevD.22.2860,Wetterich:1981bx}.
The first version of this mechanism for the vev of a Higgs doublet was considered by \cite{Ma:2000cc} in the context of a two-Higgs-doublet model with a soft mass term $\mu^2 \eta^\dagger H$ and was then applied to Dirac neutrinos in \cite{Davidson:2009ha} where  one introduces a new doublet $\eta$ that is a copy of the SM like doublet, but charged under $\text{U(1)}_\text{D}$ so that it couples to $\nu_R$
\begin{align}
    \mathcal{L}^\text{(II)} = Y_\nu \; \overline{L} \tilde{\eta} \nu_R  + \text{h.c.}\;.
\end{align}
This scenario was UV-completed in  \cite{Gu:2006dc,Bonilla:2016zef,Bonilla:2017ekt,Gu:2019ird}
by adding a singlet scalar $\sigma$ whose vev generates the soft mass term: The scalar potential can contain the following two terms
\begin{align}\label{eq:phasedependent}
    V^\text{(II)} \subset \begin{cases}   \kappa\;   \sigma H \eta^\dagger + \text{h.c.} \quad &\text{if} \quad Q_\text{D}[\nu_R] = -1, \\ \lambda_4\; \sigma^2 H \eta^\dagger + \text{h.c.}  \quad &\text{if} \quad Q_\text{D}[\nu_R] = -1/2, \end{cases} 
\end{align}
where we fixed $Q_D[\sigma] = 1$ in both cases and one can observe that the scalar potential is identical to the two possible choices for the DFSZ axion model \cite{Zhitnitsky:1980tq,Dine:1981rt}.
Note that these terms do not induce a mass for the imaginary component of $\sigma$, which becomes evident after diagonalizing the mass matrix for all pseudoscalars \cite{Bonilla:2016zef}. In the following we focus on the term $\propto \kappa$ which is linear in $\sigma$, because this allows us to discuss all Dirac Seesaw variants via the same effective operator \eqref{eq:WeinbergDirac}. All required fields  and charges can be found in table \ref{tab:charges-repsD}. If we assume that the positive $\mu_\eta^2  $ is the largest scale in the scalar potential then the aforementioned trilinear term will induce a tiny vev 
\begin{align}
    v_\eta \simeq  \frac{v_H}{\sqrt{2}}  \frac{\kappa\; v_\sigma}{\mu_\eta^2}
\end{align}
that lies  below the electroweak scale for $\kappa v_\sigma \ll \mu_\eta^2$ and can naturally accommodate the observed neutrino mass scale
\begin{align}\label{eq:TypeIISeesaw}
    m_\nu^\text{(II)} \simeq Y_\nu   \frac{\kappa \;v_\sigma }{2 \mu_\eta^2} v_H
\end{align}
without needing a small Yukawa coupling $Y_\nu$. A corresponding Feyman diagram can be found on the right hand side of \ref{fig:DiracSeesawI-II}. 
The Type II Seesaw scheme comes with an additional suppression factor $\kappa/\mu_\eta$ when compared to the fermionic Type I Seesaw (for $M_N \simeq \mu_\eta$), which is why for the same $v_\sigma$ the neutral component of the $\eta$-doublet can be made lighter than the vectorlike neutrinos $N$. An estimate illustrates the interplay of the various scales
\begin{align}
    m_\nu^\text{(II)} \simeq \SI{0.05}{\electronvolt} \cdot Y_\nu\cdot \left(\frac{\kappa}{\SI{100}{\giga\electronvolt}}\right) \cdot \left(\frac{v_\sigma}{\SI{400}{\tera\electronvolt}}\right)\cdot \left(\frac{10^{10}\;\text{GeV}}{\mu_\eta}\right)^2.
\end{align}
Again the one-loop correction to the active neutrino masses comes from closing the Higgs-singlet loop via an insertion of the mixed quartic coupling $\lambda_{\sigma H}$ 
\begin{align}
     \delta m^\text{(II)}_{\nu} = \frac{\lambda_{\sigma H}}{4\pi} \cdot   m_\nu^\text{(II)}
\end{align}
and it is subleading for perturbative values of $\lambda_{\sigma H}$.
This scenario is the most straightforward to further UV-complete in terms of a gauged  $\text{U(1)}_\text{B-L}$ since all gravitational and cubic anomalies will vanish, if one adds no other fermions than three generation of $\nu_R$ with the same canonical lepton number $Q_\text{B-L}[L] = -1$ as $L$. Renormalizable and effective Majorana masses for $L,\;\nu_R$ can then be forbidden if the scalar $\varphi$ dominantly responsible for the  $\text{U(1)}_\text{B-L}$-breaking has  a charge $|Q_\text{B-L}[\varphi]| > 2$  \cite{Ma:2014qra}.

\subsubsection{Type III}\label{sec:typeIII}

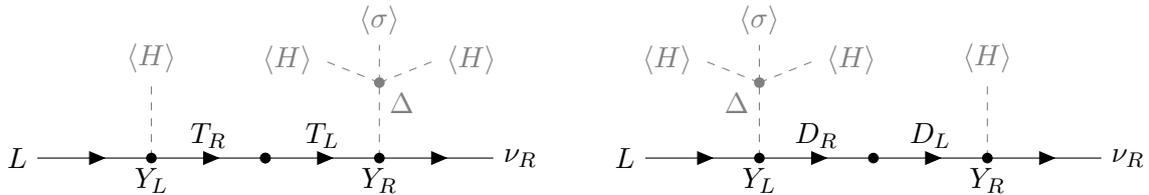
\begin{figure}[t]
 \centering
  \tikzset{
  blackline/.style={thin, draw=black, postaction={decorate},
    decoration={markings, mark=at position 0.6 with {\arrow[black]{triangle 45}}}},
     grayline/.style={thin, draw=gray, postaction={decorate},
    decoration={markings, mark=at position 0.6 with {\arrow[gray]{triangle 45}}}},
    blueline/.style={thin, draw=blue, postaction={decorate},
    decoration={markings, mark=at position 0.6 with {\arrow[blue]{triangle 45}}}},
    redline/.style={thin, draw=red, postaction={decorate},
    decoration={markings, mark=at position 0.6 with {\arrow[red]{triangle 45}}}},
    greenline/.style={thin, draw=green, postaction={decorate},
    decoration={markings, mark=at position 0.6 with {\arrow[green]{triangle 45}}}},    
    graydashed/.style={dashed, draw=gray, postaction={decorate},
    decoration={markings}},
   yellowdashed/.style={dashed, draw=orange, postaction={decorate},
    decoration={markings}},
    photon/.style={decorate, draw=red,
    decoration={coil,amplitude=12pt, aspect=0}},
  gluon/.style={dashed, decorate, draw=black,
    decoration={coil, segment length=5pt, amplitude=8pt}}
  line/.style={thick, draw=black, postaction={decorate},
    decoration={markings}}
}

\begin{tikzpicture}[node distance=1cm and 1cm]

%Case a
%particles 
\coordinate[label = left: $L$] (start1);
\coordinate[right=6cm of start1,label=right: $\nu_R$] (end);

%vertices
\coordinate[right=1.5cm of start1, label=below: $Y_L$] (H1);
\coordinate[right=1.5cm of H1] (sigma1);
\coordinate[right=1.5cm of sigma1, label=below: $Y_R$] (H2);

%vev insertions 
\coordinate[above=1cm of H1,label=above: $\color{gray} \braket{H}$] (vevH1);
\coordinate[above=1cm of H2, label = below right: $\color{gray} \Delta$] (vevH2);
\coordinate[above=0.5cm of vevH2,label=above: $\color{gray} \braket{\sigma}$] (vevH3);
\coordinate[left = 0.75 cm of vevH3] (vevH4);
\coordinate[right = 0.75 cm of vevH3] (vevH5);
\coordinate[below = 0.2  cm of vevH4, label= left: $\color{gray} \braket{H}$] (vevH44);
\coordinate[below = 0.2  cm of vevH5, label=  right: $\color{gray} \braket{H}$] (vevH55);
\coordinate[below=1cm of sigma1,label=below: ] (vevS1);

%axis
\draw[blackline] (start1)   -- (H1);
\draw[blackline] (H1)   -- (sigma1);
\draw[blackline] (sigma1)   -- (H2);
\draw[blackline] (H2)   -- (end);

%vev insertions
\draw[graydashed] (H1)   -- (vevH1);
\draw[graydashed] (H2)   -- (vevH2);
\draw[graydashed] (vevH2)   -- (vevH3);
\draw[graydashed] (vevH2)   -- (vevH44);
\draw[graydashed] (vevH2)   -- (vevH55);

%particle labels
\coordinate[right=0.75cm of H1, label=above: $T_R$];
\coordinate[left=0.75cm of H2, label=above: $T_L$];

%dots
\fill (H1) circle (2pt);
\fill (H2) circle (2pt);
\fill (sigma1) circle (2pt);
\fill  (sigma1) circle (2pt);
\fill[gray] (vevH2) circle (2pt);

%Case b
%particles 
\coordinate[right = 2 cm of end, label = left: $L$] (start2);
\coordinate[right=6cm of start2,label=right: $\nu_R$] (end2);

%vertices
\coordinate[right=1.5cm of start2, label=below: $Y_L$] (H1b);
\coordinate[right=1.5cm of H1b] (sigma1b);
\coordinate[right=1.5cm of sigma1b, label=below: $Y_R$] (H2b);

%vev insertions 
\coordinate[above=1cm of H1b,label = below left: $\color{gray} \Delta$] (vevH1b);
\coordinate[above=1cm of H2b,label=above: $\color{gray} \braket{H}$] (vevH2b);
\coordinate[below=1cm of sigma1b,label=below: ] (vevS1b);
\coordinate[above=0.5cm of vevH1b,label=above: $\color{gray} \braket{\sigma}$] (vevH3b);
\coordinate[left = 0.75 cm of vevH3b] (vevH4b);
\coordinate[right = 0.75 cm of vevH3b] (vevH5b);
\coordinate[below = 0.2  cm of vevH4b, label= left: $\color{gray} \braket{H}$] (vevH44b);
\coordinate[below = 0.2  cm of vevH5b, label=  right: $\color{gray} \braket{H}$] (vevH55b);

%axis
\draw[blackline] (start2)   -- (H1b);
\draw[blackline] (H1b)   -- (sigma1b);
\draw[blackline] (sigma1b)   -- (H2b);
\draw[blackline] (H2b)   -- (end2);

%vev insertions
\draw[graydashed] (H1b)   -- (vevH1b);
\draw[graydashed] (H2b)   -- (vevH2b);
\draw[graydashed] (vevH1b)   -- (vevH3b);
\draw[graydashed] (vevH1b)   -- (vevH44b);
\draw[graydashed] (vevH1b)   -- (vevH55b);

%particle labels
\coordinate[right=0.75cm of H1b, label=above: $D_R$];
\coordinate[left=0.75cm of H2b, label=above: $D_L$];

%dots
\fill (H1b) circle (2pt);
\fill (H2b) circle (2pt);
\fill (sigma1b) circle (2pt);
\fill[gray] (vevH1b) circle (2pt);

\end{tikzpicture}
  \caption{Diagrammatic representation of the dimension 5 operators for the Type III-a  Dirac Seesaw $\textit{(left)}$ involving new triplet fermions  and Type III-b Dirac Seesaw $\textit{(right)}$ involving new doublet fermions, both giving rise to Dirac masses for the active neutrinos.}
  \label{fig:DiracSeesawIII}
\end{figure}

The Type III Dirac Seesaw scenario is defined by the presence of a hypercharge-less scalar iso-triplet $\Delta$.
In the first case, which we call Type III-a, $\Delta$ has no couplings to SM fields and only interacts with $\nu_R$ and the right chiral component of a vector-like fermion triplet $T_{L,R}$ \cite{Popov:2022ldh} (see also \cite{Penedo:2022gej})
\begin{align}
     \mathcal{L}^\text{(III-a)} = Y_L \;\overline{L}  H  T_R + Y_R \; \overline{T_L}\Delta \nu_R + M_T\; \overline{T_L}T_R +  \text{h.c.}\;,
\end{align}
which was shown on the left of figure \ref{fig:DiracSeesawIII}.
Alternatively for the Type III-b model depicted on the right of \ref{fig:DiracSeesawIII},  one may consider a direct coupling of $\Delta$ to $L$ by introducing vector-like doublet leptons $D_{L,R}$ \cite{Borah:2022obi}
\begin{align}
     \mathcal{L}^\text{(III-b)} = Y_L \;\overline{L} \Delta D_R + Y_R \; \overline{D_L} \tilde{H} \nu_R + M_D\; \overline{D_L}D_R  + \text{h.c.}\;.
\end{align}
For both models all fields and charges where compiled in \ref{tab:charges-repsD}.
Since an iso-triplet scalar will spoil the custodial symmetry of the SM scalar potential, its vev modifies the SM $\rho$-parameter \cite{ROSS1975135,Veltman:1976rt} to be  \cite{Popov:2022ldh,Borah:2022obi}
\begin{align}
    \rho -1 = 8 \frac{v_\Delta^2}{v_H^2}.
\end{align}
The observed masses of the electroweak gauge bosons force  $\rho \equiv m_W^2 /( m_Z^2 \cos(\theta_W)^2)$ to be close to one, which requires $v_\Delta$ at or below the GeV-scale \cite{Kanemura:2012rs}.
Indeed one of the motivations for this scenario is the observed deviation in the $W$-boson mass  reported by the CDF collaboration \cite{CDF:2022hxs}, which can be stated in terms of the electroweak precision observable \cite{PhysRevLett.65.964,PhysRevD.46.381}  known as the  $T$-parameter  \cite{Borah:2022obi}
\begin{align}
    T \equiv \frac{\rho-1}{\alpha} = 0.17 \pm 0.020899.
\end{align}
The CDF-tension can then be explained by \cite{Popov:2022ldh}
\begin{align}
    v_\Delta \simeq \SI{4}{\giga\electronvolt}.
\end{align}
Such a low vev typically implies that the electrically neutral component of $\Delta$ has to be below the weak scale as well, which can phenomenologically challenging\footnote{If we assume that the neutral component of $\Delta$ is the Saxion, then the bound from Saxion isocurvature fluctuations would even require $m_{\Delta^0}\ll v_\Delta$ by many orders of magnitude.} similarly to the light scalar in neutrino-philic Two-Higgs-doublet models \cite{Wang:2006jy,Gabriel:2006ns,Davidson:2009ha,Sher:2011mx,Zhou:2011rc}. On top of that the pesudoscalar component of $\Delta$ playing the role of the Diraxion can typically have anomalous couplings to  the weak $\text{SU(2)}_\text{L}$ gauge bosons, implying a different phenomenology compared to the case, where the Diraxion comes from an SM singlet. To avoid these problems we add an additional scalar singlet $\sigma$. An attractive way to generate a small $v_\Delta$ while keeping the components of $\Delta$ ultra-heavy is the inclusion of a Type II suppression for  $v_\Delta$, completely analogous to the mechanism for a small $v_\eta$ in the previous paragraph \ref{ec:typeII}. The relevant scalar potential reads
\begin{align}
    V^\text{(III)} \subset \lambda_4\; \sigma^* H^\dagger\Delta H + \text{h.c.}
\end{align}
and one finds \cite{Borah:2022obi}
\begin{align}\label{eq:deltmass}
    v_\Delta \simeq \frac{\lambda_4}{2} \frac{v_H^2 v_\sigma}{\mu_\Delta^2},
\end{align}
which is suppressed compared to the electroweak scale as long as $\lambda_4 v_H v_\sigma \ll \mu_\Delta^2$. The required parameters turn out to be 
\begin{align}\label{eq:delta}
    v_\Delta \simeq \SI{4}{\giga\electronvolt}\cdot \left(\frac{\lambda_4}{0.3}\right) \cdot \left(\frac{v_\sigma}{10^8 \;\text{GeV}}\right)  \cdot \left(\frac{\SI{470}{\tera\electronvolt}}{\mu_\Delta}\right)^2.
\end{align}
Thus these versions of the Type III Dirac Seesaw correspond to the class of \enquote{nested Seesaws} \cite{Grimus:2009mm,Gu:2009hu,Gu:2019yvw,Gu:2019ogb,Gu:2019ird} with a built in double suppression from the heavy $T$ or $D$ masses and the tiny value of $v_\Delta$
\begin{align}
    m_\nu^\text{(III)} \simeq Y_L Y_R \; \frac{v_\Delta}{2 M_F} v_H, \quad \text{with} \quad M_F = \begin{cases}M_T \quad \text{for case (a)}\\M_D \quad \text{for case (b)} \end{cases}.
\end{align}
The heavy fermion masses $M_F$ can be estimated from 
\begin{align}
     m_\nu^\text{(III)} \simeq \SI{0.05}{\electronvolt} \cdot Y_L Y_R \cdot \left(\frac{v_\Delta}{\SI{4}{\giga\electronvolt}}\right) \cdot \left(\frac{10^{13}\;\text{GeV}}{M_F}\right)
\end{align}
and as expected the vector-like iso-mulitplet fermions $F=D,T$ are lighter than the singlet fermions $N$ from the Type I Dirac Seesaw due to the additional suppression from the scalar sector. If we make no assumption about the value of $v_\Delta$ we find that the observed neutrino mass scale would require $M_F\simeq \mu_\Delta \simeq 10^8\;\text{GeV}$ for order one couplings $Y_L, Y_R,\lambda_4$ and $v_\sigma =10^6\;\text{GeV}$. The one loop correction is found from closing the Higgs-loop via the $\lambda_H$ coupling or by using the Higgs-singlet loop from the mixed quartic $\lambda_{H\sigma}$. We obtain  a finite contribution of
\begin{align}
      \delta m^\text{(III)}_{\nu} =  \frac{\lambda_H+\lambda_{\sigma H}}{4\pi} \cdot   m_\nu^\text{(III)},
\end{align}
which is below the tree-level mass for perturbative values of $\lambda_{H\sigma}$.
A straightforward way to UV complete these models in terms of a gauged $\text{U(1)}_\text{B-L}$ is to assume that the heavy fermions are vector-like under both the SM gauge group as well as B-L.

\subsection{Saxion} 
The singlet scalar field has a renormalizable potential
\begin{align}
    V_\sigma = \lambda_\sigma \left( \left|\sigma\right|^2 -v_\sigma^2\right)^2,
\end{align}
where we minimized the potential by  balancing the tachyonic mass squared $-|\mu_\sigma|^2<0$ against the quartic coupling $\lambda_\sigma$ to define the nontrivial vev $v_\sigma \equiv |\mu_\sigma| /\sqrt{2 \lambda_\sigma}$.  One can decompose the scalar field as
\begin{align}
    \sigma = \frac{S + v_\sigma}{\sqrt{2}}e^{i\frac{\theta}{N}}, \quad \text{with} \quad  \frac{\theta}{N} \equiv \frac{a}{\braket{S}},
\end{align}
where we anticipated that  during inflation the radial mode will be displaced from its true vacuum  $\braket{S} = v_\sigma$ 
\begin{align}
  \braket{S}   = \begin{cases}  \gg v_\sigma \quad &\text{early universe}\\ v_\sigma \quad &\text{today} \end{cases}.
\end{align}
Apart from the SM and $\nu_R$ the particle spectrum will contain a massive  scalar state $S$ called the Saxion, whose mass squared reads
\begin{align}\label{eq:bare}
   m_S^2 \equiv 2\lambda_\sigma \;  v_\sigma^2.
\end{align}
The Saxion couplings were summarized in appendix \ref{sec:couplings}.

\subsection{Diraxion}\label{sec:Diraxion}
The most important ingredient of our scenario is the pseudoscalar Nambu-Goldstone-Boson (NGB)  $a$ of the $\text{U(1)}_\text{D}$ symmetry,  also known as the Diron \cite{Roncadelli:1983ty} or Diracon \cite{Bonilla:2016zef,Bonilla:2017ekt}, which can be understood of the Dirac equivalent of the well-known Majoron \cite{Gelmini:1980re,Chikashige:1980ui}. Quantum effects like instantons or classical explicit breaking of the underlying global symmetry can generate a mass for $a$ turning it into a Pseudo-NGB (PNGB), provided that its mass is parametrically below $v_\sigma$ and $m_S$.
In analogy to arguably the most well studied PNGB, the QCD-axion \cite{Peccei:1977hh,Weinberg:1977ma,Wilczek:1977pj,Kim:1979if,Shifman:1979if,Zhitnitsky:1980tq,Dine:1981rt}, we will call this  CP-odd scalar the \enquote{Diraxion}.
This PNGB will not couple to any SM fermions and does not have any anomalies with the SM gauge group. Global symmetries like our $\text{U(1)}_\text{D}$ are expected to be broken by non-perturbative quantum gravitational effects \cite{COLEMAN1988643,GIDDINGS1988854,GILBERT1989159}, as can be seen from wormhole arguments. Heuristically these effects are encoded in Planck-scale suppressed non-renormalizable operators, that explicitly violate a given symmetry. Furthermore, if we also want to use the same operator to induce a \enquote{kick} in the angular Diraxion direction from the Saxion oscillation, we need an operator with dimension larger than five \cite{Co:2020jtv}. While we normalized the number of units by which the vev of $\sigma$ breaks the global symmetry to be $Q_\text{D}[\sigma]=1$, the explicit breaking will in general occur with a different number of units and consequently we have to deal with a degenerate vacuum.
The most straight-forward way to see this, is to consider a potential of the form 
\begin{align}\label{eq:cos}
  c_N \sigma^N + \text{h.c.} = \left|c_N\right| \frac{S^N}{\sqrt{2}^N} \cos\left(\frac{N\;a}{v_\sigma}+\delta\right), \quad \text{with} \quad \delta \equiv \text{Arg}\left(c_N\right)
\end{align}
which implies that the Diraxion decay constant is not $v_\sigma$ but instead
\begin{align}
    f_a \equiv \frac{v_\sigma}{N},
\end{align}
where $N$ is the vacuum degeneracy factor also known as the Domain-Wall number and we define
\begin{align}
    \theta\equiv \frac{a}{f_a}.
\end{align}
One can deduce that the above interaction is invariant under an unbroken residual discrete $\mathcal{Z}_N$ symmetry, where $\sigma$ transforms as $\omega$ with $\omega^N =1$. 
This mismatch between spontaneous and explicit breaking will have considerable implications for cosmology since it can lead to long-lived domain walls that might overclose the universe  \cite{Press_1980,PhysRevLett.48.1156}.
However domain walls can be attached to cosmic strings from e.g. the $\text{U(1)}_\text{D}$ breaking and the resulting hybrid defect will be unstable for $N=1$ \cite{PhysRevD.24.2082,DAVIS1986225}. Furthermore even for $N>1$ one can use the same effective operator responsible for the Diraxion mass to induce domain wall decays to Diraxions \cite{PhysRevLett.48.1156}.  We begin with the simplest effective operator imaginable with more than five singlet fields
\begin{align} 
    N=d>5:\quad  V_\slashed{D} =  c_d \frac{\sigma^d}{M_\text{Pl.}^{d-4}} + \text{h.c.}\;,\quad m_a^{2} \equiv \frac{ d^4}{\sqrt{2}^{d-2}}\left|c_d\right|   \left(\frac{d\; f_a}{M_\text{Pl.}}\right)^{d-4} f_a^2,\label{eq:Diraxionmass}
\end{align}
and one can see that the Diraxion mass will automatically be suppressed compared to $f_a$ without needing a small dimensionless coefficient $c_d^{(i)}$ as long as $f_a \ll M_\text{Pl.}$ and $d>4$.
This approach comes with the downside of having a potentially large number of domain walls $N=d>5$. In section \ref{sec:defect} it will be explained why our setup is also viable for domain wall numbers larger than one. Hence for simplicity we will only consider the operator in equation \eqref{eq:Diraxionmass} for the rest of this paper as it relates two of the free parameters $N$ and $d$. More operators and a way to generate a specific value of $N$ can be found in appendix \ref{app:dirOp}. 
One can estimate e.g. for $N=6$ that
\begin{align}
    m_a \simeq \SI{9}{\kilo\electronvolt}\cdot \sqrt{c_6} \cdot \left(\frac{N}{6}\right)\cdot \left(\frac{N f_a}{10^6\;\text{GeV}}\right)^2,
\end{align}
however it turns out that we need a Diraxion mass between $1\;\text{meV}$ and $1\;\text{eV}$, which would require a tiny coefficient $c_6$. An exponential suppression of the Wilson-coefficient due to a large wormhole action \cite{Kallosh:1995hi,Alonso:2017avz} might accomplish this. Alternatively we illustrate  in appendix  \ref{app:dirOp} how the required effective operator could arise from a second scalar field $\varphi$, that we assume is charged under a gauged $\text{U(1)}_\text{B-L}$, which is responsible for the absence of Majorana masses. One such operator is
\begin{align}
    c_8 \frac{\sigma^6 \; \varphi^{*2}}{M_\text{Pl.}^{4}} + \text{h.c.}
\end{align}
and the required charges read $Q_\text{B-L}[\sigma]=3,\; Q_\text{B-L}[\varphi]=9$.
Here we assume that $\varphi$ has no direct couplings to the Seesaw messenger fields. For the Diraxion mass we find in this scenario that
\begin{align}
    m_a \simeq \SI{0.1}{\electronvolt}\cdot \sqrt{c_8}\cdot \left(\frac{N}{6}\right) \cdot \left(\frac{v_\text{B-L}}{10^9\;\text{GeV}}\right) \cdot \left(\frac{N f_a}{10^6\;\text{GeV}}\right)^2.
\end{align}
If we also charge $\sigma$ under B-L then we have to use more complicated chiral charge assignments than just $\pm 1$ for the fermion fields in order forbid Majorana masses and to avoid gauge anomalies. Since this will typically involve additional anomaly cancelling fermions, we refrain from writing down the explicit UV-completions.

\subsection{Diraxion couplings}
In the Type I Dirac Seesaw the Diraxion only has tree-level derivative couplings to $\nu_R$ as well as the heavy BSM fermions and scalars. In the Type II and III cases there are also couplings to SM states via mixing in the scalar sector. We focus on the low energy  couplings to SM particles and $\nu_R$.

\subsubsection{Derivative Couplings to neutrinos}
From the Saxion kinetic term we find that the Diraxion-Saxion coupling reads
\begin{align}\label{eq:derivDirax}
    \mathcal{L} = \frac{(S+N f_a)^2}{2} \partial_\mu \theta \partial^\mu \theta.
\end{align}
We  then remove the  phase of $\sigma$ by a field definition of the fields charged under $\text{U(1)}_\text{D}$, which leads to derivative couplings from their kinetic terms. At low energies only $\nu_R$ is in the plasma so we focus on its couplings
\begin{align}
    \mathcal{L}=  c_{\nu_R}\;\partial_\mu \theta\; \overline{\nu_R}\gamma^\mu \nu_R,
\end{align}
which read
\begin{align}\label{eq:deriv}
    c_{\nu_R} = \begin{cases}  \frac{1}{N}-\alpha^2 \quad &\text{Type I, II, III-a}\\
    \frac{1}{N} \quad &\text{Type III-b}
    \end{cases},
    \quad \text{with} \quad 
    \alpha^2 \equiv  
    \begin{cases}
    \frac{Y_R^2}{2} \frac{v_\sigma^2}{M_N^2} \quad &\text{Type I}\\
    2 \;\frac{v_\eta^2}{v_H\; N f_a }\quad &\text{Type II}\\
    \frac{Y_R^2}{2} \frac{v_\Delta^2}{M_T^2} \quad &\text{Type III-a}
    \end{cases}.
\end{align}
The couplings for the Type I and Type III-a Seesaw are reduced \cite{Co:2020xlh}, because here the $\nu_R$ with charge $Q_\text{D}[\nu_R]=-1$ mix  with $N_R$ and $T_R$ of charge $Q_\text{D}[N_R]=Q_\text{D}[T_R]=0$. For the Type II Seesaw there are no additional fermions for $\nu_R$ to mix with and the $D_R$ for a Type III-b Seesaw have the same charge $Q_\text{D}[T_R]=-1$ as $\nu_R$. In the Type II case however there is mixing in the scalar sector (see \eqref{eq:beta} in the next subsection) leading to the reduction in the coupling.
However for  Type III-b there is also the mass mixing between the neutral components of $D_R$ with $Q_\text{D}[D_R]=-1$ and  $L$ with $Q_\text{D}[L]=0$ that induces a derivative coupling $2 c_{\nu_L}= Y_R^2  v_\Delta^2/M_D^2$ for $L$.  It is evident that the parameters $\alpha^2\ll1$ due to the heavy messenger fermions and scalars, which is why in practise we will take $c_{\nu_R}\simeq 1/N$ for all Seesaws and $c_{\nu_L}\simeq 0$ for  Type III-b.
The effective coupling for the production of a single  Diraxion from a neutrino line is given by
\begin{align}\label{eq:axnu}
    g_{a\nu}= c_{\nu_R} \frac{m_\nu}{f_a}\simeq 5\times 10^{-17}\cdot \left(\frac{m_\nu}{\SI{0.05}{\electronvolt}}\right)\cdot   \left(\frac{10^6\;\text{GeV}}{N\;f_a}\right).
\end{align}
Limits from SN1987A on the energy loss and deleptonization exclude  $10^{-12}\lesssim g_{a\nu} \lesssim 10^{-5}$  \cite{Kachelriess:2000qc,Farzan:2002wx,Heurtier:2016otg} and for larger couplings the Diraxion would remain trapped inside the supernova. Meson decays exclude  only $g_{a\nu}\gtrsim 10^{-3}$ \cite{Pasquini:2015fjv}. It is obvious from \eqref{eq:axnu} that our setup is compatible with these bounds.

\subsubsection{Induced couplings to the visible sector}\label{sec:DirLoop}
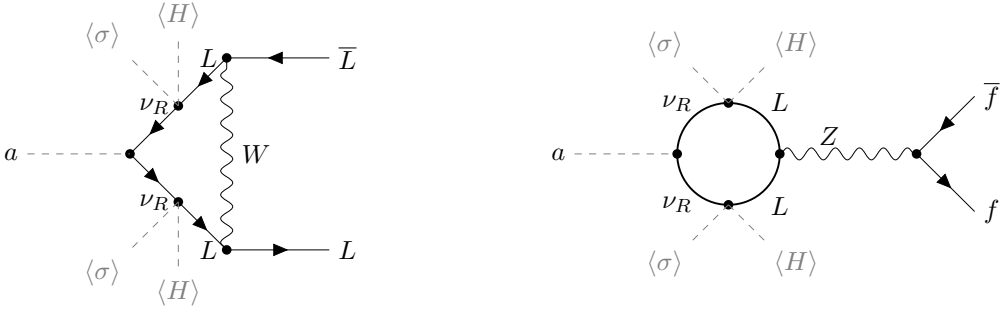
\begin{figure}[t]
    \centering
     \scalebox{0.9}{\tikzset{
  blackline/.style={thin, draw=black, postaction={decorate},
    decoration={markings, mark=at position 0.6 with {\arrow[black]{triangle 45}}}},
    graydashed/.style={dashed, draw=gray, postaction={decorate},
    decoration={markings}},
    photon/.style={decorate, decoration={snake}, draw=black},
     line/.style={thick, draw=black, postaction={decorate},
    decoration={markings}}
}

\NewDocumentCommand\semiloop{O{black}mmmO{}O{above}}
{%
\draw[#1] let \p1 = ($(#3)-(#2)$) in (#3) arc (#4:({#4+180}):({0.5*veclen(\x1,\y1)})node[midway, #6] {#5};)
}
% Syntax
%\semiloop[fermion][<draw options>]{<first node>}{<second node>}{<angle>}[<label>][<below, default: above>];

\begin{tikzpicture}[node distance=1cm and 1cm]

%vertex correction
\coordinate[label=left: $a$](start2);
\coordinate[right = 1.5 cm of start2](vertex2);
\coordinate[below right =1cm of vertex2,label=left: $\nu_R$] (Int1);
\coordinate[below right =1cm of Int1,label=left: $L$] (Int11);
\coordinate[above right =1cm of vertex2,label=left: $\qquad \qquad \nu_R$] (Int2);
\coordinate[above right =1cm of Int2,label=left: $\qquad \qquad L$] (Int22);

\coordinate[above   =1cm of Int2,label=above: $\textcolor{gray}{\braket{H}}$] (SS1);
\coordinate[above left =1cm of Int2,label=above left: $\textcolor{gray}{\braket{\sigma}}$] (HH1);
\coordinate[below   =1cm of Int1,label=below: $\textcolor{gray}{\braket{H}}$] (SS2);
\coordinate[below left =1cm of Int1,label=below left: $\textcolor{gray}{\braket{\sigma}}$] (HH2);

\fill (vertex2) circle (2pt);
\fill (Int1) circle (2pt);
\fill (Int2) circle (2pt);
\fill (Int11) circle (2pt);
\fill (Int22) circle (2pt);

\draw[graydashed] (start2)   -- (vertex2);
\draw[blackline] (vertex2) -- (Int1);
\draw[blackline] (Int1) -- (Int11);
\draw[blackline] (Int2) -- (vertex2);
\draw[blackline] (Int22) -- (Int2);
\draw[graydashed] (Int2)   -- (SS1);
\draw[graydashed] (Int2)   -- (HH1);
\draw[graydashed] (Int1)   -- (SS2);
\draw[graydashed] (Int1)   -- (HH2);

\coordinate[right=1.5cm of Int22,label=right: $\overline{L}$] (L2);
\coordinate[right=1.5cm of Int11, label= right: $L$] (H2);
\draw[blackline] (L2) -- (Int22);
\draw[blackline] (Int11) --(H2);

\coordinate[above= 1 cm of Int11] (sigma2);
\coordinate[right = 1.5 cm of vertex2,label=right: $W$] (vevS2);

\coordinate[left = 0.4cm of sigma2] (redhelp);
\coordinate[above = 1.5cm of redhelp] (red1);
\coordinate[below = 1.5cm of redhelp] (red2);
\draw[photon] (Int11)   -- (Int22);

% self energy 
\coordinate[right = 5 cm of vevS2, label=left: $a$] (start3);
\coordinate[right = 1.5cm of start3] (vertex3);
\draw[graydashed] (start3) -- (vertex3);
\fill (vertex3) circle (2pt);

\coordinate[right=1.5cm of vertex3] (vertex4);
\fill (vertex4) circle (2pt);
\semiloop[line]{vertex3}{vertex4}{0}[][above];
\semiloop[line]{vertex4}{vertex3}{180}[][below];
\coordinate[above right= 1.06cm of vertex3, label=above left: ] (vertexup);
\coordinate[below right= 1.06cm of vertex3, label=above left: ] (vertexdown);
\fill (vertexup) circle (2pt);
\fill (vertexdown) circle (2pt);

\coordinate[above left= 0.75cm of vertexup, label=above left: $\textcolor{gray}{\braket{\sigma}}$] (HH3);
\coordinate[above right= 0.75cm of vertexup, label=above right: $\textcolor{gray}{\braket{H}}$] (SS3);
\coordinate[below left= 0.75cm of vertexdown, label=below left: $\textcolor{gray}{\braket{\sigma}}$] (HH4);
\coordinate[below right= 0.75cm of vertexdown, label=below right: $\textcolor{gray}{\braket{H}}$] (SS4);

\coordinate[above=0.5cm of vertex3, label= above: $\nu_R$] (label1);
\coordinate[below=0.5cm of vertex3, label= below: $\nu_R$] (label2);

\coordinate[above=0.5cm of vertex4, label= above: $L$] (label3);
\coordinate[below=0.5cm of vertex4, label= below: $L$] (label4);

\draw[graydashed] (vertexup) -- (HH3);
\draw[graydashed] (vertexup) -- (SS3);
\draw[graydashed] (vertexdown) -- (HH4);
\draw[graydashed] (vertexdown) -- (SS4);

\coordinate[right= 1cm of vertex4, label=above left: $Z$] (sigma3);
\coordinate[right= 1cm of sigma3] (vertex5); 
\coordinate[below= 0.6 cm of sigma3, label=below:] (vevS3);

\draw[photon] (vertex4) -- (vertex5);
\fill (vertex5) circle (2pt);

\coordinate[above right =1.2cm of vertex5,label=right: $\overline{f} $] (L3);
\coordinate[below right =1.2cm of vertex5,label=right: $f$] (H3);

\draw[blackline] (vertex5) -- (H3);
\draw[blackline] (L3) -- (vertex5);

\end{tikzpicture}}  

     \caption{One loop diagrams for the coupling of the Diraxion to SM fermions involving two insertions of the Dirac Weinberg operator. The first diagram generates a coupling to the lepton doublet only, whereas the second one involves all fermions $f=L,e_R, Q, u_R, d_R$.}
     \label{fig:loops}
\end{figure}

For the Type I case our Diraxion only couples to SM quarks and leptons via the one-loop diagrams in  \ref{fig:loops} and here we recast the results of \cite{Garcia-Cely:2017oco,Heeck:2019guh} obtained for a Majoron in a conventional Type I Seesaw. For us the pseudo-scalar coupling to electrons is the most relevant and it turns out to be 
\begin{align}
    g_{ae} \simeq \frac{1}{16\pi^2} \frac{m_e}{v_H} \frac{m_\nu^2}{v_H N f_a} \simeq 10^{-37}\cdot \left(\frac{m_\nu}{\SI{0.05}{\electronvolt}}\right)^2 \cdot \left(\frac{10^6\;\text{GeV}}{N f_a}\right).
\end{align}
Stellar cooling arguments from the sun as well as red giants exclude $g_{ae}\gtrsim 10^{-13}$ \cite{Viaux:2013hca, Giannotti:2017hny,2018MNRAS.478.2569I,2018arXiv180210357S}, which is not in conflict with our scenario.
By closing the charged fermion line and attaching two photons to the one-loop diagrams depicted in \ref{fig:loops}, we obtain the two-loop Diraxion-photon coupling. Since our choice of $\text{U(1)}_\text{D}$ symmetry has no mixed anomalies with the SM gauge sector, the resulting coupling will  be proportional to $m_a^2/m_\psi^2$ \cite{Garcia-Cely:2017oco}, where $m_\psi$ is the mass of the fermion running in the loop. Consequently the dominant contribution comes from the electrons being the lightest SM fermion and it reads \cite{Garcia-Cely:2017oco,Heeck:2019guh}
\begin{align}\label{eq:gammaI}
   \left| g_{a\gamma\gamma}^{(\text{I})} \right| \simeq \frac{\alpha}{64\pi^3\;N f_a} \left(\frac{m_\nu}{v_H }\right)^2 \left(\frac{m_a}{m_e }\right)^2 \simeq  \frac{10^{-52}}{\text{GeV}}\cdot \left(\frac{m_\nu}{\SI{0.05}{\electronvolt}}\right)^2 \cdot \left(\frac{10^6\;\text{GeV}}{N f_a}\right)\cdot  \left(\frac{m_a}{\SI{10}{\milli\electronvolt}}\right)^2.
\end{align}
In contrast to the case of a QCD axion the above has no contribution from mixing with the pions, as the Type I Diraxion has no tree-level coupling to quarks. Due to their dependence on the tiny neutrino masses squared we find that these couplings are below all relevant constraints.
In the Type II Dirac Seesaw case the Diraxion mixes with the  would-be-Nambu Goldstone component of the SM-like Higgs $H$ that becomes the longitudinal mode of the $Z$-boson via the trilinear term in the scalar potential \eqref{eq:phasedependent}. This leads to flavor-independent derivative couplings to the electrically charged fermions 
\begin{align}
    \mathcal{L}= \partial_\mu \theta\: \sum_{\psi= e_i,u_i,d_i } c_\psi  \overline{\psi}\gamma^\mu \psi,
\end{align}
which read \cite{Bonilla:2016zef}
\begin{align}\label{eq:beta}
    c_u = - c_d = -c_e =  2\frac{v_\eta^2}{v_H\;    N f_a}\simeq 2 \cdot 10^{-29} \cdot \left(\frac{v_\eta}{m_\nu}\right)^2 \cdot \left(\frac{m_\nu}{\SI{0.05}{\electronvolt}}\right)^2 \cdot \left(\frac{10^6\;\text{GeV}}{N f_a}\right).
\end{align}
For completeness we mention that the coupling to two photons is found at one loop in analogy with the singlet-triplet Majoron model \cite{Schechter:1981cv,Choi:1989hj} and one obtains \cite{Bazzocchi:2008fh}
\begin{align}\label{eq:gammaII}
\left| g_{a\gamma\gamma}^{(\text{II})}\right| \simeq \frac{\alpha}{12\pi\;N f_a} \left(\frac{v_\eta}{v_H }\right)^2 \left(\frac{m_a}{m_e }\right)^2 = \frac{16 \pi^2}{3} \cdot \left(\frac{v_\eta}{m_\nu}\right)^2 \cdot  \left| g_{a\gamma\gamma}^{(\text{I})} \right|.
\end{align}
These couplings are larger than the previous loop-induced ones -  especially for $m_\nu \ll v_\eta$ - but in general they are too tiny to be of phenomenological relevance. Considering the Type III cases we expect the previous results to hold under the replacement $v_\eta \rightarrow v_\Delta$ implying significantly larger couplings, however we refrain from a detailled analysis since the Type III cases will not be compatible with our intended cosmology (see the discussion between equation \eqref{eq:thermmass}).

\section{Two-field dynamics}\label{sec:Dynamics}
\subsection{Initial Saxion field value}
We consider a radial mode whose vev is displaced during inflation from its minimum as $S_i\gg N f_a$ and treat $S_i$ as the intial condition for the evolution of the Saxion. 
Here we present two ways to induce the displacement of the Saxion field value during inflation, and in appendix \ref{app:initial} we also discuss a way to avoid or relax the isocurvature constraints. In the following we take the initial field value $S_i$ as a free parameter and $m_S(S_i)^2\simeq 3\lambda_\sigma S_i^2$ denotes the effective mass squared for a quartic potential
\begin{align}
    m_S(S_i)^2 \equiv \frac{3}{2} m_S^2  \left(\frac{S_i}{N f_a}\right)^2 \label{eq:Saxionmass},
\end{align}
where we defined the Saxion mass squared in our vaccum today to be $m_S^2$. The complex singlet scalar is assumed to be a spectator field during inflation.

\subsubsection{Quantum fluctuations}\label{app:quant}
It has been long known \cite{Bunch:1978yq,Hawking:1981fz,Vilenkin:1982de,Vilenkin:1982wt,Linde:1982uu,Starobinsky:1982ee,Vilenkin:1983xp,Linde:1983mro,Linde:1985gh,Enqvist:1987au} that scalar fields with very shallow potentials undergo large field excursions during inflation as a consequence of quantum fluctuations. We can take this potentially large field value as an initial condition $S_i$ for the Saxion field. When the Hubble rate during inflation $H_I$ is much larger than the effective mass, quantum fluctuations grow the expectation value of $S$.
The field excursion of $S$ can be determined from the variance $\braket{S^2}$ of the Starobinsky-Yokoyama distribution \cite{Starobinsky:1994bd}
\begin{align}
    S_i \simeq \sqrt{\braket{S^2}} = \left(\frac{1}{4\pi^2}\right)^\frac{1}{4} \sqrt{\frac{N f_a}{m_S}} H_I,
\end{align}
where $m_S$ is the bare mass defined in \eqref{eq:bare}. From this it becomes apparent, why a large field excursion $S_i$ demands a very flat potential, meaning a small $m_S$ (or equivalently $H_I \gg m_S(S_i)$). The abvove estimate assumes that only a single field is excited in this way \cite{Allahverdi:2012ju}. It requires $N_e\gtrsim (H_I/m_S(S_i))^2$ many e-folds of inflation for $S$ to get pushed to the value $S_i$ \cite{Starobinsky:1994bd}.
One finds that the correlation of the associated fluctuations scales as \cite{Linde:1983mro}
\begin{align}
    l \simeq H_I^{-1}\cdot \exp{\left(\frac{3 H_I^2}{m_S(S_i)^2}\right)}
\end{align}
and it is much larger than the horizon size $H_I^{-1}$ as long as $H_I \gg m_S(S_i)$, justifying the treatment in terms of a homogenous Saxion condensate.

\subsubsection{Non-minimal coupling to gravity}\label{app:R}
While quantum fluctuations required $m_S(S_i)\ll H_I$ a non minimal coupling to gravity can lead to a Hubble dependent mass 
$m_S(S_i)\simeq H_I$. This kind of mass term was first suggested in the context of supersymmetry breaking from finite energy density effects \cite{Dine:1995uk,Dine:1995kz} in the early universe. The same net effect can be generated by coupling the singlet scalar to the Ricci scalar
\begin{align}
  V \supset  c_R\; R\; |\sigma|^2,
\end{align}
whose value for a Friedmann-Lemaitre-Robertson-Walker metric depends on the equation of state of the  dominant energy density driving the cosmic expansion \cite{Opferkuch:2019zbd}
\begin{align}
    R = -3(1-3\omega) H^2 = 
    \begin{cases}
      -3 H^2\quad &\text{for MD} \quad (\omega =0)\\   
      0\quad &\text{for RD} \quad (\omega =\frac{1}{3})\\  
      -12 H^2\quad &\text{for infl.}\quad (\omega =-1)
    \end{cases}.
\end{align}
The idea is to balance this tachyonic mass squared during inflation against the quartic term $\lambda_\sigma  S^4/4$, which requires $c_R>0$. In order  to have $m_S(S_i)^2 = - H_I^2$ we take $c_R = 1/12$ and find 
\begin{align}
    S_i = \sqrt{2} \left(\frac{N f_a}{m_S}\right) H_I.
\end{align}
The Saxion remains stuck in this value even at the onset of radiation domination because of the equilibrium between the Hubble friction and the slope of the potential \cite{Kawasaki:2011zi,Kawasaki:2012qm,Kawasaki:2012rs}. 
 This can be expressed as the condition $m_S(S_i) < 3 H_I$ \cite{Co:2022kul}, which implies the bound
\begin{align}
    m_S < 120 \;\text{GeV}\cdot \left(\frac{N f_a}{10^6\;\text{GeV}}\right) \cdot \left(\frac{0.1\; M_\text{Pl.}}{S_i}\right) \cdot \left(\frac{H_I}{\SI{6e13}{\giga\electronvolt}}\right),
\end{align}
where we used the upper limit on the Hubble rate during inflation $H_I<\SI{6e13}{\giga\electronvolt}$ from the limit on the tensor-to-scalar ratio \cite{Planck:2018jri} and this limit on $m_S$ applies independently of how the initial field value is generated.
Before we close let us note that an epoch of intermediate matter domination  after inflation with $R= -3 H^2$, such as e.g. the well motivated curvaton scenario \cite{Enqvist:2001zp,Lyth:2001nq,Moroi:2001ct}, could also be used to induce the large field value, which would be entirely free from inflationary isocurvature perturbations.
The absence of fluctuations during inflation requires and effective mass during inflation that is larger than $H_I$, which could be realized via the couplings to the inflaton introduced in appendix \ref{app:initial}. 

\subsection{Saxion oscillation}
 The Hubble rate during radiation domination (RD) and  non-instantaneous reheating (RH), which corresponds to matter domination, 
\begin{align}
    H(T) = \sqrt{\frac{8\pi^3 g_*(T)}{90 M_\text{Pl.}^2}}\cdot 
    \begin{cases}
        T^2\quad& T<T_\text{RH}\\
        T^4/T_\text{RH}^2\quad&T\geq T_\text{RH}
    \end{cases}
\end{align}
in terms of the number of  relativistic degrees of freedom $g_*$. 
The radial mode starts oscillating with a frequency $m_S(S_i)$ when $3 H(T_\text{osc.}) = m_S(S_i)$ which determines for radiation domination
\begin{align}\label{eq:TOSC}
    T_\text{osc.}  &= \left(\frac{135}{8\pi^3\; g_*(T_\text{osc.})}\right)^\frac{1}{4} \sqrt{M_\text{Pl.} m_S} \sqrt{\frac{S_i}{N f_a}}
    \\
    &\simeq \SI{9.5e14}{\giga\electronvolt}\cdot \sqrt{\frac{m_S}{\SI{1}{\giga\electronvolt}}}\cdot \sqrt{\frac{10^6\;\text{GeV}}{N f_a}} \cdot  \sqrt{\frac{S_i}{0.1 \; M_\text{Pl.}}}.
\end{align}
The oscillation occurs after reheating ($T_\text{osc.} <T_\text{RH}$) as long as 
\begin{align}\label{eq:condRH}
    m_S < \frac{8\pi^3 g_*(T_\text{osc.})}{135} \frac{N f_a}{S_i} \frac{T_\text{RH}^2}{M_\text{Pl}} \simeq \SI{1.1}{\giga\electronvolt}\cdot \left(\frac{N f_a}{10^6\;\text{GeV}}\right) \cdot \left(\frac{0.1\;M_\text{Pl.}}{S_i}\right) \cdot \left(\frac{T_\text{RH}}{10^{15}\;\text{GeV}}\right)^2.
\end{align}
For a quartic potential during radiation domiantion the field value  $S \sim 1/a$ redshifts like radiation as a function of temperature for $S> N f_a$
\begin{align}
    S(T>T_S) = S_i \;\frac{T}{T_\text{osc.}}.
\end{align}
After the Saxion starts to relax to its minimum $S= N f_a$ at the temperature
\begin{align}\label{eq:TS}
    T_S \equiv T_\text{osc.}\; \frac{N f_a}{S_i} \simeq  \SI{950}{\giga\electronvolt}\cdot \sqrt{\frac{m_S}{\SI{1}{\giga\electronvolt}}}\cdot \sqrt{\frac{N f_a}{10^6\;\text{GeV}}} \cdot \sqrt{\frac{0.1\;M_\text{Pl.}}{S_i}}
\end{align}
the potential is dominated by the constant mass term in \eqref{eq:bare} and the field value of the now non-relativistic condensate redshifts as
\begin{align}
    S(T\leq T_S) = N f_a \left(\frac{T}{T_S}\right)^\frac{3}{2}.
\end{align}
For oscillations before the completion of reheating we find
\begin{align}\label{eq:TOSCRH}
    T_\text{osc.}^\text{RH} &= \left(\frac{135}{8\pi^3\; g_*(T_\text{osc.})}\right)^\frac{1}{8} \left(M_\text{Pl.} m_S  \frac{S_i}{N f_a}\right)^\frac{1}{4}\sqrt{T_\text{RH}}\\
    &\simeq   \SI{1.5e15}{\giga\electronvolt}\cdot \left(\frac{m_S}{\SI{5}{\giga\electronvolt}}\right)^\frac{1}{4}\cdot \left(\frac{10^6\;\text{GeV}}{N f_a}\right)^\frac{1}{4} \cdot  \left(\frac{S_i}{0.1\;M_\text{Pl.}}\right)^\frac{1}{4} \cdot \sqrt{\frac{T_\text{RH}}{10^{15}\;\text{GeV}}}
\end{align}
instead. During reheating we find the scaling law $S\sim 1/a \sim T^{8/3}$ because during matter domination $T\sim 1/a^{3/8}$.  For most of our parameter space $T_S$ is below $T_\text{RH}$, so the estimate in \eqref{eq:TS} applies with $ T_\text{osc.}^\text{RD}$ replaced by $T_\text{RH}$ and $S(T_\text{RH})$ obtained from the previous scaling. We can also compute the number density of the oscillating and thereby non-relativistic   Saxion condensate
\begin{align}
    n_S \equiv \frac{m_S(S)}{2} S^2,
\end{align}
which follows from its  energy density $\rho_S = V_\sigma(S) = m_S n_S$. 
We want the evolution of the cosmological background to proceed as usual, which is why we demand that there is no period of  inflation due the large Saxion field value \cite{Graziani:1988bp}
\begin{align}\label{eq:infl}
    V_\sigma(S_i) < 3 H^2(T_\text{osc.})^2 M_\text{Pl.}^2.
\end{align}
By making use of $m_S(S_i) = 3 H(T_\text{osc.})$ we find that this implies \cite{Co:2020dya,Gouttenoire:2021jhk}
\begin{align}
    S_i<2  M_\text{Pl.}.
\end{align}
The evolution of the Saxion field during radiation domination, its energy density and its equation of state parameter $\omega_S$ can be summarized by the following scaling relations
\begin{align}
    S \sim \begin{cases}a^{-1}\\ a^{-\frac{3}{2}} \end{cases}, \quad \rho_S\equiv \frac{m_S(S)^2}{2} S^2 \sim \begin{cases} a^{-4}\\ a^{-3}\end{cases}, \quad \omega_S = \begin{cases} \frac{1}{3}\quad &\text{for}\quad S\gg N f_a\\0\quad &\text{for}\quad S\simeq N f_a \end{cases}.
\end{align}
 In this work we assume that the masses of all additional particles are given by their tree-level expressions and do not get modified due to the large initial field value $S_i$. Specifically this means that 
\begin{align}\label{eq:couplings}
    S_i < 
    \begin{cases}
        M_N / Y_R \quad &\text{Type I}\\
        \mu_\eta^2 / \kappa \quad &\text{Type II}\\
        \mu_\Delta^2 / (\lambda_4 v_H) \quad &\text{Type III}\\
    \end{cases},
\end{align}
If this condition was violated $\nu_R$ and $N_L$ would form a Dirac fermion of mass $Y_R S_i$ or $H,\;\eta$ would mix strongly.  Furthermore we impose that the heavy messenger fields are not present in the plasma at any point in time
\begin{align}\label{eq:thermmass}
    \text{Max}\left(T_\text{osc.},T_\text{RH}, T_\text{max},\frac{H_I}{2\pi}\right) < 
    \begin{cases}
        M_N  \quad &\text{Type I}\\
        \mu_\eta  \quad &\text{Type II}\\
        \mu_\Delta,\; M_F \quad &\text{Type III}\\
    \end{cases},
\end{align}
where $H_I / (2\pi)$ is the Gibbons-Hawking temperature during inflation \cite{PhysRevD152738}, $T_\text{RH}$ the radiation bath temperature at the end of reheating and
\begin{align}\label{eq:max}
    T_\text{max} \equiv \beta \sqrt{\sqrt{\frac{3}{8\pi}} H_I M_\text{Pl.}}, \quad \text{with} \quad \beta \in [0,1]
\end{align}
the maximum temperature during (non-instantaneous) reheating \cite{Hertzberg:2008wr,Garcia:2017tuj}, which can be much larger than  $T_\text{RH}$. CMB observation constrain the Hubble rate during inflation to be $H_I<\SI{6e13}{\giga\electronvolt}$ \cite{Planck:2018jri} and thus puts an upper limit on  the energy scale during inflation from which one can deduce  $T_\text{RH}< T_\text{max}< \SI{6.5e15}{\giga\electronvolt}$ \cite{Co:2022kul}. Note that it will in general be hard to satisfy \eqref{eq:thermmass} for the Type III scenario due to the smallness of $\mu_\Delta$ (see equation \eqref{eq:deltmass})  unless we assume a low Hubble scale during inflation and a small reheating temperature. As long as \eqref{eq:thermmass} is satisfied the Saxion will not receive a thermal mass of the order $Y T$, where $Y$ is a Yukawa or the square root of quartic coupling from \eqref{eq:couplings}. However since  the heavy $\eta$ or $\Delta$  coupling to $\sigma$  have electroweak gauge interactions, integrating them out can potentially modify the logarithmic running of the weak gauge couplings manifesting itself in a so called thermal logarithmic potential \cite{Anisimov:2000wx,Fujii:2001zr}
\begin{align}
    V_\text{Log.}(T) = c_W \alpha_W^2 T^4 \text{log}\left(\frac{S^2}{T^2}\right).
\end{align}
Here $c_W$ is a model-dependent coefficient from the couplings of $\eta, \Delta$ to $\sigma$. Demanding that this correction is subdominant to the quartic potential, so that the Saxion oscillates around its mass in \eqref{eq:Saxionmass}, leads to the condition
\begin{align}
    S_i >  \alpha_W  \sqrt{\frac{c_W}{g_*(T_\text{osc.})}} M_\text{Pl.} \simeq 10^{16}\;\text{GeV} \cdot \sqrt{c_W}  \cdot \left(\frac{\alpha_W}{1/100}\right) \cdot  \sqrt{\frac{100}{g_*(T_\text{osc.})}}.
\end{align}
To be conservative we will take $S_i > 10^{16}\;\text{GeV}$ throughout this work. Note that this bound will be absent for the Type I Seesaw, because the heavy vectorlike neutrinos have no weak gauge couplings. In section \ref{sec:quantcorr} we investigate the one loop quantum corrections to the Saxion quartic.

\subsection{Generating the Diraxion rotation}
We assume that the global symmetry is spontaneously broken with a vev $S_i$ during inflation that is significantly larger than today. Due to this vev, Planck scale suppressed explicit symmetry breaking will be active. As longs as the mass of the angular mode (during inflation) is small compared to the radial mode, we can consider  $\text{U(1)}_\text{D}$ as an approximate symmetry (during inflation). The explicit breaking then becomes irrelevant as the vev relaxes to its true vacuum at $N f_a$ today. During inflation the Diraxion will have a mass 
\begin{align}
    m_a(S_i)^2 \equiv m_a^2 \left(\frac{S_i}{N f_a}\right)^{N-2}.
\end{align}
 In order for our description in terms of an approximate $\text{U(1)}_\text{D}$ symmetry to apply we require that the Diraxion mass during inflation is smaller than the Saxion mass  in \eqref{eq:Saxionmass} \cite{Co:2022aav}, which leads to 
\begin{align}
    m_a < m_S \left(\frac{N f_a}{S}\right)^\frac{N-4}{2} \label{eq:effdesc}.
\end{align}
For $N=4$ this would reduce to the  requirement for the  PNGB mass compared to the mass for the Higgs scalar of the underlying symmetry $m_a<m_S$. We further deduce that we can not take $N$ to be arbitrarily large in order not to spoil the previous relation.
Next one defines the charge yield or asymmetry of the scalar condensate as 
\begin{align}
    n_\theta \equiv \frac{\dot{\theta} S^2}{N^2},
\end{align}
which is conserved as long as there are no interactions that explicitly violate the underlying global $\text{U(1)}_\text{D}$ symmetry. One can understand the fact that only a rotating condensate can carry a charge compared to an oscillating one from the observation that an oscillation is a superposition of two rotations in opposite directions \cite{Allahverdi:2008pf}.
The  Planck suppressed effective operators in $V_\slashed{D}$ of section \ref{sec:Diraxion}, that are responsible for generating the Diraxion mass,  convert a part of the oscillatory motion of the radial mode into a velocity for the angle. After the radial mode has decreased via redshifting and eventually settled to its minimum, the Planck suppressed operators $\sigma^N$ become negligible for $N>5$ \cite{Co:2020jtv}.The corresponding equation of motion reads \cite{Co:2020jtv}
\begin{align}
    \dot{n}_\theta + 3 H n_\theta &= \frac{i}{N}\left(\sigma^* \frac{\partial V_\slashed{D}}{\partial \sigma^*} -\sigma \frac{\partial V_\slashed{D}}{\partial \sigma}\right),\\
    &=\frac{2 |c^{(i)}|}{\sqrt{2}^N} \frac{ S_i^N }{ M_\text{Pl.}^{N-4}}\sin(\theta+\delta).
\end{align}
Due to the redshifting of $S$ the majority of the condensate charge is produced at the start of the Saxion oscillations over one Hubble time $t\simeq 1/H(T_\text{osc.})\simeq 3/m_S(S_i)$ \cite{Co:2020dya} and we find 
\begin{align}\label{eq:chargeasym}
    n_\theta(T_\text{osc.}) \simeq \frac{\dot{n}_\theta (T_\text{osc.})}{m_S(S_i)}  = \frac{6 |c^{(i)}|}{\sqrt{2}^N} \left(\frac{S_i}{ M_\text{Pl.}}\right)^{N-4} \frac{S_i^4}{m_S(S_i)} \sin(\theta_i+\delta).
\end{align}
Here $\theta_i$ is the initial angle of the Diraxion also known as the misalignment angle. 
It is customary (see e.g. \cite{Harigaya:2014tla,Harigaya:2016hqz}) to define the  eccentricity parameter $\varepsilon$ \cite{Co:2019wyp,Co:2020jtv}  
\begin{align}\label{eq:varepsilon}
    \varepsilon \equiv  \frac{n_\theta\; m_S(S_i) }{V_\sigma(S_i)} = \frac{24}{N^2}   \left(\frac{m_a}{m_S}\right)^2 \left(\frac{S_i}{N f_a}\right)^{N-4}\sin(\theta_i+\delta) \simeq \frac{12}{N}\frac{\partial V_\slashed{D}/ \partial S}{\partial V_\sigma / \partial S},
\end{align}
where we made use of the definition of the Diraxion mass in \eqref{eq:Diraxionmass} and the Saxion mass in \eqref{eq:Saxionmass}.
The last approximate equality holds for $\sin(\theta_i+\delta)\simeq \mathcal{O}(1)$, which  suppresses axion isocurvature perturbations \cite{Co:2020jtv} (see section \ref{sec:iso}).
This parameter   $\varepsilon \leq 1$  has the following physical interpretation:
An orbit with $\varepsilon=1$ corresponds to a perfectly circular rotation, whereas one would have $\varepsilon=0$ for a pure oscillation in the Saxion direction. Our scenario is different from the case of QCD-Axiogenesis  \cite{Co:2019wyp} because here the \enquote{kick} in the angular direction and the Diraxion mass originate from the same operator, so that we can in principle choose $\varepsilon\simeq 1$. For a quartic potential and a QCD axion this is in general not possible \cite{Co:2020dya}, because of the axion quality problem \cite{PhysRevD.46.539,KAMIONKOWSKI1992137,Holman:1992us} from the Planck suppressed operators, which can shift the minimum of the QCD axion to too large angles compared to the limit from the neutron's electric dipole moment.   However for $\varepsilon=1$ two complications arise:
On the one hand we can no longer neglect $V_\slashed{D}$ compared to the quartic term when determining the Saxion mass and initial field value, so our present description based on \eqref{eq:Saxionmass} breaks down. Additionally for larger $V_\slashed{D}$ one can no longer neglect the angular gradient compared to the Hubble friction, so that the Diraxion stops being overdamped and starts to relax to the minimum of its potential given by (see also the discussion in section \ref{sec:DMiso})
\begin{align}\label{eq:mintheta}
    \frac{\partial V_\slashed{D}}{\partial \theta }\Big|_{\theta_i} \sim -m_a(S_i)^2 S_i^2 \;\sin(\theta_i + \delta) \overset{!}{=} 0.
\end{align}
If the Diraxion is trapped in such a vacuum, the rotation of the condensate will not occur or be significantly damped \cite{Co:2020jtv}, which is why we demand $\varepsilon<1$
\begin{align}\label{eq:masses2}
    m_a < \frac{N}{2\sqrt{6}} \left(\frac{N f_a}{S_i}\right)^\frac{N-4}{2} m_S.
\end{align}
This condition intuitively means that the Diraxion should not start oscillating before the Saxion \cite{Co:2020jtv}, which is nothing more than the requirement $m_a(S_i)<m_S(S_i)$ we already imposed in \eqref{eq:effdesc} to have a Pseudo-Nambu-Goldstone boson, hence the bounds being the same up to a prefactor. Both conditions automatically make sure that Diraxion is in motion \cite{Co:2022aav} since its velocity $\dot{\theta} \simeq m_S(S) $ is always larger than its mass $m_a(S)$.
In a quartic potential the angular and radial modes redshift as radiation for $S\gg N f_a$ ($\omega_S=\omega_\theta=1/3$).
For cosmological reasons that will be elaborated upon in section \ref{sec:therm}, the Saxion needs to be thermalized  and will typically loose its energy to the SM plasma. After the damping of the Saxion oscillations only the angular rotation, which is now perfectly circular, remains. 
The angular rotation is stable because of $\text{U}(1)_\text{D}$-charge conservation. It minimizes the free energy \cite{Co:2019wyp} and can loose energy to thermal bath via interactions with particles charged under $\text{U}(1)_\text{D}$ leading to the production of fermionic asymmetries. However as long as the Noether charge of the condensate $\dot{\theta} S^2$ is larger than the typically produced fermionic asymmetry $\dot{\theta} T^2$, it is energetically favoured for the rotating condensate to retain most of its charge compared to the production of fermions. This requirement can be expressed as  \cite{Laine:1998rg,Co:2019wyp} 
\begin{align}
   N  f_a \gg T_S,
\end{align}
where $T_S$ defined in \eqref{eq:TS} is the temperature at which the Saxion reaches its minimum $S= N f_a$. This condition implies that $T_\text{osc.}\ll  S_i$ for a quartic potential with the oscillation temperature from \eqref{eq:TOSC} and it can be expressed as 
\begin{align}
    \frac{m_S}{N f_a} \ll \sqrt{\frac{8 \pi^3 g_*(T_\text{osc.})}{135}} \frac{S_i}{M_\text{Pl.}} \lesssim 14.5 \cdot \sqrt{\frac{g_*(T_\text{osc})}{100}}
\end{align}
which is not a strong constraint compared to the limits from isocurvature in section \ref{sec:DMiso}. The tilt in the angular direction responsible for the Diraxion mass can also lead to washout by converting Diraxions into Saxions, who scatter of the plasma with a rate $\Gamma_S$ defined in sections \ref{sec:dissp} and \ref{sec:therm}. The resulting rate for the angular mode is then suppressed by a factor of $m_a(S_i)^4/m_S(S_i)^4$ leading to a rate of \cite{Co:2022aav}
\begin{align}
    \Gamma_a \simeq \left(\frac{m_a}{m_S}\right)^4 \left(\frac{N f_a}{S_i}\right)^{2(N-4)} \Gamma_S,
\end{align}
which is too suppressed to matter compared to $\Gamma_S$.
Once it thermalizes and  reaches  $S=N f_a$ the Saxion behaves as non-relativistic matter ($\omega_S=0$) and the Diraxion has the equation of state of kination $\omega_\theta=1$, because only the kinetic energy of the remaining circular rotation is left, which is why then $\varepsilon=1$. We can summarize the evolution of the angular field by the following scaling relations
\begin{align}\label{eq:sclingtheta}
    \dot{\theta} \sim \begin{cases}a^{-1}\\ a^{-3} \end{cases}, \quad \rho_\theta \equiv \frac{\dot{\theta}^2}{2} S^2 \sim \begin{cases} a^{-4}\\ a^{-6}\end{cases}, \quad \omega_\theta = \begin{cases} \frac{1}{3}\quad &\text{for}\quad S\gg N f_a\\1\quad &\text{for}\quad S\simeq N f_a \end{cases}.
\end{align}
%During the time when $S$ approaches $N f_a$ and redshifts as $S\sim 1/a^{3/2}$ before settling in its minimum we have a constant $\dot{\theta}\sim a^0$.
We further define the conserved charge yield $n_\theta /s$ for an oscillation starting during radiation domination with $T_\text{osc.}<T_\text{RH}$
\begin{align}\label{eq:yieldRD}
    Y_\theta^\text{RD} \equiv \frac{n_\theta(T_\text{osc.})}{s(T_\text{osc.})}.
\end{align}
If the oscillation starts before the end of reheating $T_\text{osc.}>T_\text{RH}$ we need a different adiabatic invariant than $n_\theta/s$, because entropy is not conserved during reheating. Instead one normalizes $n_\theta \sim 1/a^{3}$ to the energy density of the non-relativistic inflaton $\rho_\text{inf.}\sim n_\text{inf.}\sim 1/a^3$ \cite{Co:2020dya,Co:2020jtv}
\begin{align}\label{eq:yieldRH}
    Y_\theta^\text{RH} \equiv \frac{n_\theta(T_\text{RH})}{s(T_\text{RH})} = \frac{n_\theta}{\rho_\text{inf.}}\Big|_{T=T_\text{osc.}} \cdot \frac{\rho_\text{inf.}}{s}\Big|_{T=T_\text{RH}} = \frac{n_\theta(T_\text{osc.})}{s(T_\text{RH})}\cdot \left(\frac{T_\text{RH}}{T_\text{osc.}}\right)^8.
\end{align}
After Saxion thermalization described in section \ref{sec:therm}, where $\varepsilon\rightarrow 1$, and while  the Saxion still approaches its minimum $N f_a$ one finds that  the equations of motion for $\theta$ reduce to the balance between the radial potential gradient and the centripetal force $S(\dot{\theta}/N)^2$ \cite{Co:2022kul}
\begin{align}
    \dot{\theta}^2 = \frac{N^2}{S} \frac{\partial V(S)}{\partial S}= N^2 m_S(S)^2.
\end{align}
Note that the energy density in the rotation $\rho_\theta=\varepsilon \rho_S$ (with $\rho_S= V_\sigma(S_i)$) does not affect the condition of having less energy in the condensate than the radiation bath leading to \eqref{eq:infl}, because the Diraxion energy  originates from the Saxion oscillation, whose energy  decreases to $(1-\varepsilon)\rho_S$ after the kick in the angular direction. The $1/a^6$ scaling of the Diraxion's energy density after Saxion thermalization is reminiscent of kination. For a quartic potential starting out during radiation domination it was argued in \cite{Gouttenoire:2021jhk} that no era of kination dominance can occur without prior intermediate matter domination: This follows from the fact that the thermalization of the radial mode produces a radiation bath of energy density $(1-\varepsilon)\rho_S$, which is comparable in size to the energy density in Diraxions $\varepsilon \rho_S$. Even if the Diraxion contribution is initially larger than the radiation bath, it will redshift faster  and quickly become subdominant.\\
\\
As a consequence of the  oscillations for $\varepsilon\ll 1$ the quantities $S$ and $\dot{\theta}$ will not be constant in time and oscillate themselves around their minimum and maximum values during one cycle 
\begin{align}\label{eq:maxmin}
    S_\text{min} \equiv \varepsilon S, \qquad \dot{\theta}_\text{max}= \frac{N m_S(S)}{\varepsilon},
\end{align}
which can be obtained from the conservation of charge and energy
\begin{align}
    n_\theta= \varepsilon m_S(S) S_\text{max}^2,\quad \rho_S+\rho_\theta= m_S(S)^2  S_\text{max}^2,
\end{align}
where $S_\text{max}$ denotes the maximum field value during a cycle and we identify 
\begin{align}
    S_\text{max}(T_\text{osc.})=S_i.
\end{align}
 The Diraxion velocity is as large as $\dot{\theta}_\text{max}$ for a time scale $\Delta t \simeq 1/\dot{\theta}_\text{max}$. If we compute the cycle average of  $\dot{\theta}$ over a time scale $t_S \equiv 1/m_S(S)$ we find that \cite{Co:2020jtv} 
\begin{align}
    \braket{\dot{\theta}} \simeq\;  \frac{\Delta t}{t_S}\; \dot{\theta}_\text{max}  = N  m_S(S).
\end{align}
More refined analytical \cite{Gouttenoire:2021jhk} and numerical \cite{Co:2020jtv}  calculations have confirmed that $\braket{\dot{\theta}}= N m_S(S)$ is indeed an attractor solution, meaning that $\braket{\dot{\theta}}$ is independent of $\varepsilon$ in practise \cite{Co:2020jtv}.  In the above expression for $m_S$ we slightly abuse notation by denoting the root-mean-square $\sqrt{\braket{S^2}}$ of the Saxion amplitude with the same symbol as its (oscillating) field value $S$. If parametric resonance from Saxion oscillations occurs (see section \ref{sec:parRes}) the average of $\dot{\theta}$ gets reduced and one finds \cite{Co:2020jtv}
\begin{align}
    \braket{\dot{\theta}}= \varepsilon N m_S(S).
\end{align}
Note that reference \cite{Co:2020jtv} employs a slightly different parameterization of $\varepsilon$ 
\begin{align}
    \varepsilon \equiv \frac{n_\theta}{N \omega S_\text{max}^2}, \quad \text{with} \quad \omega \equiv \sqrt{\frac{1}{S}\frac{\partial V(S)}{\partial S }}\Big|_{S_\text{max}} \quad \text{and} \quad n_S = \frac{\dot{S}^2}{\omega}\Big|_{S_\text{max}}.
\end{align}
This parameterization reduces to our choice of $\varepsilon$ in \eqref{eq:varepsilon}   for $\varepsilon\ll1$.  We assume that $\dot{\theta}$ and consequently  $\dot{\theta}_\text{max}$  changes only adiabatically, meaning slower than the time scale of the plasma $1/T$, which is why we  impose  $\dot{\theta}_\text{max}<T_\text{osc.}$ \cite{Co:2020jtv}  leading to 
\begin{align}\label{eq:cond1}
    m_a >   0.24\; \sqrt{N} \cdot
    \begin{cases}
    g_*(T_\text{osc.})^\frac{1}{8} m_S \left(\frac{m_S}{M_\text{Pl.}}\right)^\frac{1}{4}  \left(\frac{N f_a}{S_i}\right)^\frac{2N-9}{4}&\quad \text{RD}\\
    g_*(T_\text{osc.})^\frac{1}{16} \sqrt{N} \frac{m_S^\frac{11}{8}}{M_\text{Pl.}^\frac{1}{8}T_\text{RH}^\frac{1}{4}} \left(\frac{N f_a}{S_i}\right)^\frac{4N-17}{8}&\quad \text{RH}
    \end{cases}.
\end{align}
Additionally we assume that the Saxion field value is always the largest energy scale in the plasma  $S_\text{min}= \varepsilon S_i >T_\text{osc.}$ \cite{Co:2020jtv} from which  we find
\begin{align}\label{eq:cond2}
     m_a >    0.19\; N \cdot 
    \begin{cases}
        \frac{1}{ g_*(T_\text{osc.})^\frac{1}{8}}\frac{m_S^\frac{5}{4} M_\text{Pl.}^\frac{1}{4}}{\sqrt{S_i}}     \left(\frac{N f_a}{S_i}\right)^\frac{2N-9}{4}&\quad \text{RD}\\
       \frac{1}{ g_*(T_\text{osc.})^\frac{1}{16}} \frac{m_S^\frac{9}{8} M_\text{Pl.}^\frac{1}{8} T_\text{RH}^\frac{1}{4}}{\sqrt{S_i}}  \left(\frac{N f_a}{S_i}\right)^\frac{4N-19}{8} &\quad \text{RH}
    \end{cases}
\end{align}
and one observes that both bounds are comparable in strength.
These bounds are weakest for $N=5$ due to an approximate cancellation in the exponent of $N f_a /S_i$, and get stronger for larger $N$. For reference we will take $N=6$ throughout this paper.

\section{Dirac-Lepto-Axiogenesis}\label{sec:Dirac}
\subsection{Dissipation coefficient from Dirac Weinberg operator}\label{sec:dissp}
The Saxion only couples to the thermal bath via its mixed quartic with ther Higgs or the non-renormalizable Dirac Weinberg operator. A proper determination of the relevant dissipation coefficient \cite{Mukaida:2012qn,Mukaida:2012bz} for our scenario would involve two-loop thermal self energy diagrams \cite{Turner:2018mwh}, which is why we resort to dimensional analysis. The Saxion can scatter via the reactions $H S \rightarrow \overline{L} \nu_R, \quad L S\rightarrow H^\dagger \nu_R$ or decay as $S\rightarrow \overline{L} H^\dagger \nu_R$, where we assumed a vanishing thermal abundance of $\nu_R$ and put the non-thermal $S$ in the initial state. Due to time-dilation effects \cite{Kolb:1979qa} the decay will only be relevant for $m_S(S)<T$, which is why we add both contributions as
\begin{align}\label{eq:dissp}
    \Gamma_S(S) \equiv  \frac{1}{16 \pi} \left(\frac{\sum m_\nu^2}{v_H^2}\right)\left(\frac{S}{N f_a}\right)^2 \left(2T+m_S(S)\right).
\end{align}
The scaling is similar to the interaction of two Higgsinos with two Saxions considered in \cite{Barnes:2022ren} and can be understood as follows: The matrix element involves one insertion of $S$ and the factor $m_\nu /(v_H N f_a)$ from the coefficient of the Dirac Weinberg operator. After squaring the matrix element to obtain the rate one can use dimensional analysis to find the remaining factor of $2 T$ for the two scattering modes or $m_S(S))$ for the decay. One should keep in mind that the rate depends on the time-dependent field value $S$ and not on its amplitude $S_\text{max}$ \cite{Kozow:2022whq}.
After thermalization of $S$ the rate reduces to 
\begin{align}\label{eq:disspT}
        \Gamma_L(T) \equiv  \frac{1}{16 \pi} \left(\frac{\sum m_\nu^2}{v_H^2}\right) \frac{2T^3+m_S^3}{(N f_a)^2},
\end{align}
which is now slower since $T\ll S$. The overall rate for $T>m_S(S)$ differs only by a factor of $v_H^2/(N f_a)^2$ from the result for Majorana neutrinos. We always find that $m_S(S)\ll T$ for $S\gg N f_a$ because $m_S(S_i)= 3 H(T_\text{osc.}) \ll T_\text{osc.}$, which is why the Saxion decay is negligible above $T_S$. After the Saxion has reached its minimum at $T_S$ defined in \eqref{eq:TS} and we replace it by its vev, we recover the Yukawa interaction between the neutrinos and the Higgs, which never thermalizes due to the smallness of $m_\nu/v_H$ \cite{Dick:1999je}.

\subsection{Chemical potentials and Boltzmann equation}\label{sec:Boltz}
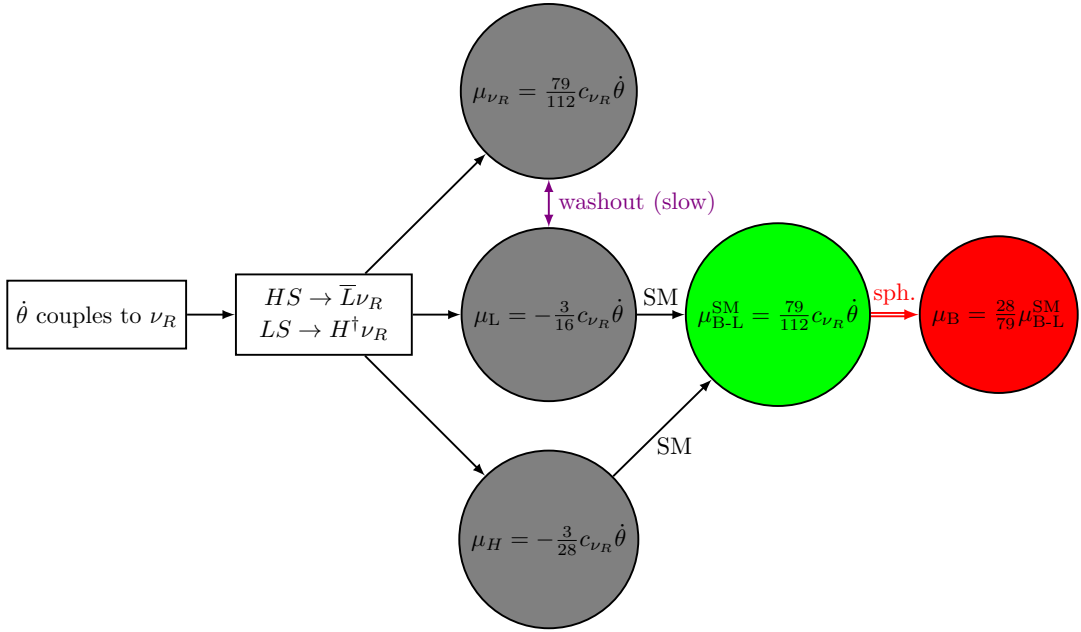
\begin{figure}
    \centering
     \scalebox{0.9}{\tikzset{
  blackline/.style={thin, draw=black, postaction={decorate},
    decoration={markings, mark=at position 0.6 with {\arrow[black]{triangle 45}}}}
}

\begin{tikzpicture}[font=\small,thick]
 
% Start block
 \node[draw,
    minimum width=1cm,
    minimum height=1cm,
] (block1) {$\dot{\theta}$ couples to $\nu_R$};

\node[draw,
    minimum width=1cm,
    minimum height=1cm,
    right=0.7cm of block1,
    fill=white
] (block2) {\begin{tabular}{c} $H S \rightarrow \overline{L} \nu_R$ \\ $L S\rightarrow H^\dagger \nu_R$ \end{tabular}};

 \node[draw,
   circle,
   minimum size = 0.3 cm,
   right= 0.7cm of block2,
   fill=gray
] (block11) { $\mu_\text{L}  = -\frac{3}{16} c_{\nu_R} \dot{\theta}$};

 \node[draw,
   circle,
   minimum size = 0.3 cm,
   above  = 0.7 cm of block11,
   fill=gray
] (block12) {$\mu_{\nu_R} = \frac{79}{112} c_{\nu_R} \dot{\theta}$};

 \node[draw,
   circle,
   minimum size = 0.3 cm,
   below  = 0.7 cm of block11,
   fill=gray
] (block13) {$\mu_H = -\frac{3}{28} c_{\nu_R} \dot{\theta}$};

 \node[draw,
   circle,
   minimum size = 0.3 cm,
   right  = 0.7 cm of block11,
   fill=green
] (block15) {$\mu_\text{B-L}^\text{SM} =\frac{79}{112} c_{\nu_R} \dot{\theta}$};

 \node[draw,
   circle,
   minimum size = 0.3 cm,
   right = 0.7 cm of block15,
    fill=red
] (block14) { $\mu_\text{B} =  \frac{28}{79} \mu_\text{B-L}^\text{SM} $};

\draw[-latex ] (block1) -- (block2)  node[right,below,midway]{};
\draw[-latex ] (block2) -- (block11) node[right,above,midway]{};
\draw[-latex ] (block2) -- (block12) node[right,above,midway]{};
\draw[-latex ] (block2) -- (block13) node[right,above,midway]{};
\draw[-latex ] (block11) -- (block15) node[right,above,midway]{SM};
\draw[-latex ] (block13) -- (block15) node[left,below,midway]{$\quad$SM};

\draw[red,-latex,double] (block15) -- (block14) node[above,midway]{sph.};
\draw[violet,latex-latex ] (block11) -- (block12)  node[right,midway]{washout  (slow)};

\end{tikzpicture}}  

    \caption{Schematic representation of the reaction chain for generating a chemical potential in baryons. The arrows labelled \enquote{SM} indicate the network of equilibrated Standard Model Yukawa interactions. The red arrow labelled \enquote{sph.} stands for the B+L violating electroweak sphaleron process. Time progresses from left to right.}
    \label{fig:sketch}
\end{figure}

The derivative coupling of the spatially homogeneous Diraxion  $\partial_\mu \theta \overline{\nu_R}\gamma^\mu\nu_R$ defined in \eqref{eq:deriv} can be thought of as inducing an effective chemical potential for the right chiral neutrinos \cite{Cohen:1987vi,Cohen:1988kt,Cohen:1993nk}
\begin{align}
    \mu_{\nu_R}^\text{eff.} \equiv c_{\nu_R} \dot{\theta}.
\end{align}
As pointed out in \cite{Dolgov:1996qq,Dolgov:1997qr,Arbuzova:2016qfh} and more recently in \cite{Dasgupta:2018eha} however this derivative coupling does not lead to actual splittings in the single particle energies of particles and anti-particles, which is why it is not an actual chemical potential. A more accurate way to think about the effect of $\dot{\theta}$ is to treat it as a background field that shifts the dispersion relation of the particles coupling to it $ E=\sqrt{p^2+m^2} \rightarrow \sqrt{p^2+m^2}  \mp  c_{\nu_R} \dot{\theta}$  \cite{Dolgov:1996qq,Dolgov:1997qr,Arbuzova:2016qfh} with a (minus) plus sign for (anti-)particles  in analogy to explicit CPT-violation in the form of different masses for particles and antiparticles, see e.g. \cite{Bossingham:2017gtm}. Since the background field $\dot{\theta}\neq0$ spontaneously violates CPT in the plasma, this scenario is known as spontaneous baryogenesis  \cite{Cohen:1987vi,Cohen:1988kt} and we do not need to worry about the third Sakharov condition \cite{Sakharov:1967dj}, which assumes CPT-conservation.
The time dependent Diraxion background is also responsible for the CP violation encoded in the second Sakharov condition \cite{Sakharov:1967dj}. If the Dirac Weinberg operator leads to  reactions $H S \rightarrow \overline{L} \nu_R, \quad L S\rightarrow H^\dagger \nu_R$  in equilibrium, then  the chemical potentials satisfy
\begin{align}\label{eq:chempot}
    \mu_L - \mu_H + \mu_{\nu_R} +  c_{\nu_R} \dot{\theta}= 0,
\end{align}
where we used that the real valued Saxion has no chemical potential.
The same result can be obtained without making reference to the effective chemical potential $\mu_{\nu_R}^\text{eff.}$: Suppose we work in the basis, where we do not remove the phase of the Dirac Weinberg operator $m_\nu /(v_H N f_a) e^{i \theta(t)}$ and hence have no derivative couplings of the Diraxion to $\nu_R$. The time dependent phase in the aforementioned coupling does not enter the matrix element but rather it modifies the delta distributions encoding energy-momentum-conservation by also taking the effect of the slowly varying background field into account \cite{Ibe:2015nfa}. For e.g. $H S \rightarrow \overline{L} \nu_R$ one would find $E_H + E_S = E_L + E_{\nu_R} - c_{\nu_R} \dot{\theta}$ \cite{Dolgov:1996qq,Dolgov:1997qr,Arbuzova:2016qfh}  and via the collision term of the Boltzmann equation\footnote{The energy and chemical potential enter the phase space distribution functions  with opposite signs e.g. $f_i\sim\exp\left(\left(-E_i + \mu_i\right)/ T\right)$  for a non-relativistic particle $i$.}  one is then lead to the same equation for the chemical potentials as in \eqref{eq:chempot}. We now also impose the conservation of the total B-L, which might be gauged in the UV and is responsible for the Dirac nature of neutrinos, 
\begin{align}
    2 \mu_Q + \mu_u  + \mu_d-2 \mu_L - \mu_e -\mu_{\nu_R} \equiv  \mu_\text{B-L}^\text{SM}  -\mu_{\nu_R} = 0.
\end{align}
Here  the actual chemical potential $\mu_{\nu_R}$ appears, because as previously discussed $c_{\nu_R} \dot{\theta}$ only contributes to  scattering processes involving $\nu_R$. Next, if we impose the conservation of hypercharge, take all SM Yukawa and gauge interactions to be fast (see e.g. appendix C in \cite{Berbig:2022pye}) and treat all flavors the same, then  the network of chemical reaction implies that in equilibrium
\begin{align}
    \mu_\text{B}^\text{eq.}= \frac{28}{79}\mu_\text{B-L}^\text{SM eq.}, \quad \text{with} \quad \mu_\text{B-L}^\text{SM eq.} = \mu_{\nu_R}^\text{eq.} = \frac{79}{112} c_{\nu_R} \dot{\theta}.
\end{align}
The relation between $ \mu_\text{B}^\text{eq.}$ and $\mu_\text{B-L}^\text{SM eq.}$ is given by the standard value $28/79$  \cite{Harvey:1990qw,Khlebnikov:1996vj} for the sphaleron redistribution coefficient. Note that the Yukawa interactions for all three generations of SM fermions are only in equilibrium below about $10^6\;\text{GeV}$ \cite{Nardi:2005hs,Bodeker:2019ajh}, but we do not expect this detail to change the sphaleron redistribution coefficient by more than $10\%$.  By employing the relation $n_i = \mu_i/6 T^2$ between the asymmetry and chemical potential for a fermionic 
particle $i$  and using the principle of detailed balance we can immediately write down the Boltzmann equation for $n_\text{B-L}^\text{SM}$
\begin{align}\label{eq:boltz}
    \frac{\text{d}}{\text{d}t}\; n_\text{B-L}^\text{SM} + 3 H \; n_\text{B-L}^\text{SM} = -\Gamma_S(S) \left(n_\text{B-L}^\text{SM} - \frac{79}{672}c_{\nu_R}  \dot{\theta} T^2\right)
\end{align}
in terms of the dissipation coefficient $\Gamma_S(S)$ defined in \eqref{eq:dissp}. 
Our mechanism is similar to the case of Dirac-Leptogenesis \cite{Dick:1999je} in the sense that total lepton number vanishes and we produce equal   asymmetries ($ n_\text{B-L}^\text{SM} = n_{\nu_R}$) in the left- and right-chiral leptons. These leptonic asymmetries are never equilibrated because in the following we take $\Gamma_S$ to be slow for $S\gg N f_a$. Once $S\simeq N f_a$ we recover the usual Yukawa interaction with the Higgs from the Dirac Weinberg-operator, which is always slow due to the smallness of the effective Yukawa coupling $m_\nu /v_H$ \cite{Dick:1999je}.  The B+L-violating sphaleron transition only acts on the doublet of left-chiral leptons and converts $n_\text{B-L}^\text{SM}$ into the observed baryon asymmetry. A sketch of the scenario can be found in figure \ref{fig:sketch}. Conventional Lepto- and Baryogenesis from scattering was covered in \cite{Bento:2001rc,Cui:2011ab}. 

\subsection{Baryon Asymmetry}\label{sec:barasym}
We rewrite \eqref{eq:boltz} in terms of the dimensionless yield $n_\text{B-L}^\text{SM}/s$, which is conserved in the comoving volume as long as entropy is conserved 
\begin{align}\label{eq:boltz2}
    \frac{\text{d}}{\text{d}t}\; \frac{n_\text{B-L}^\text{SM}}{s} =  -\Gamma_S(S) \left(\frac{n_\text{B-L}^\text{SM}}{s} - \frac{79}{672} \frac{c_{\nu_R}  \dot{\theta} T^2}{s}\right)
\end{align}
Solving \eqref{eq:boltz2} is in general complicated by the fact that our problem has two intrinsic time-scales: The expansion rate of the universe and the oscillation frequency $m_S(S)$. Since our dissipation coefficient is given in terms of an effective operator, it will be UV-dominated and hence we expect the baryon asymmetry to be predominantly produced at $T_\text{osc.}$. By definition we have $3H(T_\text{osc.})= m_S(S_i)$ at this point in time, which simplifies the problem. The last complication is the fact that the production rate $\Gamma_S(S)$ depends on the oscillating field value, which is why we need to average over one cycle of oscillations. Reference \cite{Co:2020jtv} solved \eqref{eq:boltz2} numerically (see appendix E of the aforementioned reference) and provided analytical arguments to understand the resulting abundances given by 
\begin{align}\label{eq:sols}
    \left\langle \frac{n_\text{B-L}^\text{SM}}{s} \right\rangle \simeq  \frac{79}{672} \frac{c_{\nu_R} N \braket{\dot{\theta}} T_\text{osc.}^2}{s} \cdot
    \begin{cases}
        1 \quad& \Gamma_S(S_i)\gg \frac{m_S(S_i)}{\varepsilon^3}\\
        \varepsilon \left(\frac{\Gamma_S(S_i)}{m_S(S_i)}\right)^\frac{1}{3} \quad&  m_S(S_i)\ll \Gamma_S(S_i) \ll \frac{m_S(S_i)}{\varepsilon^3}\\
        2 \varepsilon\quad&  \Gamma_S(S_i) \ll m_S(S_i)
    \end{cases}.
\end{align}
The first result can be understood as follows \cite{Co:2020jtv}: In order for $\braket{n_\text{B-L}^\text{SM}}/s$ to track its equilibrium value, $\dot{\theta}$ needs to be close to its maximum $\dot{\theta}_\text{max}= m_S(S_i)/\varepsilon > m_S(S_i)$. However due to the conservation of the condensate charge $n_\theta = \varepsilon m_S(S_i) S_i^2$ we find that this only occurs when $S$ is at its minimum for a cycle given by $S_\text{min} = \varepsilon S_i$ (see \eqref{eq:maxmin}). For a field-dependent interaction it was found in \cite{Co:2020jtv}, that the charge transfer from the condensate to the plasma is only efficient if the transfer rate is larger than the frequency of the coherent motion, which for this case reads 
\begin{align}
    \Gamma_S(S_\text{min}) \sim \varepsilon^2 S_i^2\; \gg\; \dot{\theta}_\text{max} = \frac{m_S(S_i)}{\varepsilon}
\end{align}
and reduces to the condition $ \Gamma_S(S_i)\gg m_S(S_i)/\varepsilon^3$ displayed in \eqref{eq:sols}. However if we take e.g. $\varepsilon=0.1$ then this implies that $\Gamma_S(S_i)$ would have to be a thousand times faster than the Hubble rate at the beginning of the oscillations. We discuss in section  \ref{sec:earlytherm} how this is potentially problematic. 
The next regime $m_S(S_i)\ll \Gamma_S(S_i) \ll  m_S(S_i)/\varepsilon^3$ suffers from the same drawback. Since $\Gamma_S(S_\text{min})$ is now too small to track $\dot{\theta}_\text{max}$, the transfer can only be relevant when $S$ is close to its maximum value $S_\text{max}(T_\text{osc.})=S_i$. Once $S$ decreases from its maximum during a cycle and $\dot{\theta}$ increases, $\braket{n_\text{B-L}^\text{SM}/s}$ stops tracking $\dot{\theta}T^2/s$ as soon as $\dot{\theta}$ reaches the value $\dot{\theta}_d\simeq \varepsilon \Gamma_S(S_i)^\frac{2}{3} m_S(S_i)^\frac{1}{3}$  at which it stays for a time of $\Delta t_d\simeq \Gamma_S(S_i)^{-\frac{1}{3}} m_S(S_i)^{-\frac{2}{3}}$ so that the average over a time scale $t_S=1/m_S(S_i)$ becomes  \cite{Co:2020jtv}
\begin{align}
    \braket{\dot{\theta}}= \frac{\Delta t_d}{t_S} \dot{\theta}_d = \varepsilon m_S(S_i) \left(\frac{\Gamma_S(S_i)}{m_S(S_i)}\right)^\frac{1}{3}.
\end{align}

\subsubsection{Rotation during Radiation domination}

To be conservative we focus on the regime  $\Gamma_S(S_i)\ll m_S(S_i)\simeq 3 H(T_\text{osc.})$, which can be thought of as Freeze-In production \cite{Hall:2009bx} in contrast to the previous two cases, which are akin to thermal Freeze-Out.
We average the Boltzmann equation in \eqref{eq:boltz2} over a time scale larger than $t_S=1/m_S(S_i)$ and factor out the $S$-dependence of the rate by employing \eqref{eq:dissp} and \eqref{eq:disspT} to rewrite $\Gamma_S= \Gamma_L \cdot (S/T)^2$  leading to
\begin{align}\label{eq:BoltzAv}
        \left\langle\frac{\text{d}}{\text{d}t}\; \frac{n_\text{B-L}^\text{SM}}{s}\right\rangle =  -\Gamma_L(T) \left(\left\langle\frac{n_\text{B-L}^\text{SM}}{s}\right\rangle \frac{\braket{S^2}}{T^2} - \frac{79}{672} \frac{c_{\nu_R}  \braket{\dot{\theta}S^2}}{s}\right).
\end{align}
Because  $\theta$ and $S$ move together coherently in the broken phase, we have to average them together as $\braket{\dot{\theta} S^2}$ (unless parametric resonance occurs). For $\Gamma_S(S_i)\ll m_S(S_i)$ it follows that $S$ evolves much faster than $n_\text{B-L}^\text{SM}/s$, which is why we can set $\left\langle\frac{\text{d}}{\text{d}t}\; \frac{n_\text{B-L}^\text{SM}}{s}\right\rangle=0$ \cite{Co:2020jtv} and thus solve for $\braket{n_\text{B-L}^\text{SM}/s}$ from the vanishing of the parenthesis above. To do so we make use of the charge conservation \cite{Co:2020jtv}
\begin{align}\label{eq:conv}
  N n_\theta=  \frac{1}{N}\braket{\dot{\theta}S^2} = \varepsilon n_S = \varepsilon m_S(S) \braket{S^2}
\end{align}
and obtain $\braket{n_\text{B-L}^\text{SM}/s}\sim \braket{\dot{\theta}S^2}/\braket{S^2}= \varepsilon N m_S(S_i)$, which explains the factor of $\varepsilon$ in the last line of \eqref{eq:sols} and the factor of two comes from matching to the previous two solutions.  However $\left\langle\frac{\text{d}}{\text{d}t}\; \frac{n_\text{B-L}^\text{SM}}{s}\right\rangle=0$ can also be obtained from $\Gamma_S\rightarrow 0$ and we expect \cite{Hall:2009bx}  the Freeze-In abundance to vanish if the production rate goes to zero (there is only an upper limit on $\Gamma_S$ in the third line of \eqref{eq:sols}). Therefore we write
\begin{align}\label{eq:asym}
    \left\langle \frac{n_\text{B}}{s} \right\rangle &\simeq \frac{c_{\nu_R} N}{24} \cdot  \frac{\Gamma_S(S_i)}{H(T_\text{osc.})} \cdot   \frac{ \varepsilon m_S(S_i) T_\text{osc.}^2}{s(T_\text{osc.})}\\
   & = \frac{c_{\nu_R} N^2}{24} \cdot  \frac{\Gamma_S(S_i)}{H(T_\text{osc.})} \cdot \left(\frac{T_\text{osc.}}{S_i}\right)^2\cdot Y_\theta^\text{RD},
\end{align}
where in the last line we used the charged yield during radiation domination defined in \eqref{eq:yieldRD}.
The same result can also be found from dropping the back-reaction term $\propto \braket{n_\text{B-L}^\text{SM} /s}$ in \eqref{eq:BoltzAv} and considering the amount of asymmetry produced $\braket{n_\text{B-L}/s}\simeq 1/H\cdot  \text{d}/\text{d}t\; \braket{n_\text{B-L}/s}   $ over one Hubble time $1/H$ together with the charge conservation in \eqref{eq:conv}. One can observe that the resulting asymmetry is suppressed by both $\varepsilon<1$ and $\Gamma_S/H\ll 1$ compared to the scenario for a very efficient charge transfer $\Gamma_S(S_i) \gg  m_S(S_i)/\varepsilon^3$, which also justifies  neglecting the back-reaction.
By using the scaling relations for the redshift of $S$ one can deduce that during radiation domination we have  $m_S(S)\sim S\sim T$ and  $\Gamma_S/H \sim S^2 /T \sim T$ leading to $\braket{n_B /s }\sim T$, which means that the produced asymmetry is indeed UV dominated. We deviate from \cite{Co:2020jtv} by writing our result in terms of $m_S(S_i)$ instead of $\braket{\dot{\theta}}$ for the following reason: Typically one has $\braket{\dot{\theta}}=m_S(S_i)$ at the beginning of oscillations, except when parametric resonance from Saxion oscillations takes place (see section \ref{sec:parRes}), where one obtains $\braket{\dot{\theta}}=\varepsilon m_S(S_i)$ instead. However our asymmetry is sourced by $\braket{\dot{\theta}S^2}$ and not just $\braket{\dot{\theta}}$, so no additional factor of $\varepsilon$ arises. The net effect of parametric resonance is that the radial and angular fields will not move coherently anymore due to the non-thermal symmetry restoration, which simply means that we can separate the averages $\braket{\dot{\theta}S^2}=\braket{\dot{\theta}}\braket{S^2}$ \cite{Co:2020jtv}.\footnote{Using this together with the charge conservation \eqref{eq:conv} then readily implies $\braket{\dot{\theta}}=\varepsilon m_S(S_i)$.} 
The baryon asymmetry differs from the result  for Majorana Lepto-Axiogenesis \cite{Co:2020jtv} in two aspects: On the one hand our asymmetry always depends on $\varepsilon$ due to the linear coupling to $S$ and on the other hand the ratio $\Gamma_S/H$ is parametrically different from the suppression factor $\Gamma_L/H$ found in the Majorana scenario. The order of magnitude and sign of the asymmetry will be determined by the first oscillation \cite{Kusenko:2014lra,Kusenko:2014uta,Yang:2015ida}. If we plug in the out-of-equilibrium condition for successful Freeze-In  $ \Gamma_S(S_i) \ll m_S(S_i)=3 H(T_\text{osc.})$ we obtain an upper limit on the produced asymmetry
\begin{align}\label{eq:asym2}
    \left\langle \frac{n_\text{B}}{s} \right\rangle \ll \frac{ c_{\nu_R} N^2}{8}   \cdot \left(\frac{T_\text{osc.}}{S_i}\right)^2  \cdot Y_\theta,
\end{align}
in terms of the charge yield of the condensate. Since the above is much less\footnote{This assumes that $S_i > T_\text{osc}/\sqrt{N}$, which is valid throughout our parameter space and we used $c_{\nu_R}\simeq 1/N$.} than $n_\theta /s$, it was valid to neglect the back-reaction of the asymmetry production on the condensate; Leptogenesis will not evaporate the condensate. The produced leptonic asymmetry is not washed out, because the Dirac Weinberg-operator is the only coupling connecting $\nu_R$ to the bath and we take it to be slow throughout the evolution of the universe. Once the Saxion settles at its minimum $N f_a$ the Dirac Weinberg operator reduces to the Yukawa coupling of the neutrinos with the SM like Higgs, which never equilibrates \cite{Dick:1999je}. To estimate the baryon asymmetry we eliminate $f_a$ via the condition $\Gamma_S(S_i) /H(T_\text{osc})<1$
\begin{align}\label{eq:fa1}
    f_a = \SI{2e6}{\giga\electronvolt}\cdot \left(\frac{0.1 }{\Gamma_S(S_i)/H(T_\text{osc})}\right)^\frac{2}{3}\cdot \left(\frac{\sum m_\nu^2}{(0.05\;\text{eV})^2}\right)^\frac{2}{3}\cdot \left(\frac{S_i}{0.1\;M_\text{Pl.}}\right) \cdot \left(\frac{\SI{100}{\mega\electronvolt}}{m_S}\right)^\frac{1}{3}
\end{align}
and obtain the following $Y_\text{B} = n_\text{B}/s$ for $N=6$ 
\begin{align}\label{eq:leptoaxio1}
   \frac{ Y_B}{ \SI{8e-11}{}} \simeq  \left(\frac{m_a}{10^{-3}\;\text{eV}}\right)^2 \cdot \left(\frac{\SI{100}{\mega\electronvolt}}{m_S}\right)^\frac{2}{3} \cdot \left(\frac{\Gamma_S(S_i) /H(T_\text{osc})}{0.1}\right)^\frac{8}{3}\cdot \left(\frac{(\SI{0.05}{\electronvolt})^2}{\sum m_\nu^2}\right)^\frac{5}{3}.
\end{align}
The baryon asymmetry only depends on the ratio $S_i/f_a$, so the dependence on $S_i$ divides out when inserting \eqref{eq:fa1}. This parameter choice is compatible with the bound from $\varepsilon<1$ in \eqref{eq:masses2}.
Appendix \ref{sec:CPI} explains the absence of the chiral hypermagnetic instability for our scenario.

\subsubsection{Rotation during Reheating}

In the above we assumed that the oscillation occurs during radiation domination. For the opposite case of $T_\text{osc.}>T_\text{RH}$ we have production during reheating, where entropy is not conserved. Thus $n_B /s$ is not the correct adiabatic invariant anymore and we have to use $\braket{n_B} / \rho_\text{inf}$ evaluated at production multiplied with $\rho_\text{inf.}/s$ evaluated at reheating. In this context $\rho_\text{inf}$ denotes the energy density of the non-relativistic inflaton. During the matter dominated reheating phase we have $T\sim 1/ a^{3/8}$ and a Hubble rate of $H\sim T^4$. By using the Friedmann equation $\rho_\text{inf.}=3 M_\text{Pl.}^2 H^2$ we find that $\rho_\text{inf.}\sim T^8$. The scaling of the asymmetry could be different depending on whether the Saxion settles to its minimum $N f_a$ during ($T_\text{osc.}>T_S> T_\text{RH}$) or after reheating $(T_\text{osc.}> T_\text{RH}>T_S$). For $T_\text{RH}>T_S$ we find that $ m_S(S)\sim S\sim 1/a \sim T^{8/3}$ and $\Gamma_S(S)/H \sim S^2/T^3 \sim T^{7/3}$ implying $\braket{n_B}/\rho_\text{inf.}\sim 1/T$.  In the second regime $T_\text{RH}<T_S$ we have $S=N f_a=\text{const.}$ so conservation of the Noether charge $\dot{\theta} N^2 f_a^2$ implies $\braket{\dot{\theta}}\sim 1/a^{3}\sim T^8$. Further we find $\Gamma_S/H \sim T f_a^2 /T^4\sim 1/T^3$ resulting in $\braket{n_B}/\rho_\text{inf.}\sim 1/T$. Both regimes redshift the same, because they   depend on $\braket{\dot{\theta}S^2}\sim n_\theta $, which always redshifts as $1/a^3$ no matter if $S$ is at or above $N f_a$. During reheating the asymmetry is always IR-dominated and gets predominantly generated at $T_\text{RH}$, which is why we can   evaluate $\braket{n_B/s}$ at reheating. 
 This is unlike the usual case for  Lepto-Axiogensis, where one would find $\braket{n_B}/\rho_\text{inf.}\sim T$ for $T_S> T_\text{RH}$ \cite{Co:2020jtv}.
We arrive at
\begin{align}\label{eq:asymRH}
    \left\langle\frac{n_B}{s}\right\rangle
   &\simeq \frac{c_{\nu_R} N}{24} \cdot  \frac{\Gamma_S(S_\text{RH})}{H(T_\text{RH})} \cdot   \frac{ \varepsilon m_S(S_\text{RH}) T_\text{RH}^2}{s(T_\text{RH})}\\
   &=  \frac{c_{\nu_R} N^2}{24} \cdot  \frac{\Gamma_S(S_i)}{H(T_\text{osc.}^\text{RH})} \cdot \left(\frac{T_\text{osc.}^\text{RH}}{S_i}\right)^2\cdot \left(\frac{T_\text{osc.}^\text{RH}}{T_\text{RH}}\right)\cdot Y_\theta^\text{RH},
\end{align}
where in the second line we used the charge yield for reheating defined in \eqref{eq:yieldRH} as well as the scaling laws to relate  the field value at reheating $S_\text{RH}$ and $S_i$, which restores the factor of $\Gamma_S(S_i)/H(T_\text{osc.})\ll1$ \footnote{For $T_\text{RH}<T_S$ the scattering rate is IR dominated and we should demand $\Gamma_S(S_\text{RH})/H(T_\text{RH})\ll1$ instead. However in practise we typically find that $T_\text{RH}>T_S$ so we ignore this additional complication.}. Note that the above is not enhanced by $T_\text{osc.}^\text{RH}>T_\text{RH}$ because $Y_\theta^\text{RH}\sim T_\text{RH}^8/T_\text{osc.}^{\text{RH}\;8}$.
We eliminate $f_a$ via the condition  $\Gamma_S(S_i) /H( T_\text{osc.}^\text{RH})<1$
\begin{align}\label{eq:fa2}
    f_a &= \SI{9.9e5}{\giga\electronvolt}\cdot \left(\frac{0.1 }{\Gamma_S(S_i)/H(T_\text{osc}^\text{RH})}\right)^\frac{4}{5}\cdot \left(\frac{\sum m_\nu^2}{(0.05\;\text{eV})^2}\right)^\frac{4}{5}\nonumber\\
    &\cdot \left(\frac{S_i}{0.1\;M_\text{Pl.}}\right) \cdot \left(\frac{\SI{100}{\mega\electronvolt}}{m_S}\right)^\frac{3}{5}\cdot \left(\frac{ T_\text{RH}}{10^{13}\;\text{GeV}}\right)^\frac{2}{5}
\end{align}
to find 
\begin{align}\label{eq:leptoaxio2}
   \frac{ Y_B}{ \SI{8e-11}{}} \simeq  \left(\frac{m_a}{10^{-3}\;\text{eV}}\right)^2 \cdot \left(\frac{\SI{40}{\mega\electronvolt}}{m_S}\right)^\frac{6}{5} \cdot \left(\frac{\Gamma_S(S_i) /H(T_\text{osc}^\text{RH})}{0.1}\right)^\frac{12}{5}\cdot \left(\frac{(\SI{0.05}{\electronvolt})^2}{\sum m_\nu^2}\right)^\frac{7}{5}.
\end{align}

\section{Dark Matter}\label{sec:Matter}

\begin{figure}
    \centering
    \includegraphics[width=0.7\textwidth]{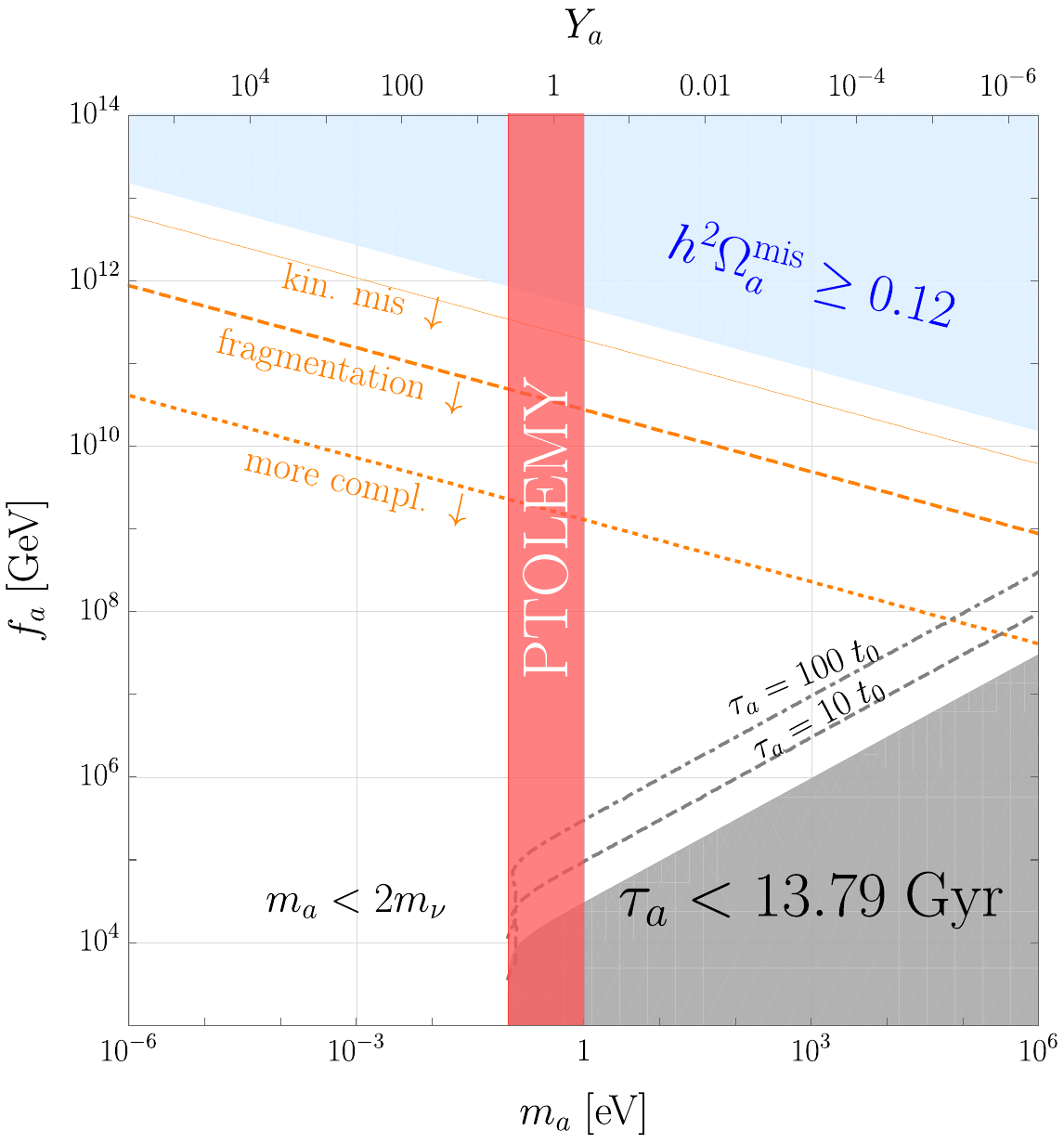}
    \caption{Allowed parameter space for Diraxion dark matter in terms of its mass $m_a$ and decay constant $f_a$, that is stable enough on cosmological time scales and potentially detectable with PTOLEMY. Kinetic Misalignment operates below the orange line and the fragmentation of the Diraxion condensate becomes important below the dashed orange line (see below \eqref{eq:condss} for a discussion).}
    \label{fig:DMoverview}
\end{figure}

\subsection{Dark Matter decay}
The Diraxion is only a good dark matter candidate if it is sufficiently long-lived. Its main decay mode is to neutrinos, followed by a subleading mode to photons via a two-loop coupling and the width is given by \cite{Reig:2019sok}
\begin{align}\label{eq:decaAxion}
    \Gamma\left(a\rightarrow  \overline{\nu}  \nu\right) \simeq \frac{m_a}{16\pi} \frac{\sum_\nu m_{\nu}^2}{N^2 f_a^2}\; \sqrt{1-4 \frac{\sum_\nu m_\nu^2}{m_a^2}},
\end{align}
where we estimated the phase space suppression in the single flavor approximation. Since the coupling is dependent on the absolute neutrino masses  $\sum_\nu m_{\nu}^2 $ we use the sum of the observed mass splittings as a reference \cite{ParticleDataGroup:2022pth}
\begin{align}
    \sum_\nu m_{\nu}^2 \simeq 
    \begin{cases}
    (\SI{0.05}{\electronvolt})^2&\quad \text{for NH}\\
    (\SI{0.1}{\electronvolt})^2&\quad \text{for IH}
    \end{cases},
\end{align}
where NH (IH) denotes the normal (inverted) hierarchy of neutrino masses. Throughout this work we focus on the normal hierarchy. Normalized to the age of the universe $t_0 \simeq 13.79\;\text{Gyr}$ we obtain
\begin{align}\label{eq:tauDM}
    \tau= 1/ \Gamma\left(a\rightarrow  \overline{\nu}  \nu\right) \simeq 30 \;t_0 \cdot \left(\frac{(\SI{0.05}{\electronvolt})^2}{\sum_\nu m_\nu^2}\right)\cdot \left(\frac{1\;\text{eV}}{m_a}\right)\cdot \left(\frac{N f_a}{10^{6}\;\text{GeV}}\right)^2.
\end{align}
There is no appreciable bound from the Diraxion to photon decay via the couplings in \eqref{eq:gammaI},\eqref{eq:gammaII}, as the rate would be suppressed by the free factors $(m_a/v_H)^2$, $(m_\nu/ N f_a)^4$ and $(m_a/m_e)^4$.
The limit on the life-time of Diraxion dark matter was depicted in black in figure \ref{fig:DMoverview}. Experiments such as PTOLEMY \cite{Betts:2013uya,PTOLEMY:2018jst} aimed at detecting the cosmic neutrino background could potentially detect \cite{McKeen:2018xyz,Chacko:2018uke} the neutrinos produced in decays of dark matter with masses between $\SI{0.1}{\electronvolt}$ and $\mathcal{O}(\SI{1}{\electronvolt})$ for lifetimes between $(10-100)\;t_0$ \cite{Reig:2019sok}. We show the corresponding parameter region colored in red in  figure \ref{fig:DMoverview}. While standard misalignment and decays of topological defects (blue region in \ref{fig:DMoverview}) works only for too large values of $f_a$ compared to the detectable region, we will see that  kinetic misalignment (see the orange lines in the aforementioned plot together with \ref{sec:kinmis}) and parametric resonance in \ref{sec:parRes} could both reproduce the relic abundance for much smaller $f_a$ and could potentially be probed by PTOLEMY. 

\subsection{Standard Misalignment}\label{sec:mis}
First we discuss the conventional misalignment mechanism, which operates when the kinetic energy of the Diraxion is smaller than its potential barrier. We will state the precise conditions for this at the end of this subsection and in the next one. During inflation the Diraxion is trapped in its initial angle $\theta_{inf.}=\pi/2-\delta$ by the Hubble friction (see section \ref{sec:DMiso}). The QCD axion dark matter literature usually employs an axion potential of the form $1-\cos(\theta)$ unlike our choice of $\cos(\theta)$ from \eqref{eq:Diraxionmass}. The constant piece is irrelevant for dark matter and the correct sign can be obtained by defining $\theta_i = \theta_\text{inf.}\pm \pi$, where we take the plus sign for concreteness. However since the Diraxion is light during inflation, its initial angle will receive corrections from quantum fluctuations giving rise to isocurvature perturbations \cite{Hertzberg:2008wr}
\begin{align}\label{eq:thetaeff}
    \braket{\theta_i^2} = \left(\delta -\frac{3\pi}{2}\right)^2 + \left(\frac{H_I}{2\pi S_i}\right)^2.
\end{align}
Using $H_I \simeq \SI{6e13}{\giga\electronvolt}$ and $10^{16}\;\text{GeV}<S_i< 10^{19}\;\text{GeV}$ one can deduce that $10^{-5}<H_I/(2\pi)<10^{-2}$, indicating that we can not take the initial misalignment angle to be arbitrarily small. However parametric resonance occurs (see \ref{sec:parRes}) for a large part of our parameter space and it has the effect of randomizing the initial misalignment angle, so that the estimate from post-inflationary symmetry breaking with an average angle of $\pi/\sqrt{3}$ applies \cite{Co:2020jtv}. Since this angle is much larger than the contribution from the fluctuations we take 
\begin{align}
    \sqrt{ \braket{\theta_i^2}} \simeq \frac{\pi}{\sqrt{3}}
\end{align}
in the following.
The Diraxion oscillates around the minimum of its cosine potential \eqref{eq:cos}, which reduces to the Diraxion mass $m_a^2$ in the small angle limit. During radiation domination oscillations start at the temperature 
\begin{align}\label{eq:formation}
    T^a_\text{ osc.}=  \left(\frac{45}{4\pi^3\; g_*(T_\text{osc.})}\right)^\frac{1}{4} \sqrt{M_\text{Pl.} m_a}\simeq \SI{27}{\tera\electronvolt} \cdot \left(\frac{100}{g_*(T_\text{osc.})}\right)^\frac{1}{4} \cdot  \sqrt{\frac{m_a}{1\;\text{eV}}} 
\end{align}
defined via $m_a= 3 H(T^a_\text{ osc.})$. The energy density from these coherent oscillations with an amplitude $\sqrt{\braket{\theta_i^2}} f_a$ is 
\begin{align}\label{eq:misalign}
    \rho_a^\text{mis.}(T^a_\text{ osc.}) = \frac{\braket{\theta_i^2} m_a^2 f_a^2}{2},
\end{align}
which normalized to the entropy density reads \cite{Co:2020xlh}
\begin{align}
    \frac{\rho_a^\text{mis.}(T^a_\text{ osc.})}{s(T^a_\text{ osc.})} \simeq \frac{2.4}{g_*(T^a_\text{ osc.})^\frac{1}{4}}
   \frac{ \sqrt{m_a} \braket{\theta_i^2}f_a^2}{M_\text{Pl.}^\frac{3}{2}}
\end{align}
and the relic abundance is found to be
\begin{align}
    \Omega_a^\text{mis.}\;h^2 &= \frac{\rho_a^\text{mis.}(T^a_\text{ osc.})}{s(T^a_\text{ osc.})} \cdot \frac{s_0}{\rho_c} h^2\\
    &\simeq 0.12\cdot \sqrt{\frac{m_a}{1\;\text{eV}}}\cdot \left(\frac{\braket{\theta_i^2}}{\pi^2/3}\right) \cdot \left(\frac{f_a}{\SI{5e11}{\giga\electronvolt}}\right)^2 \cdot\left(\frac{100}{g_*(T_\text{osc.}^a)}\right)^\frac{1}{4}
\end{align}
in terms of the  the critical density $\rho_c = 3/(8\pi G_N) H_0^2$, the Hubble rate today $H_0 \equiv  h\cdot  100\;\text{km/s}\;\text{Mpc}^{-1}$ with $h\simeq 0.7$ and  the entropy density today $s_0$ related via $\rho_c/(s_0 h^2)\simeq \SI{3.6}{\electronvolt}$.
The corresponding parameter space was displayed in figure \ref{fig:DMoverview}. 
Since we are typically interested in $f_a = \mathcal{O}(10^5-10^6\;\text{GeV})$, one can deduce that standard misalignment can not be responsible for the observed dark matter relic abundance.

\subsection{Topological Defect decay}\label{sec:defect}
Cosmic strings are formed via the Kibble mechanism \cite{Kibble:1976sj,Kibble:1982dd,Kibble:1980mv}, when  the  underlying  $\text{U}(1)_\text{D}$ symmetry is spontaneously broken down to $\mathcal{Z}_N$.  Detailed numerical simulations of the cosmic strings decaying to axions e.g. \cite{Chang:1998tb,Hagmann:1998me,Hagmann:2000ja,Martins:2000cs,Yamaguchi:2002sh,Moore:2001px,Hiramatsu:2010yu,Hiramatsu:2012gg,Hiramatsu:2012sc,Fleury:2015aca,Klaer:2017ond,Gorghetto:2018myk,Kawasaki:2018bzv,Vaquero:2018tib,Buschmann:2019icd,Klaer:2019fxc,Gorghetto:2020qws,Buschmann:2021sdq,OHare:2021zrq} typically have large uncertainties and often  disagree with each other, see e.g. \cite{Dine:2020pds} for a recent overview. This is why  we resort to simple order of magnitude estimates. Cosmic strings are characterized by their string tension or energy by unit length \cite{Kawasaki:2014sqa}
\begin{align}
    \mu_\text{str.} = \pi (N f_a)^2 \text{Log}\left(\frac{m_S}{\sqrt{\xi} H}\right),
\end{align}
where $\xi$ is a dimensionless length parameter that needs to be determined from numerical simulations. Following reference \cite{Hiramatsu:2012gg} we take $\xi\simeq 1$ and note that this parameter could very well be larger. Cosmic strings have a core diameter $d\sim 1/m_S$ and typical distances of $L\sim 1/H$, which acts as a regulator for global strings. We obtain the energy density via dividing the string tension by an area, which is given by the typical Hubble volume $4/3 \pi \cdot  1/H^3$ over the string separation $1/H$ to be
\begin{align}
    \rho_\text{str.}\simeq \frac{3}{4}\pi\;\text{Log}\left(\frac{m_S}{\sqrt{\xi} H}\right)\; H^2 (N f_a)^2.
\end{align}
Apart from cosmic strings there is a second type of defect that might be formed:  When the Diraxion oscillates at $T^a_\text{osc.}$  and settles into one of its $N$ degenerate minima the residual $\mathcal{Z}_N$ is explicitly broken.  Consequently domain walls, whose energy density interpolates between the different vacua, will be formed  around the time of $T^a_\text{osc.}$. A similar argument can be made for the case of kinetic misalignment: Since the Diraxion start its oscillation near the top of the potential, even a small fluctuation could lead it to fall into one of the available minima \cite{Co:2020jtv}. However these effects typically only occur for scenarios where the global $\text{U}(1)_\text{D}$ symmetry is broken after inflation: The observable universe consists of many patches with randomly distributed, different values of $\theta_i$ \cite{Turner:1990uz} and domain walls separating different patches. Our scenario on the other hand involves (spontaneous and explicit \cite{Dine:2004cq}) symmetry breaking during inflation, so one would expect a uniform $\theta_i$ throughout the observable universe \cite{Turner:1990uz}. However this does not guarantee the absence of domain walls.  Parametric resonance (see section \ref{sec:parRes}) grows  fluctuations and the resulting large fluctuations facilitate a non-thermal restoration of the $\text{U}(1)_\text{D}$  symmetry \cite{Tkachev:1995md,Kofman:1995fi,Kasuya:1996ns,Kasuya:1997ha,Kasuya:1998td,Tkachev:1998dc,Kasuya:1999hy} so that the angular field becomes randomized instead of being stuck in its value $\theta_\text{inf.}$ (see the beginning of \ref{sec:mis}). Once the fluctuations are smaller than $f_a$, the $\text{U}(1)_\text{D}$ symmetry is broken again with different initial angles for each Hubble patch leading to domain wall formation at $T^a_\text{osc.}$ \cite{Co:2020jtv}. If the condensate was thermalized before parametric resonance can occur, this outcome could be avoided. However as we argue in section \ref{sec:earlytherm}, such an early thermalization is not possible from the Dirac-Weinberg operator, as this UV-dominated rate would dampen the oscillation too much before a rotation could be induced. The only way out is to consider the Seesaw messenger fields as bath particles, which is the scenario sketched in \ref{app:earlysketch}.
Since the symmetry is restored only to be broken again, cosmic strings will also form and the resulting hybrid network of decays can decay for $N=1$ \cite{Vilenkin:1984ib,DAVIS1986225}. A second reason to expect domain walls is  that the isocurvature fluctuations can experience a power law growth \cite{Co:2020dya}: Due to the isocurvature fluctuations the different patches that make up our visible universe start with different initial velocities $\dot{\theta}_i$ and evolve to different field values. By the time  $T^a_\text{osc.}$, when the Diraxion mass becomes cosmologically relevant, domain walls will form to interpolate between the  different $\theta$. The surface tension or energy via unit area of a domain wall reads \cite{Hiramatsu:2012sc}
\begin{align}
    \sigma_\text{DW} = 8 m_a f_a^2 
\end{align}
and one finds the energy density from the ratio of the surface density over the typical length scale. Here we assume $\mathcal{O}(1)$ domain walls in our Hubble volume. For a Hubble volume $4/3 \pi \cdot  1/H^3$ and a typical surface area of $4\pi^2 \cdot 1/H^2$  the result is
\begin{align}\label{eq:domain}
    \rho_\text{DW} \simeq 24 H m_a f_a^2.
\end{align}
Domain walls would start to dominate a previously radiation dominated universe 
at a temperature of 
\begin{align}\label{eq:domDW}
    T_\text{DW}^\text{dom.} \simeq \SI{21}{\electronvolt}\cdot \sqrt{\frac{2}{g_*(T_\text{DW}^\text{dom.} )}} \cdot  \sqrt{\frac{m_a}{\SI{1}{\electronvolt}}} \cdot \left(\frac{f_a}{10^5\;\text{GeV}}\right).
\end{align}
The network collapses under its tension when \cite{Kawasaki:2014sqa}
\begin{align}
    H_\text{dec.} = \frac{\mu_\text{str.}}{\sigma_\text{DW} }  = \frac{8}{\pi N^2} \text{Log}\left(\frac{m_S}{H_\text{dec.}}\right)^{-1} m_a.
\end{align}
As long as $\text{Log}\left(m_S/ H_\text{dec.}\right) \gtrsim 8/(3\pi N^2)$  we find that $H_\text{dec.} \lesssim 3 m_a$ which means that the axions produced from the hybrid defect decay are produced at a time comparable to the conventional misalignment scenario \cite{Kawasaki:2014sqa}. For a QCD axion it was found that the logarithm can be as large as 70 \cite{Dine:2020pds}, so the previous conclusion is likely to hold true. Since the domain walls are expected to form at   $H\simeq 3 m_a$ and decay around the same epoch, they do not have enough time to dominate the energy budget of the universe (compare the formation and domination temperatures in \eqref{eq:formation} and \eqref{eq:domDW}). 
We find that the total energy density of the produced Diraxions 
\begin{align}\label{eq:strwall}
    \rho_\text{str.Wall} = \frac{48}{\pi}\left(4+\frac{1}{\pi}\right) \text{Log}\left(\frac{m_S}{ H_\text{dec.}}\right)^{-1} m_a^2 (N f_a)^2
\end{align}
is comparable to the conventional misalignment contribution \ref{eq:misalign}. 
The typical energy of Diraxions radiated by the network is about three times their mass \cite{Kawasaki:2014sqa} , so they are cold dark matter.\\
\\
For  $N\neq 1$ we have more than one domain wall. For odd dimensional operators responsible for the Diraxion mass we sketch in appendix \ref{app:dirOp} how to construct such operators while maintaining a domain wall number of one. Since we are however interested in an even dimensional operator with $N=6$, we have to invoke an additional operator with a dimension that is co-prime with $N=6$, explicitly breaking the residual $\mathcal{Z}_6$ that protects the domain wall \cite{Reig:2019sok}. This additional bias term will then lead to the annihilation of domain walls in neighboring vacua  \cite{PhysRevLett.48.1156}. For group theoretical solutions to the domain wall problem see \cite{LAZARIDES198221,BARR1982227}. We use the dimension seven operator from \eqref{eq:Diraxionmass} to define $\Delta m_a$ which is much smaller than $m_a$ from the dimension six operator
\begin{align}\label{eq:split}
    \frac{\Delta m_a}{m_a}  \simeq 8.4\times 10^{-7}\cdot \sqrt{\frac{c_7}{c_6}} \cdot \sqrt{\frac{N \;f_a}{10^6\;\text{GeV}}},
\end{align}
as long as we  assume $c_6\simeq c_7$. It is evident that the shift in $\varepsilon$ defined in \eqref{eq:varepsilon}  due to such values of  $\Delta m_a / m_a$ is negligible. The explicit symmetry breaking potential or bias term $V_7$  induces a different energy for each of the $N$ vacua and the  pressure difference between neighboring vacua  is proportional to
\begin{align}
    \Delta V = V_7(\theta=0)-V_7\left(\theta=\frac{2\pi}{N}\right) = \Delta m_a^2 f_a^2 \left(\cos(\delta_7) - \cos\left(\delta_7 + \frac{2\pi}{N}\right)\right),
\end{align}
where $\delta_7$ is the phase of the coefficient $c_7$ and  we will take the cosine term in brackets to be of order one from now on. The domain walls collapse if the pressure difference is larger than the domain wall tension implying 
\begin{align}
     H_\text{bias} = \frac{\Delta V}{\sigma_\text{DW}}  = \frac{\Delta m_a^2}{8 m_a}. 
\end{align}
Since  $H_\text{bias} \ll m_a$ the walls decay long after the onset of coherent oscillations. Reference \cite{Reig:2019sok} obtained that the domain walls annihilate before they dominate the energy budget as long as 
\begin{align}
    \frac{\Delta m_a}{m_a} \gg 3.3\times 10^{-12}\cdot \left(\frac{N\; f_a}{10^6\;\text{GeV}}\right),
\end{align}
which is compatible with \eqref{eq:split}. For our benchmark $m_a=\SI{0.1}{\electronvolt}$ and $\Delta m_a/m_a$ from \eqref{eq:split}, we find that the domain walls disappear before the onset of BBN at $T\simeq \SI{1}{\mega\electronvolt}$. 
There is an upper bound on the bias term from demanding that the vacua on both sides of the domain wall percolate sufficiently, which reads $\Delta m_a/ m_a < 0.2$ \cite{PhysRevD.39.1558,Hiramatsu:2010yz,Hiramatsu:2010yn} and is always satisfied for our scenario. The energy density of Diraxions is given by \eqref{eq:domain} evaluated at $H_\text{bias}$ and multiplied by $N$ to take the multiple walls into account. In terms of the energy density one finds  \cite{Reig:2019sok}
\begin{align}
    \rho_\text{bias} = 3 N \left(\frac{\Delta m_a}{m_a}\right)^2 m_a^2 f_a^2,
\end{align}
which corresponds to a relic abundance today of 
\begin{align}
     \Omega_a^\text{DW}\;h^2 \simeq 5\times 10^{-10}\cdot  \left(\frac{m_a}{\Delta m_a} \right)\cdot \sqrt{\frac{m_a}{\SI{0.1}{\electronvolt}}} \cdot \left(\frac{f_a}{10^6\;\text{GeV}}\right)^2,
\end{align} 
that can be much larger than the previous abundance \eqref{eq:strwall} and the misalignment contribution \eqref{eq:misalign} due to $\Delta m_a \ll m_a$. The numerical simulations of reference  \cite{Kawasaki:2014sqa} revealed that the energy of the Diraxions from such long lived domain walls  is  about twice their rest mass, so they would be cold dark matter. Fixing the relic abundance would require 
\begin{align}
    \frac{\Delta m_a}{m_a} \simeq 1.4\times 10^{-9}\cdot \sqrt{\frac{m_a}{\SI{0.1}{\electronvolt}}} \cdot   \left(\frac{f_a}{10^6\;\text{GeV}}\right)^2 \cdot \left(\frac{0.12}{ \Omega_a^\text{DW}\;h^2}\right),
\end{align}
which is typically too small to be realized in our case, see \eqref{eq:split} with $c_6\simeq c_7$. In other words we expect the Diraxions produced from long-lived domain walls to contribute at most $\Omega_a ^\text{DW}\;h^2 \simeq \mathcal{O}(10^{-4})$ due to \eqref{eq:split}.
The amount of gravitational radiation produced in the domain wall decay computed by \cite{Saikawa:2017hiv} turns out to be negligible for our range of parameters.
\\
\\
Note that string-wall networks with $N>1$ can have spherically collapsing closed walls leading to the formation of potentially over-abundant primordial black holes (PBH) \cite{Ferrer:2018uiu,Gelmini:2022nim,Gelmini:2023ngs}. Reference \cite{Gelmini:2023kvo} points out that more numerical simulations are needed in order to conclusively treat this effect and further mentions that the presence of angular momentum could lead to a sufficient deviation from the spherical symmetry impeding the  PBH formation. Since  here the rotating Diraxion background constitutes a source of angular momentum our scenario could be safe from this effect, but in any case a detailed study would be needed.\\
\\
We conclude that topological defects do not pose a cosmological problem in our scenario. Since we focus on decay constants in the $\mathcal{O}(10^5\;\text{GeV}-10^6\;\text{GeV})$ range, we find that the contributions from coherent oscillations and topological defect decay are negligible compared to the Kinetic Misalignment and Parametric resonance scenarios encountered in the next sections.

\subsection{Kinetic Misalignment and fragmentation}\label{sec:kinmis}
If the Diraxion velocity $\dot{\theta}$ is larger than $2 m_a$ \cite{Co:2020xlh} the rolling pesudoscalar will get trapped in its cosine potential later than for conventional misalignment. As a consequence of its large kinetic energy it does not just probe the harmonic part of its potential (small angles) but instead can start near the hilltop. Due to these  effects one obtains the dark matter yield \cite{Co:2019jts} from \eqref{eq:yieldRD} and \eqref{eq:yieldRH}
\begin{align}\label{eq:kinmisabund}
    Y_a = 2 Y_\theta,
\end{align}
where the factor of two, found numerically in \cite{Co:2019jts} and analytically in \cite{Eroncel:2022vjg},  encodes the enhancement from the anharmonicity. This mechanism reproduces the observed relic abundance for \cite{Co:2020xlh}
\begin{align}\label{eq:yieldrelic}
    & \Omega_a^\text{KM}\;h^2 = 0.12 \cdot \left(\frac{m_a}{1\;\text{eV}}\right)\cdot \left(\frac{Y_\theta}{0.4}\right)\\
     &\simeq0.12\cdot
     \begin{cases}
         \left(\frac{m_a}{\SI{0.01}{\electronvolt}}\right)^3 \cdot   \left(\frac{\SI{270}{\mega\electronvolt}}{m_S}\right)^\frac{5}{2}\cdot   \left(\frac{S_i}{0.1\;M_\text{Pl.}}\right)^\frac{7}{2}\cdot   \left(\frac{10^7\;\text{GeV}}{f_a}\right)^\frac{3}{2}\quad &\text{RD}\\
         \left(\frac{m_a}{\SI{0.01}{\electronvolt}}\right)^3 \cdot \left(\frac{\SI{140}{\mega\electronvolt}}{m_S}\right)^3\cdot   \left(\frac{S_i}{0.1\;M_\text{Pl.}}\right)^3\cdot \left(\frac{10^7\;\text{GeV}}{f_a}\right)\cdot \left(\frac{T_\text{RH}}{10^{13}\;\text{GeV}}\right)\quad &\text{RH},
     \end{cases}
\end{align}
where we used $N=6$ together with \eqref{eq:yieldRD} and \eqref{eq:yieldRH} for our estimate.
We show the required value of $Y_\theta$ as a function of $m_a$ on the upper axis in \ref{fig:DMoverview}. To take into account the constraint on $\varepsilon$ in \eqref{eq:masses2} we eliminate $m_S$ for $N=6$
\begin{align}
    m_S = \SI{222}{\giga\electronvolt} \cdot \sqrt{\frac{0.8}{\varepsilon}}\cdot \left(\frac{m_a}{\SI{0.6}{\electronvolt}}\right) \cdot \left(\frac{\SI{5e5}{\giga\electronvolt}}{f_a}\right) \cdot \left(\frac{S_i}{0.1\;M_\text{Pl.}}\right)
\end{align}
and find
\begin{align}
     \Omega_a^\text{KM}\;h^2 = 0.12\cdot
     \begin{cases}
    \left(\frac{\varepsilon}{0.8}\right)^\frac{5}{4}\cdot  \sqrt{\frac{m_a}{\SI{0.6}{\electronvolt}}} \cdot \left(\frac{f_a}{\SI{5e5}{\giga\electronvolt}}\right)\cdot \left(\frac{S_i}{0.1\;M_\text{Pl.}}\right)\quad &\text{RD}\\
     \left(\frac{\varepsilon}{0.8}\right)^\frac{3}{2}\cdot \left(\frac{f_a}{\SI{5e5}{\giga\electronvolt}}\right)^2 \cdot \left(\frac{T_\text{RH}}{\SI{9e15}{\giga\electronvolt}}\right)\quad &\text{RH}
     \end{cases}.
\end{align}
We can accommodate the parameter region $\SI{0.1}{\electronvolt}\lesssim m_a \lesssim \mathcal{O}(\SI{1}{\electronvolt})$ and $f_a\simeq 10^5\;\text{GeV}$ for detecting Diraxion dark matter with PTOLEMY via decays to neutrinos  depicted in figure \ref{fig:DMoverview} via kinetic misalignment. One can fix $m_a, f_a$ by adjusting the combination of parameters $\varepsilon, S_i$ ($\varepsilon, T_\text{RH}$) for rotations initiated during radiation domination (reheating). This typically leads to $\varepsilon\gtrsim 0.1$ once $S_i$ is determined via thermalization (see section \ref{sec:therm}) and $T_\text{RH}$ is accounted for by \eqref{eq:condRH} for oscillations before radiation domination. We picked a rather large value of $\varepsilon\simeq 0.8$ in the above estimate, to ensure the absence of parametric resonance (see \eqref{eq:ADAD} in the next section), which for the previous parameter range would overclose the universe (see \eqref{eq:yieldPR}) with too warm dark matter (consult \eqref{eq:warmness}). The used value of $m_S/(N f_a)$ 
is about a factor of three to large to comply with the bound from isocurvature fluctuations for a Hubble induced mass in \eqref{eq:bound22}. Since all of our estimates come with $\mathcal{O}(1)$ uncertainties, we treat this parameter point as right on the edge of the allowed parameter space. If (for rotations during radiation domination) we decrease $S_i$ and hence the relic abundance by a factor of three, we can make the Saxion mass compatible again.
This parameter region only produces dark radiation (see section \ref{sec:Radiation}) with $\Delta N_\text{eff.}\lesssim 0.01$ due to the large $m_S$  (see \eqref{eq:fa1} and also figure  \ref{fig:SaxionOverview}) required by the constraint on $\varepsilon=0.8$.\\
\\
Alternatively fixing the ratio $f_s/S_i$ via the relations \eqref{eq:fa1} and \eqref{eq:fa2} together with explaining the observed baryon asymmetry via Dirac Lepto-Axiogenesis (see \eqref{eq:leptoaxio1} and \eqref{eq:leptoaxio2}) allows us to estimate the required Diraxion and Saxion masses for DM from kinetic misalignment:
\begin{align}
    m_a &= 
    \begin{cases}
        \SI{0.1}{\electronvolt} \cdot \left(\frac{0.01}{\Gamma_S(S_i)/H(T_\text{osc})}\right)^\frac{7}{3}\cdot \left(\frac{\sum m_\nu^2}{(0.05\;\text{eV})^2}\right)^\frac{4}{3}\cdot \left(\frac{S_i}{0.1\;M_\text{Pl.}}\right)^\frac{2}{3} \cdot \left(\frac{0.12}{\Omega_a^\text{KM}\;h^2}\right)^\frac{1}{3}\;&\text{RD}\\
         \SI{0.66}{\electronvolt} \cdot \left(\frac{0.01}{\Gamma_S(S_i)/H(T_\text{osc}^\text{RH})}\right)^4\cdot \left(\frac{\sum m_\nu^2}{(0.05\;\text{eV})^2}\right)^2\cdot \left(\frac{S_i}{0.1\;M_\text{Pl.}}\right)^2\cdot \left(\frac{10^{13}\;\text{GeV}}{T_\text{RH}}\right) \cdot \left(\frac{0.12}{\Omega_a^\text{KM}\;h^2}\right)\; &\text{RH}
    \end{cases}\label{eq:case1KM}\\
    m_S &= 
    \begin{cases}
    \SI{11.5}{\giga\electronvolt}\cdot \left(\frac{0.01}{\Gamma_S(S_i)/H(T_\text{osc})}\right)^3\cdot\left(\frac{\sum m_\nu^2}{(0.05\;\text{eV})^2}\right)^\frac{3}{2}\cdot \left(\frac{S_i}{0.1\;M_\text{Pl.}}\right)^2 \cdot \left(\frac{0.12}{\Omega_a^\text{KM}\;h^2}\right)\; &\text{RD}\\
    \SI{14}{\giga\electronvolt}\cdot \left(\frac{0.01}{\Gamma_S(S_i)/H(T_\text{osc})}\right)^\frac{14}{3}\cdot\left(\frac{\sum m_\nu^2}{(0.05\;\text{eV})^2}\right)^\frac{13}{6}\cdot \left(\frac{S_i}{0.1\;M_\text{Pl.}}\right)^\frac{10}{3}\cdot \left(\frac{10^{13}\;\text{GeV}}{T_\text{RH}}\right)  \cdot \left(\frac{0.12}{\Omega_a^\text{KM}\;h^2}\right)^\frac{5}{3}\; &\text{RH}
    \end{cases}\label{eq:case2KM}
\end{align}
In section \ref{sec:Diss} we check this parameter range against the constraints from the conditions $\varepsilon<1$ in \eqref{eq:masses2},  $\dot{\theta}<T$  from \eqref{eq:cond1} and $S>T$ from \eqref{eq:cond2}.\\
\\
The Diraxion gets trapped in its potential at the temperature \cite{Eroncel:2022vjg}
\begin{align}
    T_* \simeq  \SI{0.23}{\giga\electronvolt}\cdot \left(\frac{100}{g_*(T_*)}\right)^\frac{1}{3} \cdot  \left(\frac{m_a}{\SI{10}{\milli\electronvolt}}\right)^\frac{2}{3} \cdot 
    \left(\frac{f_a}{10^6\;\text{GeV}}\right)^\frac{2}{3}
    \cdot \left(\frac{0.12}{\Omega_a h^2}\right)^\frac{1}{3} 
    \end{align}
defined as the time when the Diraxion's kinetic and potential energy coincide. Kinetic misalignment occurs for  $T_* < T^a_\text{ osc.}$ and   one finds that  this implies \cite{Eroncel:2022vjg}
\begin{align}
    f_a < \SI{6e11}{\giga\electronvolt} \cdot \left(\frac{g_*(T_*)}{100}\right)^\frac{1}{8} \cdot \left(\frac{\SI{10}{\milli\electronvolt}}{m_a}\right)^\frac{1}{4} \cdot \sqrt{\frac{\Omega_a h^2}{0.12}}.
\end{align}
Since the Diraxion  scans its potential for a long time, parametric resonance from the Diraxion self-interactions becomes possible leading to a fragmentation of the zero mode condensate into higher momentum excitations.
Parametric resonance was first discussed in the context of (p)reheating \cite{Kofman:1994rk,Kofman:1997yn}. 
This effect was then applied to the study of  relaxions and axions from monodromy \cite{Jaeckel:2016qjp,Berges:2019dgr,Fonseca:2019ypl} and recently to kinetic misalignment in \cite{Eroncel:2022vjg}. The basic idea is that the mass term for the higher momentum fluctuations $a_k$ is time dependent and acts like the external  force for a driven oscillator \cite{Co:2022aav}
\begin{align}
    \ddot{a}_k + \left(k^2 +m_a^2 \cos(\dot{\theta}t)\right) a_k=0.
\end{align}
In the limit $\dot{\theta}\gg m_s$ one finds a narrow resonance band around the momentum $k\simeq \dot{\theta}/2$ with a relative width $\Delta k/k \simeq m_a^2/\dot{\theta}^2$ and that the fluctuations are produced with a rate $\Gamma_{a \;\text{PR}}\simeq m_a^4/\dot{\theta}^3$ \cite{Fonseca:2019ypl}. 
It turns out that the Diraxion abundance including fluctuations coincides with the zero mode estimate in \eqref{eq:kinmisabund} \cite{Co:2021rhi},
which can be understood by noting that the characteristic energy scale of both processes is $\dot{\theta}$ \cite{Harigaya:2021txz}. 
Fragmentation can even lead to a slight enhancement of the relic abundance, because the zero mode redshifting as $\rho_\theta \sim 1/a^6$ is converted into fluctuations that redshift slower than $1/a^6$ \cite{Eroncel:2022vjg}. As a rule of thumb fragmentation is more important for smaller decay constants \cite{Eroncel:2022vjg}:
\begin{align}\label{eq:condss}
    f_a <   \left(\frac{g_*(T_*)}{100}\right)^\frac{1}{8} \cdot \left(\frac{\SI{10}{\milli\electronvolt}}{m_a}\right)^\frac{1}{4} \cdot \sqrt{\frac{\Omega_a h^2}{0.12}} \cdot 
    \begin{cases}
        \SI{8.79e10}{\giga\electronvolt} &\quad \text{weak frag.}\\
        \SI{4.39e10}{\giga\electronvolt} &\quad \text{complete frag.}\\
         2.25\times 10^{10}\;\text{GeV}&\quad \text{more complic.}
    \end{cases}
\end{align}
One can show that the condensate does not fragment completely  as long as $ f_a < \SI{8.79e10}{\giga\electronvolt}$ \cite{Eroncel:2022vjg},
meaning that most of its energy density is still in the zero mode and the fragmentation ends after the trapping has taken place. Complete fragmentation occurs before the trapping can take place \cite{Eroncel:2022vjg} and implies $f_a < \SI{4.39e10}{\giga\electronvolt} $.  The analysis of  \cite{Eroncel:2022vjg} breaks down if the backreaction of the fluctuations on the zero mode occurs before the onset of fragmentation, which can be expressed as $m_a/H(T_*) < \mathcal{O}(10^3-10^4)$  corresponding to $f_a < 2.25 \times 10^{10}\;\text{GeV}$. This does not mean that the corresponding parameter space is excluded, just that a more complicated analysis e.g. a lattice study is needed. The aforementioned regions in which kinetic misalignment and fragmentation take place are shown as orange lines in figure \ref{fig:DMoverview}.
Since the typical momenta produced during parametric resonance can be much larger than the Diraxion mass, the fluctuations will be less cold than the zero mode oscillations. We need to ensure that the dark matter redshifts enough so that it is sufficiently cold by the time of matter-radiation equality.
Reference \cite{Eroncel:2022vjg} found that the momentum modes  $m_a   a(T_*)$, with $a(T_*)\sim 1/T_*$ the scale factor at the trapping temperature, undergo the most efficient growth from parametric resonance. The warmness bound on the dark matter velocity reads \cite{Irsic:2017ixq,Lopez-Honorez:2017csg}
\begin{align}
    v_{a\;\text{eq.}} \simeq \frac{m_a \cdot a(T_*)/a_\text{eq.}}{m_a}  \lesssim 2\times 10^{-4}.
\end{align}
implying $T_* \gtrsim 5\times 10^{3}\; T_\text{eq.}\simeq \SI{5}{\kilo\electronvolt}$ and
\begin{align}\label{eq:struct}
    f_a > \SI{9.3e-2}{\giga\electronvolt}\cdot  \sqrt{\frac{g_*(T_*)}{100} }\cdot \left(\frac{\SI{10}{\milli\electronvolt}}{m_a}\right) \cdot \sqrt{\frac{\Omega_a h^2}{0.12}},
\end{align}
which is hardly constraining . One should also take into account that, if the field is still not trapped at the time of Big Bang Nucleosynthesis (BBN) then its energy density scaling as $\rho_\theta \sim 1/a^6$, which is different from ordinary dark radiation $\sim 1/a^4$, could modify the expansion history of the universe and alter the produced light element abundances. This condition  $T_* \gtrsim  \SI{20}{\kilo\electronvolt}$  leads to the bound \cite{Eroncel:2022vjg}
\begin{align}
    f_a >  \SI{8.1e-2}{\giga\electronvolt}\cdot  \left(\frac{\SI{10}{\milli\electronvolt}}{m_a}\right) \cdot \left(\frac{\Omega_a h^2}{0.12}\right),
\end{align}
comparable to the structure formation one.

\subsection{Parametric Resonance from Saxion oscillations}\label{sec:parRes}
\begin{figure}
    \centering
    \includegraphics[width=0.7\textwidth]{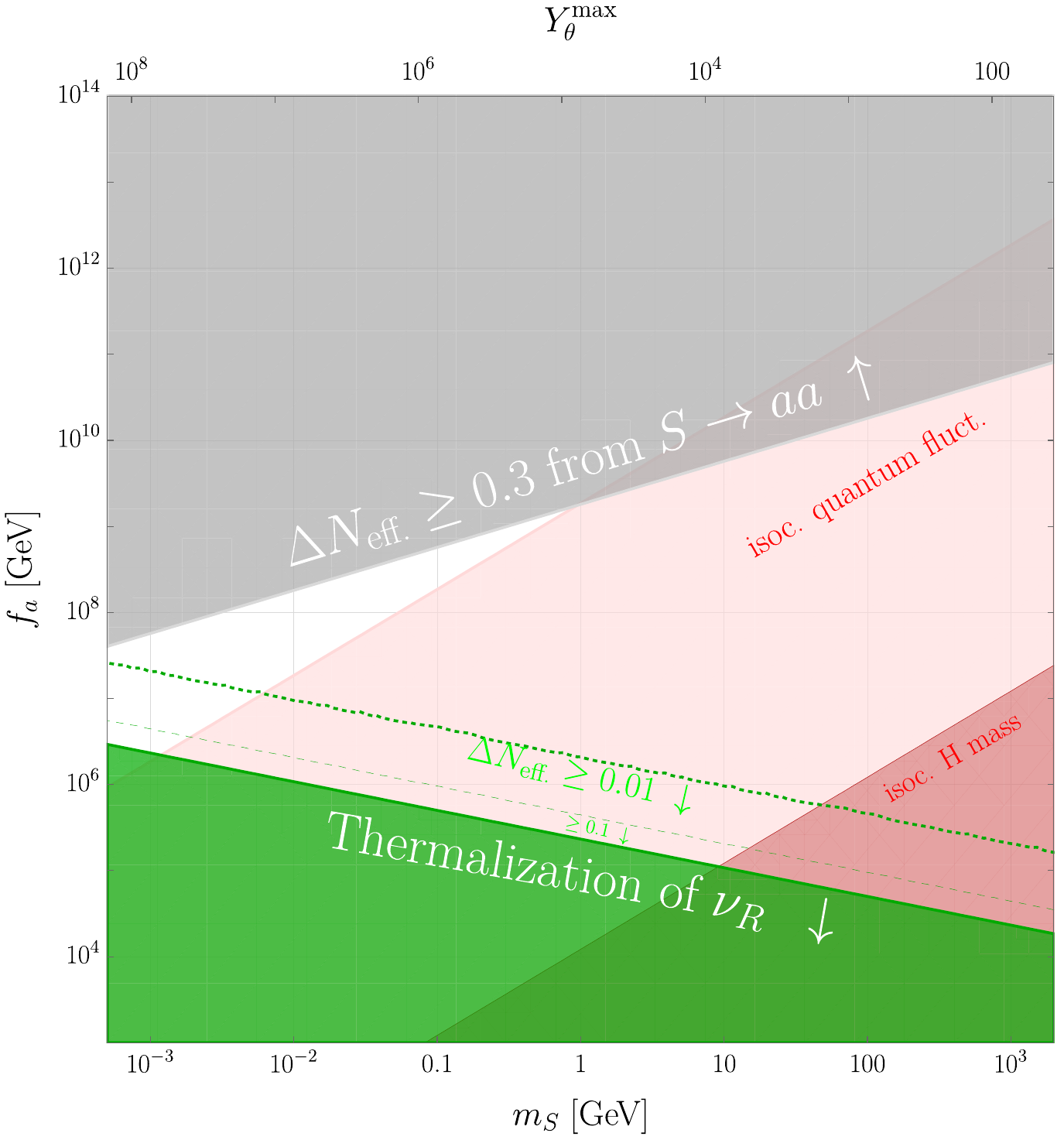}
    \caption{Saxion parameter space in terms of its mass $m_S$ and decay constant $f_a$ for the isocurvature and dark radiation constraints. The gray region illustrates the amount of dark radiation produced from out-of-equilibrium Saxion decays. The green lines denote the amount of $\nu_R$ dark radiation produced from scattering with the Dirac Weinberg operator. The dotted (dashed) line corresponds to $\Delta N_\text{eff.} \geq 0.01\;(0.1)$. In the region labelled \enquote{Thermalization} the freeze-in approximation breaks down as the $\nu_R$ would thermalize (see the discussion before and after \eqref{eq:nurtherm}).}
    \label{fig:SaxionOverview}
\end{figure}

The non-perturbative process of Parametric resonance  \cite{Kofman:1994rk,Kofman:1997yn} is analogous to stimulated emission and can also be driven by the Saxion oscillations \cite{Harigaya:2015hha,Co:2017mop}. One finds that the oscillating zero mode of the Saxion field is rapidly converted into higher momentum Saxion and Diraxion fluctuations.
Production of QCD axion dark radiation from Saxion parametric resonance was studied in \cite{Ballesteros:2016euj,Ballesteros:2016xej}. The fluctuations in the angular mode can be produced because both equations of motions in appendix \ref{app:eom} are coupled as long as $S> N f_a$.
Since the energy density in the fluctuations is comparable to the one in the zero mode, these large fluctuations will lead to an effective mass squared for the radial mode that is positive, hence the original  $\text{U}(1)_\text{D}$ symmetry  gets non-thermally restored \cite{Tkachev:1995md,Kofman:1995fi,Kasuya:1996ns,Kasuya:1997ha,Kasuya:1998td,Tkachev:1998dc,Kasuya:1999hy}. Once the amplitude of the fluctuations redshifts below $f_a$ the symmetry is broken again. Saxion fluctuations will later be thermalized and only the rotation together with the Diraxion fluctuations will remain. It was found that the transfer from the condensate into the higher momentum modes begins shortly after the start of the oscillations and for a quartic potential during radiation (matter) domination it ends at around $S\simeq 10^{-2} (10^{-4}) S_i$ \cite{Kasuya:1998td,Kasuya:1999hy,Kawasaki:2013iha}, when the backreaction  from scattering of the fluctuations with the condensate and themselves becomes important. 
Parametric resonance from Saxion oscillations requires a violation of the adiabaticity condition 
$|\dot{m_S}(S_i)/m_S(S_i)^2|<1 $ which can be translated into the condition \cite{Co:2020dya}
\begin{align}\label{eq:ADAD}
    \varepsilon<0.8\;.
\end{align}
During parametric resonance comparable amounts of Saxion and Diraxion fluctuations are produced \cite{Co:2020dya}
\begin{align}
    n_S\sim n_\theta \sim \frac{V_\sigma(S_i)}{m_S(S_i)},
\end{align}
which is why the abundance of Diraxion fluctuations is formally equivalent to the result for kinetic misalignment  under the replacement $\varepsilon\rightarrow 1/2$ in the definitions of the yields \eqref{eq:yieldRD} and  \eqref{eq:yieldRH}. This is also the reason why parametric resonance becomes the leading production channel for dark matter for $\varepsilon<1/2$. We estimate  that 
\begin{align}\label{eq:yieldPR}
   \Omega_a^\text{PR}\;h^2 &\simeq  0.12   \cdot
     \begin{cases}
         \left(\frac{m_a}{\SI{0.01}{\electronvolt}}\right) \cdot   \sqrt{\frac{\SI{500}{\mega\electronvolt}}{m_S}}\cdot   \left(\frac{S_i}{0.1\;M_\text{Pl.}}\right)^\frac{3}{2}\cdot   \sqrt{\frac{10^7\;\text{GeV}}{f_a}}\quad &\text{RD}\\
         \left(\frac{m_a}{\SI{0.01}{\electronvolt}}\right) \cdot \left(\frac{\SI{500}{\mega\electronvolt}}{m_S}\right)\cdot   \left(\frac{S_i}{0.1\;M_\text{Pl.}}\right)\cdot \left(\frac{10^7\;\text{GeV}}{f_a}\right)\cdot \left(\frac{T_\text{RH}}{10^{13}\;\text{GeV}}\right)\quad &\text{RH},
     \end{cases}
\end{align}
where we used $N=6$ together with \eqref{eq:yieldRD} and \eqref{eq:yieldRH}. We can determine the Diraxion and Saxion mass by eliminating $f_a/S_i$ via \eqref{eq:fa1}, \eqref{eq:fa2} and demanding successful Dirac-Lepto-Axiogenesis:
\begin{align}
    m_a &= 
    \begin{cases}
        \SI{0.11}{\electronvolt} \cdot \left(\frac{0.01}{\Gamma_S(S_i)/H(T_\text{osc})}\right)^3\cdot \left(\frac{\sum m_\nu^2}{(0.05\;\text{eV})^2}\right)^2\cdot \left(\frac{S_i}{0.1\;M_\text{Pl.}}\right)^2 \cdot \left(\frac{0.12}{\Omega_a^\text{PR}\;h^2}\right)\; &\text{RD}\\
        \SI{0.05}{\electronvolt} \cdot \left(\frac{0.01}{\Gamma_S(S_i)/H(T_\text{osc})}\right)^\frac{12}{5}\cdot \left(\frac{\sum m_\nu^2}{(0.05\;\text{eV})^2}\right)^\frac{16}{10}\cdot \left(\frac{S_i}{0.1\;M_\text{Pl.}}\right)^\frac{6}{5}\cdot \left(\frac{T_\text{RH}}{10^{13}\;\text{GeV}}\right)^\frac{1}{5}\cdot \left(\frac{0.12}{\Omega_a^\text{PR}\;h^2}\right)^\frac{3}{5}\;&\text{RH}
    \end{cases}\label{eq:case1PR}\\
    m_S &= 
    \begin{cases}
    \SI{14}{\giga\electronvolt}\cdot \left(\frac{0.01}{\Gamma_S(S_i)/H(T_\text{osc})}\right)^5\cdot\left(\frac{\sum m_\nu^2}{(0.05\;\text{eV})^2}\right)^\frac{7}{2}\cdot \left(\frac{S_i}{0.1\;M_\text{Pl.}}\right)^6\cdot \left(\frac{0.12}{\Omega_a^\text{PR}\;h^2}\right)^3\; &\text{RD}\\
    \SI{11.5}{\giga\electronvolt}\cdot \left(\frac{0.01}{\Gamma_S(S_i)/H(T_\text{osc})}\right)^2\cdot\left(\frac{\sum m_\nu^2}{(0.05\;\text{eV})^2}\right)^\frac{3}{2}\cdot \left(\frac{S_i}{0.1\;M_\text{Pl.}}\right)^2\cdot \left(\frac{T_\text{RH}}{10^{13}\;\text{GeV}}\right)\cdot \left(\frac{0.12}{\Omega_a^\text{PR}\;h^2}\right)\; &\text{RH}
    \end{cases}\label{eq:case2PR}
\end{align}
In section \ref{sec:Diss} we check this parameter range against the constraints from the conditions $\varepsilon<1$ in \eqref{eq:masses2},  $\dot{\theta}<T$  from \eqref{eq:cond1} and $S>T$ from \eqref{eq:cond2}.\\
\\
Owing to the fact that the relativistic fluctuations are produced with  momenta $k\sim \dot{\theta}(S_i)\sim m_S(S_i)$ warmness constraints become relevant again:
Even though the fluctuations are produced with much larger momenta compared to the fragmentation from Diraxion self-interactions, the non-perturbative production occurs much earlier at $T_\text{osc.}\gg T_*$ so in principle the fluctuations can have  enough time to redshift sufficiently. The constraint on the velocity is \cite{Irsic:2017ixq,Lopez-Honorez:2017csg}
\begin{align}
    v_a(T=1\;\text{eV}) \simeq \frac{k_{a}(T=1\;\text{eV})}{m_a}  \lesssim 2\times 10^{-4}
\end{align}
and signals from the cosmic 21-cm lines are expected to test $ v_a(T=1\;\text{eV})\gtrsim 10^{-5}$ \cite{Sitwell:2013fpa}. We estimate the momentum  $k_{a}(T=1\;\text{eV})$ at matter-radiation equality $T_\text{eq.}\simeq 1\;\text{eV}$  following \cite{Co:2020dya,Co:2020jtv}: Since $n_a\sim k_a^3\sim 1/a^3$ one deduces that $n_a/k_a^3$ is an adiabatic invariant. From $k_a \sim m_S(S)$ and $n_a \sim m_S(S) S^2$ we obtain
\begin{align}
    \frac{n_a}{k_a^3} \simeq \frac{1}{3} \left(\frac{N f_a}{m_S}\right)^2.
\end{align}
With this expression we trade the momenta for $n_a(T=1\;\text{eV})$ from the yield \eqref{eq:yieldrelic} that reproduces the dark matter relic abundance
and find
\begin{align}\label{eq:warmness}
    m_S < \SI{0.28}{\mega\electronvolt}\cdot   \left(\frac{v_a(T=1\;\text{eV})}{2\times10^{-4}}\right)^\frac{3}{2} \cdot \left(\frac{m_a}{\SI{10}{\milli\electronvolt}}\right)^2 \cdot \sqrt{\frac{0.12}{\Omega_a^\text{PR}h^2}} \cdot \left(\frac{N f_a}{10^6\;\text{GeV}}\right).
 \end{align}
One can see that the parameter region for DM from parametric resonance 
and Dirac-Lepto-Axiogenesis requires larger $m_S$ and is thus only viable with early thermalization (see section \ref{sec:earlytherm}).\\
\\
If we only try to explain the dark matter relic abundance detectable via PTOLEMY (see figure \ref{fig:DMoverview}), we can eliminate $m_S$ via \eqref{eq:warmness} for cold enough DM and  fix $m_a, f_a$ together with $N=6$ to obtain a lower limit of
\begin{align}
      \Omega_a^\text{PR}\;h^2 >  0.12 \cdot \left(\frac{2\times10^{-4}}{v_a(T=1\;\text{eV})}\right)   \cdot \left(\frac{S_i}{0.033 M_\text{Pl.}}\right)^2 \cdot 
     \begin{cases}
         1 \quad &\text{RD}\\
         \left(\frac{\SI{0.6}{\electronvolt}}{m_a}\right)^2 \cdot \left(\frac{T_\text{RH}}{\SI{5.7e14}{\giga\electronvolt}}\right)^2 \quad &\text{RH}
     \end{cases}.
\end{align}
Here for the rotation during radiation domination we find only one free parameter $S_i$.
The above parameter choice corresponds to a Saxion mass of
\begin{align} 
    m_S < \SI{60}{\giga\electronvolt}\cdot   \left(\frac{v_a(T=1\;\text{eV})}{2\times10^{-4}}\right)^\frac{3}{2} \cdot \left(\frac{m_a}{\SI{0.6}{\electronvolt}}\right)^2 \cdot \sqrt{\frac{0.12}{\Omega_a^\text{PR}h^2}} \cdot \left(\frac{ f_a}{10^7\;\text{GeV}}\right)
 \end{align}
and we had to pick a value of $f_a>10^5\;\text{GeV}$, outside of the band potentially detectable with PTOLEMY, to ensure that $\varepsilon$ stays below unity (see \eqref{eq:masses2})
\begin{align}
    \varepsilon> 3\times 10^{-3} \cdot \left(\frac{\Omega_a^\text{PR}}{0.12}\right) \cdot \left(\frac{2\times10^{-4}}{v_a(T=1\;\text{eV})}\right)^3 \cdot \left(\frac{\SI{0.6}{\electronvolt}}{m_a}\right)^2 \cdot \left(\frac{10^7\;\text{GeV}}{f_a}\right)^4 \cdot \left(\frac{S_i}{0.033\;M_\text{Pl.}}\right)^2.
\end{align}
Cold enough Diraxion dark matter from parametric resonance is not detectable with PTOLEMY via decays to neutrinos. Due to the larger $f_a$ we expect $\Delta N_\text{eff.}<0.01$ for this range of $m_S$ (see section \ref{sec:Radiation} together with  \eqref{eq:fa1} and the plot \ref{fig:SaxionOverview}).

\section{Dark Radiation}\label{sec:Radiation}

The Standard Model prediction for the  amount of  non-photonic radiation  is conventionally expressed as an effective number of neutrinos given by $ N_\text{eff.} =3.0432 \pm 0.0002$ \cite{Gnedin:1997vn,Mangano:2005cc,deSalas:2016ztq,Froustey:2020mcq,Akita:2020szl,Bennett:2020zkv,Cielo:2023bqp}.
Additional dark radiation would shift this number by \cite{Luo:2020fdt}
\begin{align}\label{eq:neff}
    \Delta N_\text{eff.}(T) =   \frac{4}{7}\; g_{*\rho}(T) \left(\frac{10.75}{g_{*S}(T)}\right)^\frac{4}{3}  \frac{\rho_\text{DR}(T)}{\rho_\text{SM}(T)}\quad \text{with} \quad \rho_\text{SM}(T) = \frac{\pi^2}{30} g_{*\rho}(T) T^4.
\end{align}
Current limits from BBN and the Planck data together with baryon accoustic oscillations (BAO) read \cite{Planck:2018vyg}
\begin{align}
    N_\text{eff}^\text{BBN} = 2.97^{+0.58}_{-0.54}, \quad N_\text{eff}^\text{Planck+BAO} = 2.99^{+0.34}_{-0.33},
\end{align}
which amounts to $\Delta N_\text{eff.}^\text{Planck+BAO} \simeq 0.28$.
A sensitivity of $\Delta N_\text{eff.}\lesssim 0.12$ could be reached by next generation experiments such as COrE \cite{CORE:2017oje}, Euclid \cite{2011arXiv1110.3193L}, SPT-3G  \cite{SPT-3G:2014dbx} or the Simons observatory \cite{SimonsObservatory:2018koc}. CMB-S4 \cite{Abazajian:2019eic,annurev-nucl-102014-021908} and PICO  \cite{NASAPICO:2019thw}  could even reach down to $\Delta N_\text{eff.}\simeq  0.06$ and CMB-HD might probe $\Delta N_\text{eff.}\simeq 0.014$ \cite{CMB-HD:2022bsz}.

\subsection{Freeze-In scattering from Dirac Weinberg operator}
We estimate the energy density of $\nu_R$ as follows
\begin{align}
    \rho_{\nu_R} \simeq \braket{E_{\nu_R}}\cdot 2 \cdot 3\cdot n_{\nu_R} 
\end{align}
where the factor of  two counts the helicity states and the three comes from the number of generations. The typical energy of a $\nu_R$ produced out of thermal equilibrium reads
\begin{align}
    \braket{E_{\nu_R}} = \text{Max}\left(2.5\;T, m_S(S)\right).
\end{align}
Freeze-in from a thermal bath leads to typical momenta of $2.5\; T$ \cite{Heeck:2017xbu}, whereas production from an non-thermal condensate involves momenta of the order of the  oscillation frequency $m_S(S)$ (see section \ref{sec:parRes}).
Since we have $m_S(S)\ll T$ (see section \ref{sec:dissp}) we take $ \braket{E_{\nu_R}}\simeq 2.5\;T$. As argued in section \ref{sec:barasym} the Dirac-Weinberg operator will only operate in the Freeze-in regime, which is why we can simplify the Boltzmann equation for the right handed neutrino abundance in terms of the thermally averaged cross sections $\braket{\sigma |v|}$ and the equilibrium number densities 
\begin{align}
    \frac{\text{d}}{\text{d}t}n_{\nu_R} + 3 H n_{\nu_R} &\simeq \left( \braket{\sigma |v|}_{L S\rightarrow H^\dagger \nu_R} n_L^\text{eq.}  + \braket{\sigma |v|}_{H S\rightarrow \overline{L} \nu_R} n_H^\text{eq.}  \right) n_S 
    \simeq \frac{7}{6}\; \Gamma_S(S) n_L^\text{eq.},
\end{align}
where we made use of the definition of the dissipation coefficient \eqref{eq:dissp} and used $n_H^\text{eq.}/n_L^\text{eq.}=4/3$. It is straightforward to find the number density frozen-in  \cite{Luo:2020sho,Luo:2020fdt} over a Hubble time $1/H$ and from that
\begin{align}
    \rho_{\nu_R}(T) \simeq \frac{35}{2}\cdot \frac{\Gamma_S(S)}{H(T)}\cdot n_L^\text{eq.}(T) \;T.
\end{align}
The scaling relations from \ref{sec:barasym} reveal that the energy density     UV-dominated during radiation domination since $\rho_{\nu_R}/\rho_\text{SM}$ scales like
$T$. Consequently the production is  peaked at the beginning of the oscillations. For production during radiation domination we obtain
\begin{align}\label{eq:neffRD}
    \Delta N_\text{eff.} 
    &\simeq 0.01 \cdot \left(\frac{\sum_\nu m_\nu^2}{(\SI{0.05}{\electronvolt})^2}\right)\cdot \sqrt{\frac{\SI{100}{\mega\electronvolt}}{m_S}} \cdot \left(\frac{S_i}{\SI{5e15}{\giga\electronvolt}}\right)^\frac{3}{2}\cdot \left(\frac{10^5\;\text{GeV}}{N f_a}\right)^\frac{3}{2}
\end{align}
If the oscillations occur before the completion of reheating, we have to take into account that the entropy of the decaying inflaton heats the SM plasma compared to the decoupled\footnote{We assume no coupling of the inflaton to $\nu_R$.} $\nu_R$. Another way to come to the same conclusion is the fact that 
$\rho_\text{SM}$ is only conserved after reheating.
Since $\rho_{\nu_R}\sim 1/a^4$ and the non-relativistic inflaton energy density redshifts as $\rho_\text{inf.}\sim 1/a^3$ we are lead to consider $\rho_{\nu_R}/\rho_\text{inf.}^{4/3}$ as an invariant, which scales as $(1/T^{13/3}, 1/T^{29/3})$ for $(T_\text{osc.}> T_\text{RH}>T_S,\;T_\text{osc.}>T_S> T_\text{RH})$. This implies that production is IR dominated and mostly occurs at $T_\text{RH}$
\begin{align}\label{eq:hlepidity}
    \frac{\rho_{\nu_R}(T_\text{RH})}{\rho_\text{SM}(T_\text{RH})}\simeq \frac{35}{2}\cdot \frac{\Gamma_S(S_i)}{H(T^\text{RH}_\text{osc.})}\cdot \left(\frac{T_\text{RH}}{T^\text{RH}_\text{osc.}}\right)^\frac{7}{3} \cdot \frac{n_L^\text{eq.}(T_\text{RH}) \;T_\text{RH}}{\rho_\text{SM}(T_\text{RH})},
\end{align}
where we used the scaling relations to restore  $\Gamma_S(S_i)/H(T_\text{osc.})\ll1$ and focused on the regime $T_\text{osc.}> T_\text{RH}>T_S$ where  $\Gamma_S(S )/H(T)\sim T^{7/3}$, because we typically find that the Saxion reaches its minimum after reheating. The amount of dark radiation is 
\begin{align}\label{eq:neffRH}
    \Delta N_\text{eff.} 
    &\lesssim 2\times 10^{-5} \cdot \left(\frac{\sum_\nu m_\nu^2}{(\SI{0.05}{\electronvolt})^2}\right)\cdot \left(\frac{S_i}{5\times10^{15}\;\text{GeV}}\right)^2 \cdot \left(\frac{10^5\;\text{GeV}}{N f_a}\right)^2 \cdot   \left(\frac{10^{15}\;\text{GeV}}{T_\text{RH}}\right),
\end{align}
where we eliminated $m_S$ using \eqref{eq:condRH}  to make sure the oscillation starts before the completion of reheating. Comparing equations \eqref{eq:neffRD} and  \eqref{eq:neffRH} reveals that typically less dark radiation is produced for oscillations during reheating due to the entropy dilution in \eqref{eq:hlepidity}. For both scenarios we find an upper limit of 
\begin{align}\label{eq:neffmax}
  \Delta N_\text{eff.} = 0.028 \cdot \left(\frac{\Gamma_S(S_i)/H(T_\text{osc.})}{0.1}\right)\cdot \left(\frac{100}{g_*(T_\text{osc.})}\right)^\frac{4}{3},
\end{align}
where $g_*(T_\text{osc.})$ has to be replaced with $g_*(T_\text{RH})$ for production before the end of reheating. Fixing $f_a$ via equations \eqref{eq:fa1} and \eqref{eq:fa2} is equivalent to fixing the amount of dark radiation produced from Freeze-In. The above value for the dark radiation abundance is only a factor of two away from the projected  sensitivity of CMB-S4 \cite{Abazajian:2019eic,annurev-nucl-102014-021908}, and PICO  \cite{NASAPICO:2019thw}. CMB-HD \cite{CMB-HD:2022bsz} would have sufficient sensitivity to probe this scenario. 
The regions corresponding to observable $ \Delta N_\text{eff.}$ from $\nu_R$ produced via scattering can be found in green in figure  \ref{fig:SaxionOverview}, where we made use of \eqref{eq:fa1}. 
Note that naively extrapolating $\Gamma_S(S_i)/H(T_\text{osc.})\rightarrow 1$ would lead to $\Delta N_\text{eff.}\simeq 0.28$, however such large rates would thermalize the $\nu_R$ (see the region in  \ref{fig:SaxionOverview} labelled \enquote{Thermalization}).  In that case the amount of dark radiation can be found from entropy conservation and it only depends on the decoupling temperature $T_{\nu_R}^\text{FO}$ of the three generations of  $\nu_R$   \cite{Abazajian:2019oqj}
\begin{align}\label{eq:nurtherm}
    \Delta N_\text{eff.} = 3\cdot g_{\nu_R}\cdot \frac{7}{8} \cdot 0.027 \cdot \left(\frac{100}{g_{*S}(T_{\nu_R}^\text{FO})}\right)^\frac{4}{3} = 0.142 \cdot \left(\frac{100}{g_{*S}(T_{\nu_R}^\text{FO})}\right)^\frac{4}{3},
\end{align}
where $g_{\nu_R}=2$ counts the spin degrees of freedom of the right handed neutrinos and the factor of $7/8$ arises due to their fermionic nature.
As will be explained in section \ref{sec:therm}, we are not interested in the regime  $\Gamma_S(S_i)/H(T_\text{osc.})\rightarrow 1$, since this would lead to additional damping of the Saxion and a delay in the commencement of its oscillations.\\
\\
There are additional model dependent processes that can source an abundance of $\nu_R$: For the Type I Dirac Seesaw this would be $S\;S \leftrightarrow \overline{\nu_R}\;\nu_R$ via exchange of a virtual $N$ and for the Type II case one can have $ \overline{L} \;L \leftrightarrow \overline{\nu_R}\;\nu_R$ by exchange of $\eta$. The rates scale as
\begin{align}
    \Gamma_{\nu_R} \simeq 
    \begin{cases} 
        Y_R^4 \;S^3/M_N^2 \quad &\text{Type I}\\
        Y_\nu^4 \; T^5/\mu_\eta^4 \quad &\text{Type II}
    \end{cases}
\end{align}
and since they are UV-dominated we can ensure that their contribution is negligible by demanding that they are slower than the rate from the Dirac Weinberg-operator in \eqref{eq:dissp} evaluated at the largest temperature $T_\text{max}$ defined in \eqref{eq:max} and the largest field value $S_i$.
By using the Seesaw relations in \eqref{eq:TypeISeesaw} and \eqref{eq:TypeIISeesaw} we eliminate the dependence on the heavy messenger masses and find for the Type I Dirac Seesaw that
\begin{align}
    \frac{Y_R}{Y_L} < 0.1 \cdot \sqrt{\frac{0.1\;M_\text{Pl}}{S_i}} \cdot \sqrt{\frac{T_\text{max}}{10^{15}\:\text{GeV}}}
\end{align}
and for the Type II case we arrive at 
\begin{align}
    \frac{\kappa}{Y_\nu} > 10^{13}\;\text{GeV}\cdot \left(\frac{0.1\;M_\text{Pl}}{S_i}\right) \cdot \left(\frac{T_\text{max}}{10^{15}\:\text{GeV}}\right)^2,
\end{align}
which is compatible with the active neutrino mass scale for $\mu_\eta / Y_\nu \gtrsim \SI{6e15}{\giga\electronvolt}$. This is compatible with \eqref{eq:couplings} and $S_i = 0.1\; M_\text{Pl.}$ as long as $Y_\nu$ is not too small.

\subsection{Out-of-equilibrium Saxion decays}\label{sec:DRdecay}
The Saxion can decay to Diraxions via its derivative coupling in \eqref{eq:derivDirax} with a width \cite{Mazumdar:2016nzr} 
\begin{align}
    \Gamma(S\rightarrow  a a) = \frac{1}{32 \pi}\frac{m_S^3}{(N f_a)^2}
\end{align}
and to neutrinos with a decay width, that is given by the expression \eqref{eq:decaAxion} for the decays of Diraxion to neutrinos under the replacement $m_a\rightarrow m_S$.
The decay to Diraxions is the dominant mode because $m_S^2 > 2 \sum_\nu m_\nu^2$.
Non-thermalized Saxions could basically produce an arbitrarily large $\Delta N_\text{eff.}$, which is why we require Saxion thermalization (see section \ref{sec:therm}).  The thermalized Saxions decouple at $T_D$ and could decay out of equilibrium to Diraxions at a later time $T_\text{dec.}<T_D$
\begin{align}
   T_\text{dec.} = \frac{0.08}{g_*(T_\text{dec.})^\frac{1}{4}} \frac{\sqrt{m_S^3 M_\text{Pl.}}}{N f_a}\simeq \SI{1.52}{\giga\electronvolt} \cdot \left(\frac{m_S}{\SI{1}{\giga\electronvolt}}\right)^\frac{3}{2}\cdot \left(\frac{10^6\;\text{GeV}}{N f_a}\right) \cdot \left(\frac{10.75}{g_*(T_\text{dec.})}\right)^\frac{1}{4}.
\end{align}
Its energy density at $T_D$ reads $\rho_S(T_D)=m_S T_D^3  \zeta(3)/\pi^2$ and since $\rho_S \sim n_S \sim 1/a^3$ it redshifts to $\rho_S(T_\text{dec.})=\rho_S(T_D) ( g_*(T_\text{dec.})\;T_\text{dec.}^3) /  (g_*(T_D)\; T_D^3 ) $.
Using this together with the definition of  $T_\text{dec.}$ one finds that \cite{Co:2020jtv}
\begin{align}
    \Delta N_\text{eff.}\simeq 0.25 \cdot \sqrt{\frac{\SI{10}{\mega\electronvolt}}{m_S}}\cdot \left(\frac{N f_a}{10^8\;\text{GeV}}\right)\cdot \left(\frac{100}{g_*(T_D)}\right)\cdot \left(\frac{10.75}{g_*(T_\text{dec.})}\right)^\frac{1}{12}.
\end{align}
The current bound $\Delta N_\text{eff.}<0.28$ \cite{Planck:2018vyg} excludes $ N f_a > 10^8\;\text{GeV}\cdot\sqrt{m_S/\SI{10}{\mega\electronvolt}}$. This bound is not very restrictive to our scenario due to the absence of stellar cooling constraints, which exclude $f_a <10^8\;\text{GeV}$ for a QCD axion. We showcase this bound as the gray area in figure \ref{fig:SaxionOverview}. However the above estimate only holds when the Saxion decouples from the thermal bath while relativistic.
In section \ref{sec:late} we show that for late thermalization the Saxion can stay in thermal equilibrium until it becomes non-relativistic and hence Boltzmann-suppressed. Furthermore one can see from  \ref{fig:SaxionOverview} that the channels for Diraxion and $\nu_R$ dark radiation exist in different regions of parameter space and only overlap for Saxion masses below the MeV-scale.

\section{Isocurvature perturbations}\label{sec:iso}
Since both the Saxion and the Diraxion are light during inflation we expect that quantum fluctuations imprint on them and lead to isocurvature modes \cite{Linde:1985yf,Lyth:1989pb,Kawasaki:1995ta,Beltran:2006sq}. For a quartic potential the typical time-scale over which $S$ relaxes to its minimum $N f_a$ is given by $t_\text{Relax} \simeq 1/m_S(S)$ \cite{Kusenko:2014lra,Yang:2015ida}. In order for the radial mode to remain stuck in its large initial field value until after inflation, we have to demand that the relaxation time is larger than age of the universe $t\simeq 1/H_I$ after inflation with a Hubble rate $H_I$. The  corresponding condition 
\begin{align}
    m_S(S) \leq H_I
\end{align}
immediately implies that there will be isocurvature fluctuations in the Saxion direction, which is typical for Affleck-Dine scenarios.

\subsection{Dark Matter and Baryon isocurvature}\label{sec:DMiso}
First we investigate dark matter and baryon isocurvature modes. The gauge invariant entropy perturbation reads \cite{Beltran:2006sq}
\begin{align}\label{eq:gauge}
    \mathcal{S}_a = \frac{\delta (n_a/s)}{n_a/s}  = \frac{\delta Y_\theta}{Y_\theta}
\end{align}
and we compute the amplitude of the isocurvature power spectrum following reference \cite{Co:2020dya} assuming that the Diraxion makes up the whole of dark matter:
\begin{align}
    \mathcal{P}_a(k_*) = \left\langle \left(\frac{\delta Y_\theta}{Y_\theta}\right)^2 \right\rangle  = \left(\frac{1}{Y_\theta }  \frac{\partial Y_\theta}{\partial \theta }\right)^2 \braket{\delta\theta_i^2}+\left(\frac{1}{Y_\theta }  \frac{\partial Y_\theta}{\partial S }\right)^2 \braket{\delta S_i^2}
\end{align}
We find the amplitude of the baryon isocurvature perturbations by an analogous calculation
\begin{align}
    \mathcal{P}_B(k_*) = \left\langle \left(\frac{\delta Y_B}{Y_B}\right)^2 \right\rangle  = \left(\frac{1}{Y_B }  \frac{\partial Y_B}{\partial \theta }\right)^2 \braket{\delta\theta_i^2}+\left(\frac{1}{Y_B }  \frac{\partial Y_B}{\partial S }\right)^2 \braket{\delta S_i^2}.
\end{align}
In the above we defined the fluctuations in the angular and radial modes in terms of the Hubble rate during inflation to be
 \begin{align}
    \sqrt{\braket{\delta\theta_i^2}} \equiv \frac{H_I}{2\pi S_i}, \quad  \sqrt{\braket{\delta S_i^2}}\equiv \frac{H_I}{2\pi}.
 \end{align}
Note that in the presence of multiple perturbations, one should add the perturbations in \eqref{eq:gauge} coherently to obtain the total amplitude. In our case the baryonic and dark matter fluctuations for $N=6$ appear with the same signs, so there is no destructive interference between them, which would give rise to compensated isocurvature  perturbations \cite{Grin:2013uya}. CMB observations \cite{Planck:2018jri} at the pivot scale $k_* = 0.05\;\text{Mpc}^{-1}$ determined the power spectrum of the adiabatic perturbations to be $P_\zeta(k_*) \simeq \SI{2.2e-9}{}$ and they constrain
\begin{align}
    \beta_\text{iso}\equiv \frac{ \mathcal{P}_a(k_*)}{ \mathcal{P}_a(k_*)+\mathcal{P}_\zeta(k_*)} < 0.038,
\end{align}
which implies 
\begin{align}\label{eq:isoDMbound}
     \mathcal{P}_a(k_*)< \frac{\beta_\text{iso}}{1-\beta_\text{iso}} \mathcal{P}_\zeta(k_*)\simeq \SI{8.7e-11}{}.
\end{align}
By rescaling the bound on $\mathcal{P}_a(k_*)$ to the observed baryon abundance and using $h^2 \Omega_B = 0.0224$ as well as $h^2 \Omega_a = 0.12$ \cite{Planck:2018vyg} we find \cite{Harigaya:2014tla}
\begin{align}
     \mathcal{P}_B(k_*) < \left(\frac{\Omega_a}{\Omega_B}\right)^2 \cdot \frac{\beta_\text{iso}}{1-\beta_\text{iso}} \mathcal{P}_\zeta(k_*)\simeq \SI{2.5e-9}{}.
\end{align}
Let us first deal with the contribution from the Diraxion fluctuations: Since only $\varepsilon$ depends on the angle $\theta$, we find that the usual and kinetic misalignment scenarios are relevant; parametric resonance depends only the initial Saxion number density. Furthermore we deduce from equations \eqref{eq:asym} and \eqref{eq:asymRH} that in our scenario the baryon asymmetry always depends on $\varepsilon$. Consequently we find for kinetic misalignment and baryogenesis that 
\begin{align}
 \frac{1}{Y_\theta }  \frac{\partial Y_\theta}{\partial \theta } =  \frac{1}{Y_B }  \frac{\partial Y_B}{\partial \theta}  = \frac{1}{\varepsilon} \frac{\partial \varepsilon}{\partial \theta} =  \cot(\theta_i+\delta)
\end{align}
where we used \eqref{eq:varepsilon} for the last equality.
If we choose $\theta_i + \delta = \pm \pi/2$ the cotangent vanishes and there will be no perturbations in the Diraxion direction \cite{DeSimone:2016ofp}. The Diraxion will be stuck in this value due to the Hubble friction. Hence we have to assume that the Diraxion is not aligned with the minimum of its potential which would instead enforce from equation \eqref{eq:mintheta} $\partial V_\slashed{D}/\partial \theta \sim \sin(\theta_i+\delta)=0$.
For conventional misalignment the amplitude of the dark matter isocurvature spectrum is given by the standard expression  \cite{Harigaya:2015hha} with $f_a$ replaced by $S_i$
\begin{align}
     \mathcal{P}_a^\text{mis.}(k_*) = \frac{4}{\braket{\theta_i^2}} \left(\frac{H_I}{2\pi S_i}\right)^2,
\end{align}
where $\braket{\theta_i^2}$ was defined in \eqref{eq:thetaeff}. Since parametric resonance will randomize the angle we take $\braket{\theta_i^2}\simeq \pi^2 /3$ \cite{Co:2020jtv} and find
\begin{align}
    \frac{m_S}{N f_a} < 
    \begin{cases}
    4.5\times 10^{-10}\quad&\text{quant. fluct.}\\
    7.5\times 10^{-5} \quad&\text{$H$-mass from $R$}
    \end{cases},
\end{align}
where the first line fixes $S_i$ via quantum fluctuations and in the second line it is induced from a Hubble-dependent mass term (see sections \ref{app:quant} and \ref{app:R} respectively).
For our parameter space with smaller $f_a$ regular misalignment can not be responsible for the majority of the dark matter relic abundance and the previous bounds disappear.\\
\\
In the following we only have to deal with the Saxion isocurvature perturbations: Using the definition of the charge yields 
in \eqref{eq:yieldRD} and \eqref{eq:yieldRH} together with the definition of $\varepsilon$ in \eqref{eq:varepsilon} we obtain for kinetic misalignment that \cite{Co:2020dya}
\begin{align}
    \mathcal{P}_a(k_*) = \frac{(N-x)^2}{4\pi^2} \frac{H_I^2}{S_i^2}, \quad \text{with} \quad x= \begin{cases}
        \frac{5}{2}\quad &\text{RD}\\
        3\quad &\text{RH}
    \end{cases}.
\end{align}
For parametric resonance we deduce from \eqref{eq:yieldRD} and \eqref{eq:yieldRH} that \cite{Co:2020dya}
\begin{align}
    \mathcal{P}_a(k_*) = \frac{1}{4\pi^2} \frac{H_I^2}{S_i^2} \cdot 
    \begin{cases}
        \frac{9}{4}\quad &\text{RD}\\
        1\quad &\text{RH}
    \end{cases}.
\end{align}
One can see that the spectra for parametric resonance correspond to the ones for kinetic misalignment under the replacement $N\rightarrow 4$, because here the relic abundance is induced from the Saxion potential $V_\sigma \sim S^4$ and not the Diraxion potential $V_\slashed{D}\sim S^N$ \cite{Co:2020dya}. For the baryonic perturbations we find using \eqref{eq:asym} and \eqref{eq:asymRH}
\begin{align}
       \mathcal{P}_B(k_*) = \frac{(N-y)^2}{4\pi^2} \frac{H_I^2}{S_i^2}, \quad \text{with} \quad y= \begin{cases}
        2\quad &\text{RD}\\
        \frac{5}{2}\quad &\text{RH}
    \end{cases}. 
\end{align}
We focus on the bound for the dark matter perturbations, since the limit is slightly stronger than for baryons. Our result for the scenario in section \ref{app:quant}, where $S_i$ originates from quantum fluctuations,  reads
\begin{align}\label{eq:bound11}
    \frac{m_S}{N f_a} < \frac{5.4\times 10^{-10}}{(N-x)^2}
\end{align}
and the case of a Hubble-dependent mass from the Ricci scalar in section \ref{app:R} is constrained to be
\begin{align}\label{eq:bound22}
    \frac{m_S}{N f_a} < \frac{8.2\times 10^{-5}}{N-x}.
\end{align}
The corresponding regions were drawn in red in figure \ref{fig:SaxionOverview} and we see that for the case of quantum fluctuations inducing $S_i$, most of the interesting parameter space would be excluded.
In section \ref{sec:quantcorr} we demonstrate under which conditions the Dirac Seesaw models do not induce radiative corrections  that violate the aforementioned bounds.  The large hierarchy between $m_S$ and $N f_a$ is the main drawback of using a quartic potential. This problem can be avoided in supersymmetric models such as \cite{Co:2020jtv,Kawamura:2021xpu,Co:2021qgl,Barnes:2022ren}, where the approximately quadratic scalar potential arises from supersymmetry breaking via soft masses in a two field model \cite{Kim:1983ia}, dimensional transmutation from the RGE running of soft masses \cite{Moxhay:1984am} or loop corrections in gauge mediated supersymmetry breaking \cite{Arkani-Hamed:1998mzz} in single field models. 

\subsection{Dark Radiation isocurvature}
We define the gauge invariant perturbation in the dark radiation fluid via the relation \cite{Adshead:2020htj}
\begin{align}
    \mathcal{S}_\text{DR}= \frac{3}{4} \frac{\delta (\rho_{\nu_R}/\rho_\text{SM})}{\rho_{\nu_R}/\rho_\text{SM}} = \frac{3}{4} \frac{\delta\; (\Delta N_\text{eff.})}{\Delta N_\text{eff.}},
\end{align}
where we made use of the fact that the photons are adiabatic with respect to the perturbations in the SM plasma to trade $\rho_\gamma$ for $\rho_\text{SM}$. The amplitude of the dark radiation isocurvature spectrum reads
\begin{align}
    \mathcal{P}_\text{DR}(k_*) = \left\langle \left(\frac{\delta\; (\Delta N_\text{eff.})}{\Delta N_\text{eff.}}\right)^2 \right\rangle  = \left(\frac{1}{\Delta N_\text{eff.}}  \frac{\partial \;\Delta N_\text{eff.}}{\partial S }\right)^2 \braket{\delta S_i^2}
\end{align}
and we already used the fact that $\Delta N_\text{eff.}$ does not depend on $\theta$ in our model, see e.g. \eqref{eq:neffRD} and \eqref{eq:neffRH}. Dark radiation isocurvature leads to a spatial variation of $\Delta N_\text{eff.}$ compared to its spatial average $\Delta \overline{N}_\text{eff.}$  \cite{Adshead:2020htj}
\begin{align}
    \Delta N_\text{eff.} \simeq \Delta \overline{N}_\text{eff.}\left( 1 +\frac{4}{3} \sqrt{\mathcal{P}(k_*)_\text{DR}}\right).
\end{align}
A spatially varying $\Delta N_\text{eff.}$ would lead to spatially varying abundances of light elements produced during BBN. Reference  \cite{Adshead:2020htj} sets constraints on this effect by making use of the $^4\text{He} /\text{D}$ data extracted from local galaxies  \cite{Cooke:2017cwo}  and measurements of $\text{D} /\text{H}$ obtained from high-redshift Lyman-$\alpha$ absorption systems \cite{Aver:2015iza}, which correspond to a pivot scale of $k_* \simeq 1 \;\text{Mpc}^{-1}$ and they obtain the limit
\begin{align}
    \mathcal{P}_\text{DR}(k_*)  <\frac{0.17}{\Delta \overline{N}_\text{eff.}^2}. 
\end{align}
CMB data with $k_*\simeq 0.1\;\text{Mpc}^{-1}$  can be used to set constraints of isocurvature modes in the left-chiral SM neutrinos, which can be recast as a bound on dark radiation \cite{Adshead:2020htj}
\begin{align}\label{eq:bound1}
    \mathcal{P}_\text{DR}(k_*) < 10^{-10} \left(\frac{N_\text{eff.}}{\Delta \overline{N}_\text{eff.}}\right)^2.
\end{align}
The authors of \cite{Ghosh:2021axu} recast CMB data with variable $\Delta N_\text{eff.}$, updating a similar study \cite{Kawasaki:2011rc} based on WMAP data, and their bound reads
\begin{align}\label{eq:bound2}
  \mathcal{P}_\text{DR}(k_*) < \frac{2\times 10^{-8}}{\Delta N_\text{eff.}^2}  
\end{align}
For simplicity we neglect the  correlation between the dark matter and dark radiation isocurvature perturbations, which both inherit the fluctuations from $S$.   By setting $N_\text{eff.}\simeq 3$ one can see that the bound \eqref{eq:bound1} is about an order of magnitude stronger than \eqref{eq:bound2}, so to be conservative we use \eqref{eq:bound1} to set limits on the amplitude of the power spectrum
\begin{align}
     \mathcal{P}_\text{DR}(k_*) = \frac{H_I^2}{S_i^2}\cdot 
    \begin{cases}
        0.032\quad &\text{RD}\\
        2.4\times10^{-3}\quad &\text{RH}
    \end{cases}.
\end{align}
From this result and for $N_\text{eff.}\simeq 3.045+\Delta N_\text{eff.}$ with $\Delta N_\text{eff.} =0.028$ from \eqref{eq:neffmax} we can deduce that for $S_i$ from quantum fluctuations (see section \ref{app:quant}) 
\begin{align}
    \frac{m_S}{N f_a} < 
    \begin{cases}
        7.5\times 10^{-6}\quad &\text{RD}\\
        10^{-4}\quad &\text{RH}
    \end{cases}
\end{align}
and for $S_i$ from a Hubble-dependent mass (see section \ref{app:R}) 
\begin{align}
    \frac{m_S}{N f_a} < 
    \begin{cases}
        9.7\times 10^{-3}\quad &\text{RD}\\
        3.5\times10^{-2}\quad &\text{RH}
    \end{cases}.
\end{align}
These limits are sub-leading to the ones from dark matter isocurvature in \eqref{eq:bound11} and \eqref{eq:bound22}.

\subsection{One Loop corrections}\label{sec:quantcorr}
Since isorcurvature perturbations typically require a tiny  Saxion quartic coupling $\lambda_\sigma$
\begin{align}\label{eq:lambdas}
    \lambda_\sigma = \frac{m_S^2}{2 N^2 f_a^2}< 
    \begin{cases}
    9.1\times 10^{-21}\quad&\text{quant. fluct.}\\
    2.1\times10^{-10}\quad&\text{$H$-mass from $R$}
    \end{cases},
\end{align}
where we used the bounds in \eqref{eq:bound11} as well as \eqref{eq:bound22} and set $N-x=3.5$ for the strongest limit, we have to check that radiative corrections to this parameter and to $m_S$ are under control. Here we write out only the finite pieces of the loop corrections and neglect logarithmic factors. The mixed quartic coupling with the SM like Higgs, the scalar doublet $\eta$ for the Type II scenario, or the triplet $\Delta$ in the Type III Seesaw induce corrections of \cite{Ballesteros:2016xej}
 \begin{align}
     \delta m_S^{(H)\;2} = -\frac{\lambda_{\sigma H}^2}{8\pi^2} (N f_a)^2, \quad \delta m_S^{(\eta)\;2} = -\frac{\lambda_{\eta\sigma}^2}{8\pi^2} (N f_a)^2, \quad \quad \delta m_S^{(\Delta)\;2} = -\frac{\lambda_{\Delta\sigma}^2}{8\pi^2} (N f_a)^2,
 \end{align}
where the minus signs take into account that at tree level $m_S^2<0$, as well as 
\begin{align}
    \delta \lambda_\sigma^{(H)} =  \frac{\lambda_{\sigma H}^2}{16\pi^2},\quad \delta \lambda_\sigma^{(\eta)} =  \frac{\lambda_{\eta\sigma}^2}{16\pi^2}, \qquad  \delta \lambda_\sigma^{(\Delta)} =  \frac{\lambda_{\Delta\sigma}^2}{16\pi^2}.
\end{align}
We find that for both kinds of corrections the mixed quartic couplings need to satisfy
\begin{align}
    |\lambda_{\sigma H}|, |\lambda_{\eta\sigma}|, |\lambda_{\Delta\sigma}| <     
    \begin{cases}
    1.2\times 10^{-9}\quad&\text{quant. fluct.}\\
    1.8\times10^{-4}\quad&\text{$H$-mass from $R$}
    \end{cases}.
\end{align}

\subsubsection{Type I-II Dirac Seesaws }
For the one-loop correction from the $N_L-\nu_R$ loop in the Type I Seesaw we can recycle the result for a Majorana Seesaw \cite{Brivio:2017dfq,Brivio:2018rzm}, because only one chirality of $N$ runs in the loop 
\begin{align}
  \delta m_S^{(\text{I})\;2} = \frac{Y_R^2}{8 \pi^2}M_N^2,\quad    \delta \lambda_\sigma^{(\text{I})} \simeq -\frac{5}{32 \pi^2}Y_R^4.
\end{align}
Both contributions come with a minus sign from the closed fermion loop, which for the two-point function was already absorbed in the definition of $m_S^2<0$.
The bound from the correction to the negative $m_S^2$ can be re-expressed by using the Type I Seesaw relation in \eqref{eq:TypeISeesaw} as 
\begin{align}\label{eq:oneloopI}
    M_N < \sqrt{Y_L} \cdot\left(\frac{N}{6}\right)\cdot \left(\frac{f_a}{10^6\;\text{GeV}}\right)\cdot  \left(\frac{(\SI{0.05}{\electronvolt})^2}{\sum m_\nu^2}\right)^\frac{1}{4} \cdot
    \begin{cases}
    \SI{4.6e8}{\giga\electronvolt}\quad&\text{quant. fluct.}\\    
    \SI{1.8e11}{\giga\electronvolt}\quad&\text{$H$-mass from $R$}
    \end{cases}
\end{align}
and the one from the quartic is given by
\begin{align}\label{eq:Iquartic}
   M_N < Y_L\cdot \left(\frac{N}{6}\right)\cdot \left(\frac{f_a}{10^6\;\text{GeV}}\right)\cdot \sqrt{\frac{(\SI{0.05}{\electronvolt})^2}{\sum m_\nu^2}} \cdot
    \begin{cases}
    \SI{8.1e14}{\giga\electronvolt}\quad&\text{quant. fluct.}\\
    \SI{3.2e17}{\giga\electronvolt}\quad&\text{$H$-mass from $R$}
    \end{cases}.
\end{align}
It is evident that the bound on $M_N$ from the correction to $m_S^2$ is the stronger one and would lead to values of $M_N$ that are in conflict with our cosmological assumptions in \eqref{eq:couplings} and \eqref{eq:thermmass}, so that the heavy $N$ could be produced from the plasma. On the other hand the bound from the correction to the quartic is compatible with our choices of e.g. $T_\text{RH}=(10^{14}-10^{15})\text{GeV}$. One way to avoid this conclusion is to assume that there is an accidental cancellation between the corrections $\delta m_S^{(\text{I})\;2}$ and $\delta m_S^{(H)\;2}$. This  would lead to 
\begin{align}
    M_N \simeq \SI{1.8e13}{\giga\electronvolt}\cdot\sqrt{Y_L}\cdot \sqrt{ \lambda_{\sigma H}} \cdot \left(\frac{N}{6}\right)\cdot \left(\frac{f_a}{10^6\;\text{GeV}}\right)  \cdot\left(\frac{(\SI{0.05}{\electronvolt})}{\sum m_\nu^2}\right),
\end{align}
which is still too small for our purposes. However we can ameliorate this problem by assuming the simultaneous presence of two or more different Dirac Seesaws.
The trilinear term connecting $\sigma, \eta$ and $H$ in the Type II Dirac Seesaw induces 
\begin{align}
   \delta m_S^{(\text{II})\;2} = -\frac{1}{16 \pi^2}\kappa^2,\quad  \delta \lambda_\sigma^{(\text{II})} \simeq \frac{1}{16\pi^2} \frac{\kappa^4}{\mu_\eta^4},
\end{align}
where we eliminate $\kappa$ by using the Type II Seesaw relation in \eqref{eq:TypeIISeesaw} and set a bound on $\mu_\eta$.
The limit from the correction to $m_S$ is given by
\begin{align}\label{eq:oneloopII}
    \mu_\eta <  \sqrt{ Y_\nu }\cdot\left(\frac{N }{6}\right)\cdot \left(\frac{f_a}{10^6\;\text{GeV}}\right)\cdot\left(\frac{(\SI{0.05}{\electronvolt})^2}{\sum m_\nu^2}\right)^\frac{1}{4}
    \begin{cases}
     3.3\times 10^{8}\;\text{GeV}\quad&\text{quant. fluct.}\\
    1.3\times10^{11}\;\text{GeV}\quad&\text{$H$-mass from $R$}
    \end{cases}
\end{align}
and is stronger than the bound from the correction to the quartic 
\begin{align}\label{eq:IIquartic}
    \mu_\eta <  Y_\nu  \cdot\left(\frac{N }{6}\right)\cdot \left(\frac{f_a}{10^6\;\text{GeV}}\right)\cdot \sqrt{\frac{(\SI{0.05}{\electronvolt})^2}{\sum m_\nu^2}}\cdot
    \begin{cases}
     5.1\cdot 10^{13}\;\text{GeV}\quad&\text{quant. fluct.}\\
    2\times10^{16}\;\text{GeV}\quad&\text{$H$-mass from $R$}
    \end{cases}.
\end{align}
We observe that the limit from $\delta m_S^{(\text{II})\;2}$ enforces values of $\mu_\eta$, which can be problematic in the context of equations  \eqref{eq:couplings} and \eqref{eq:thermmass}, just like for $M_N$ in the Type I scenario. If both the Type I and Type II seesaw are responsible for Dirac neutrino masses $m_\nu =m_\nu^{\text{(I)}}+m_\nu^{\text{(II)}}$ (before diagonalization), which is known as a Hybrid-Seesaw \cite{Wetterich:1998vh,Chen:2005jm}, their one loop corrections to $m_S^2$ could cancel each other as long as 
\begin{align}\label{eq:tuned}
    M_N \simeq \frac{1}{2} \sqrt{\frac{|m_\nu^{\text{(II)}}|}{|m_\nu^{\text{(I)}}|}}\sqrt{\frac{|Y_L|}{|Y_\nu|}}\; \mu_\eta,
\end{align}
owing to the fact that the bosonic and fermionic contributions have different signs.
In this case  the bounds from the quartic in \eqref{eq:Iquartic} and \eqref{eq:IIquartic} also get weakened. If we switch on both types of Dirac Seesaws we find a two loop correction of 
\begin{align}
    \delta m^{2}_{S\;(2)} \simeq \frac{1}{\left(16\pi^2\right)^2} \frac{\left| m_\nu^\text{(I)} M_N^2 + \left(m_\nu^{\text{(II)}}\right)^\dagger \mu_\eta^2  \right| ^2}{v_H^2 (N\;f_a)^2}.
\end{align}
By itself the Type I (II) Seesaw contribution would give a bound similar to \eqref{eq:oneloopI} (\eqref{eq:oneloopII})  with $Y_L = \mathcal{O}(1)\; \left(Y_\nu = \mathcal{O}(1)\right)$ and an additional loop factor. Owing to the fact that we have to coherently add the two loop contributions from both Seesaw schemes (there are four ways to glue together the diagrams in figure \ref{fig:DiracSeesawI-II} at two loop) there could be another  cancellation: We find that the two loop correction becomes zero if the phases of the different contributions to $m_\nu$ satisfy $\delta_\text{II}= -\delta_\text{I}-\pi$, which together with \eqref{eq:tuned}  would imply $|Y_L|\simeq 4 |Y_\nu|$.

\subsubsection{Type III Dirac Seesaw }
Analogously we find for the quartic term connecting $\sigma, \Delta$ and $H$ for both versions of the Type III Seesaw
\begin{align}
    \delta m_S^{(\text{III})\;2} = \frac{\lambda_4^2}{16 \pi^2}\left(v_H^2+v_\Delta^2\right),\quad \delta \lambda_\sigma^{(\text{III})} \simeq \frac{\lambda_4^4}{64\pi^2} \left(\frac{v_H^4}{\mu_\Delta^4} + \frac{v_\Delta^4}{m_h^4}\right),
\end{align}
where there are two contributions each, because we can either have just $H$ or   $H$ and   $\Delta$ running in the loop. With the relation \eqref{eq:delta} for $v_\Delta$ in mind one sees that the first term for the correction to $m_S^2$ and the second term for the correction of $\lambda_\sigma$ are the leading ones and the limits read 
\begin{align}\label{eq:cond41}
    \lambda_4 <  \left(\frac{N }{6}\right)\cdot \left(\frac{f_a}{10^6\;\text{GeV}}\right) \cdot 
    \begin{cases}
    3\times 10^{-5}\quad&\text{quant. fluct.}\\
    4.4\quad&\text{$H$-mass from $R$}
    \end{cases}
\end{align}
as well as 
\begin{align}\label{eq:cond4}
    \lambda_4 < \left(\frac{\SI{4}{\giga\electronvolt}}{v_\Delta}\right)\cdot 
    \begin{cases}
    1.5\times 10^{-3}\quad&\text{quant. fluct.}\\
    0.6\quad&\text{$H$-mass from $R$}
    \end{cases}.
\end{align}
As was discussed below \eqref{eq:Iquartic} and \eqref{eq:IIquartic}, the new degrees of freedom in the Type III scenario even without the aforementioned range of $v_\Delta$  are typically so light that we can not avoid their presence in the plasma anyway, which will be exploited in appendix \ref{app:earlysketch}.

\section{Thermalization}\label{sec:therm}
The Saxion only couples to the bath via the Dirac Weinberg-operator, leading to a UV-dominated rate \eqref{eq:dissp}, and the mixed quartic coupling with the Higgs, that leads to a IR-dominated rate to be discussed in the next subsection. We need to ensure that the Saxion is thermalized to avoid the overproduction of dark radiation or relic Saxions. 

\subsection{Early Thermalization}\label{sec:earlytherm}
If the Dirac Weinberg-operator is supposed to be fast at early times $T\lesssim T_\text{osc.}$ to thermalize the Saxions, it would be already fast at $T_\text{osc.}$. As we saw in section \ref{sec:barasym}, an efficient charge transfer from the condensate to the bath would require a rate $\Gamma_S(T_\text{osc.})$ that is a thousand times faster then the Hubble rate for $\varepsilon=0.1$. Such a large dissipation rate would naively lead to immediate  evaporation of the condensate. Reference \cite{Kozow:2022whq} however showed that since the dissipation rate depends on the oscillating field value (and not the amplitude) and further since it is the coefficient of $\dot{S}$ in the equation of motion (see appendix \ref{app:eom} for the  coupled equations of motion for both fields) 
\begin{align}
    \ddot{S} + \left(3H + \Gamma_S\right) \dot{S} + \frac{\partial V_\sigma}{\partial S} = S \dot{\theta}^2
\end{align}
it vanishes twice every period: Once when $\Gamma_S\sim S^2=0$ and once when $\dot{S}=0$. The authors of \cite{Kozow:2022whq} found from numerical simulations that such a term $\Gamma_S \sim S^2$ does not quickly evaporate the condensate, but instead leads to stronger damping of the amplitude than Hubble friction alone. The backreaction of particle production on an oscillating condensate can be captured by multiplying the amplitude with a factor of $\exp(-\Gamma_S/H)$ \cite{Dolgov:1994zq}.
A stronger decrease of the amplitude at  early times could however decrease the amount of angular rotation produced during the first couple oscillations in \eqref{eq:chargeasym}. We also expect that the additional friction from the coupling to the thermal bath would delay the commencement of the oscillations until $m_S(S)\simeq  \sqrt{\Gamma_S(S)\; H}$ \cite{Papageorgiou:2022prc}, which is below the usual oscillation temperature defined in \eqref{eq:TOSC}. Since the production rate $\Gamma_S$ depends on  $T$, the baryon asymmetry production would consequently be even more  inefficient. 
Studies of oscillating QCD axions with additional friction from a thermal bath \cite{Papageorgiou:2022prc,Choi:2022nlt} also find that the oscillations are  damped and delayed. We conclude that early thermalization via an effective operator is not necessarily viable and requires a dedicated numerical simulation. Reference \cite{Barnes:2022ren} considers thermalization of Saxions with Higgsinos via an effective operator similar to \eqref{eq:dissp} and avoids the aforementioned problem in the following way: The Higgsino mass gets a correction from the same coupling leading to the thermalization operator and the Higgsinos only become kinematically accessible at some time after the start of the oscillations. We sketch a scenario based on  using the potentially light triplet scalar in the Type III Dirac Seesaw (see \ref{sec:typeIII}) in appendix \ref{app:earlysketch}. Early thermalization before the Saxion has reached $N f_a$ has the appealing advantage that the Diraxion gets automatically thermalized as well \cite{Co:2020jtv} as a consequence of their coupled equations of motion in appendix \ref{app:eom}, which removes the warmness-bound for parametric resonance dark matter in \eqref{eq:warmness}.

\subsection{Late Thermalization}
\begin{figure}
    \centering
    \includegraphics[width=\textwidth]{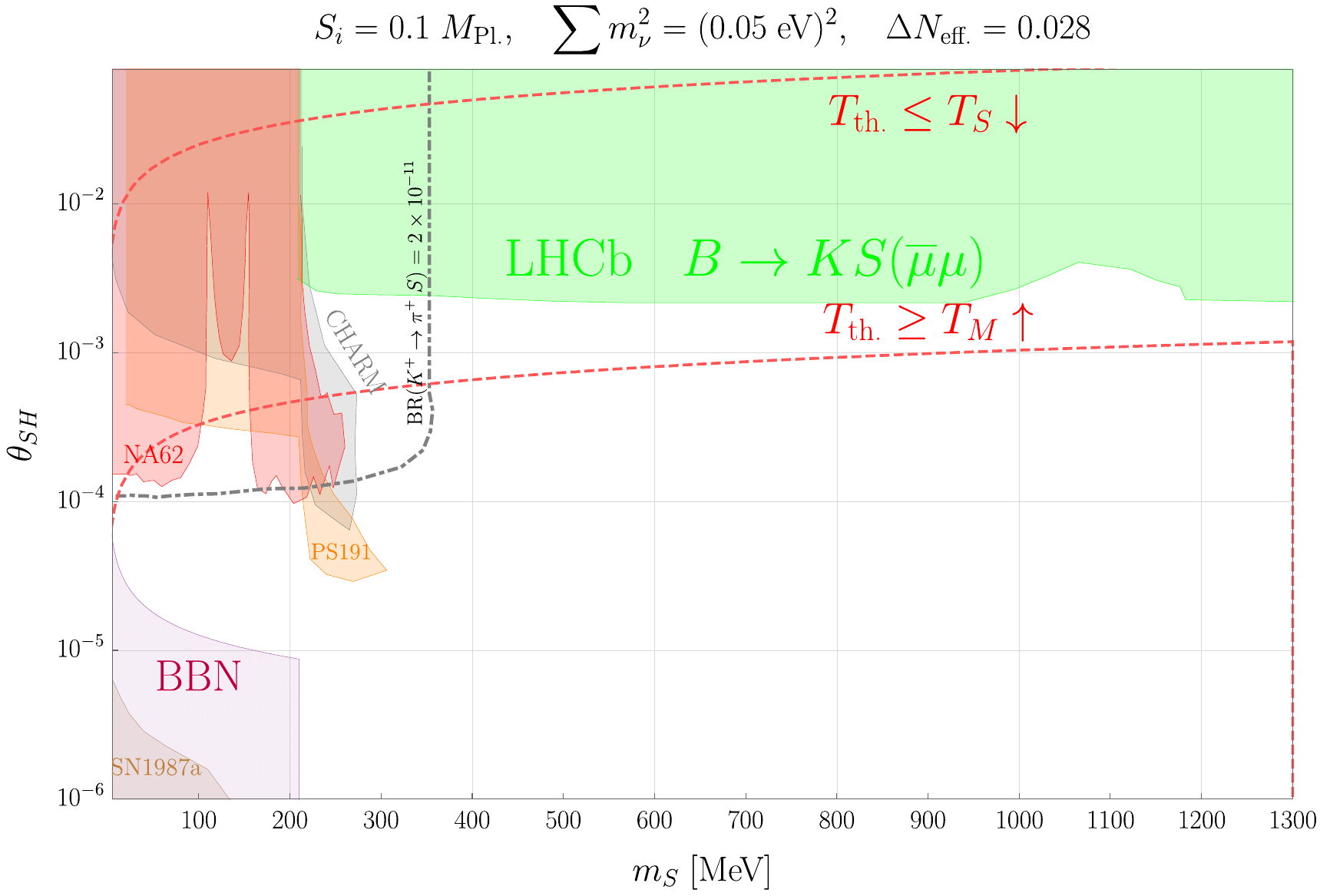}
    \caption{Parameter space for successful thermalization compared to current experimental and astrophysical limits on the production of scalars in terms of the mixing angle with the SM like Higgs $\theta_{SH}$ and the Saxion mass $m_S$. Above the line labelled $T_\text{th.}\geq T_M$ thermalization occurs before intermediate matter domination due to the Saxion and the line  would be in the   region excluded by LHCb for $S_i > 0.1 \;M_\text{Pl.}$.  Below the dotted line labelled $T_\text{th.}\leq T_S$ thermalization occurs after the Saxion has settled to the minimum of its potential and above this line a separate analysis would be needed.  If we decrease $\Delta N_\text{eff.}$  the aforementioned dotted lines move upwards.  The gray dotted line for the Kaon branching ratio was obtained from \cite{Goudzovski:2022vbt} and the LHCb constraint continues (with interruptions) until below 5 GeV. Equation \eqref{eq:BBNbound} leads to the purple BBN bound from light Saxion decays to electrons below the muon threshold.}
    \label{fig:therm}
\end{figure}

Here we adopt the thermalization scenario of references \cite{Co:2017mop,Co:2020dya,Co:2020jtv}, which utilizes the mixed quartic coupling between the singlet scalar and the SM Higgs and summarize the required parameter space. The Saxion can thermalize via the processes $S H \leftrightarrow H Z, H W$ and the corresponding rate is \cite{Co:2020dya,Co:2020jtv}
\begin{align}\label{eq:thermrate}
    \Gamma_{S,H} = \lambda_{\sigma H}^2 \;\alpha_2(T)^2\; \frac{(S+ N f_a)^2}{T},
\end{align}
where we take the weak fine structure constant to be $\alpha_2(T)\simeq 1/30$. 
Laboratory searches for light scalars constrain the mixing angle $\theta_{SH}$ between the Saxion and the Higgs. This angle is defined in \eqref{eq:mixangle} of the appendix and in the small angle approximation together with $m_S\ll m_h$
we can re-express it as 
\begin{align}\label{eq:angleS}
   \theta_{SH}  \simeq   \lambda_{\sigma H} \cdot \frac{v_H N f_a}{ m_h^2}.
\end{align}
Higgs to invisible decays constrain $\lambda_{\sigma H}<0.01$ \cite{Fernandez-Martinez:2021ypo}. We also use the limits from Kaon decays at NA62 \cite{NA62:2021zjw,NA62:2020pwi,NA62:2020xlg}, recasts of  PS191 data \cite{Gorbunov:2021ccu} as well as a recast \cite{Egana-Ugrinovic:2019wzj} of old CHARM data \cite{CHARM:1985anb}, which are relevant for Saxion masses below about 300 MeV.
Additionally we employ the bounds from displaced vertex searches of B-meson decays $B\rightarrow K S (\overline{\mu} \mu)$ at LHCb \cite{LHCb:2015nkv,LHCb:2016awg}, where $S$ decays to  $\overline{\mu} \mu $, which are sensitive below around 5 GeV. 
Note that these limits assume a 100 \% branching fraction of the Saxion to muons and including the kinematically allowed hadronic decay modes above the two pion threshold leads to  less stringent bounds \cite{LHCb:2015nkv,LHCb:2016awg}, which is why we treat the aforementioned limits as conservative constraints. For heavier singlets there exists a LEP search \cite{L3:1996ome} for the process $e^+ e^- \rightarrow Z^* S$, which only excludes $\theta_{SH}\gtrsim 0.1$. For an overview of the existing laboratory constraints consult \cite{Winkler:2018qyg,Goudzovski:2022vbt}.
The bounds from SN1987A on light scalars were recently reevaluated in \cite{Dev:2020eam}.\\
\\
One needs to check that the required values of $\lambda_{\sigma H}$ do not upset the small tree-level coupling $\lambda_\sigma$, whose value is dictated by the isocurvature constraints from section \ref{sec:quantcorr}.
Due to one loop corrections to the quartic we require that 
\begin{align}\label{eq:portal}
    \lambda_{\sigma H} <     
    \begin{cases}
    1.2\times 10^{-9}\quad&\text{quant. fluct.}\\
    1.8\times10^{-4}\quad&\text{$H$-mass from $R$}
\end{cases}.
\end{align}
In the limit $m_h \gg m_S$ one would also find that integrating out the SM Higgs would lead to the following tree-level threshold correction to the Saxion quartic \cite{Chan:1979ce,Randjbar-Daemi:2006ada,Elias-Miro:2012eoi}
\begin{align}
    \lambda_\sigma \rightarrow \lambda_\sigma - \frac{\lambda_{\sigma H}^2}{\lambda_H}.
\end{align}
The upper limit in $\theta_{SH}< 3\times 10^{-3}$ from LHCb \cite{LHCb:2015nkv,LHCb:2016awg}
(see figure \ref{fig:therm}) can be recast by using \eqref{eq:angleS} into 
\begin{align}
    \frac{\lambda_{\sigma H}^2}{\lambda_H} = 2.8\times 10^{-13}\cdot \left(\frac{10^6\;\text{GeV}}{N f_a}\right)\cdot \left(\frac{\theta_{SH}}{3\times 10^{-3}}\right)^2.
\end{align}
One can deduce that the corresponding $\lambda_{\sigma H}$ complies only with the limit for Hubble induced initia field value in \eqref{eq:portal}. Furthermore one can see from \eqref{eq:lambdas} that the tree-level correction is subdominant to the limit on $\lambda_\sigma$  only for the case of a Hubble induced mass ($\lambda_\sigma <10^{-10}$), but not for the scenario with $S_i$ from quantum fluctuations  ($\lambda_\sigma <10^{-20}$).

\subsubsection{Thermalization during radiation domination} 
Below $T_S$ the Saxion redshifts like non-relativistic matter $ S= N f_a (T/T_S)^{3/2}$ and would lead to an era of matter domination starting at a temperature
\begin{align}
    T_M &\simeq \frac{4.8}{g_*(T_M)^\frac{1}{4}} \left(\frac{S_i}{M_\text{Pl.}}\right)^\frac{3}{2} \sqrt{m_S \; N f_a} \\
    &\simeq \SI{63}{\giga\electronvolt}  \cdot \left(\frac{10}{g_*(T_M)}\right)^\frac{1}{4} \cdot \sqrt{\frac{m_S}{\SI{1}{\giga\electronvolt}}} \cdot \sqrt{\frac{N f_a}{10^6\;\text{GeV}}} \cdot \left(\frac{S_i}{0.1\;M_\text{Pl.}}\right)^\frac{3}{2},
\end{align}
if it was not thermalized beforehand. 
During radiation domination we find that $\Gamma_{S,H}/H\sim 1/T$ so thermalization is IR dominated. The thermalization temperature is found to be
\begin{align}\label{eq:Ttherm}
    T_\text{th} \simeq \frac{0.27}{g_*(T_\text{th})^\frac{1}{6}} M_\text{Pl.}^\frac{1}{3} (N f_a)^\frac{2}{3} \lambda_{\sigma H}^\frac{2}{3} \simeq \SI{45.7}{\tera\electronvolt}\cdot \left(\frac{\theta_{SH}}{10^{-3}}\right)^\frac{2}{3} \cdot \left(\frac{100}{g_*(T_\text{th})}\right)^\frac{1}{6}.
\end{align}
 To avoid complications from the oscillatory effective   masses of the Higgs for $S\gg N f_a$ we work in the limit $T_\text{th}<T_S$ \cite{Co:2020dya} with $T_S$ defined in \eqref{eq:TS}, which implies
 \begin{align}\label{eq:therm1}
      \theta_{S H} < \SI{1.1e-5}{}\; \cdot \left(\frac{m_S}{\SI{1}{\giga\electronvolt}}\right)^\frac{3}{4} \cdot  \left(\frac{N f_a}{10^6\;\text{GeV}}\right)^\frac{3}{4}
      \cdot \left(\frac{0.1\;M_\text{Pl.}}{S_i}\right)^\frac{3}{4}.
 \end{align}
 The condition $T_\text{th} > T_M$ on the other hand leads to 
 \begin{align}\label{eq:therm2}
    \theta_{S H} > \SI{2.2e-8}{}\cdot  \left(\frac{m_S}{\SI{1}{\giga\electronvolt}}\right)^\frac{3}{4}\cdot \left(\frac{N f_a}{10^6\;\text{GeV}}\right)^\frac{3}{4}\cdot
    \left(\frac{S_i}{0.1\;M_\text{Pl.}}\right)^\frac{9}{4}.
 \end{align}
 One should keep in mind that the SM like Higgs has to be as abundant as radiation for this process to work, which is why we require $T_\text{th} > m_h = \SI{125}{\giga\electronvolt}$
 \begin{align}
     \theta_{S H} > \SI{1.4e-7}{} \cdot \left(\frac{g_*(T_\text{th})}{100}\right)^\frac{1}{4}.
 \end{align}
 The Higgs could also receive a mass correction from its coupling to $S$ namely $\Delta m_H^2 \simeq 2 \lambda_{\sigma H} N f_a S$, where we dropped the $S^2$ term following \cite{Co:2020dya,Co:2020jtv}, because it is expected to be subdominant for $T<T_S$ owing to $S= N f_a (T/T_S)^\frac{3}{2} < N f_a$. Demanding that $T_\text{th}>\Delta m_H $ gives an upper bound 
 \begin{align}
     \theta_{S H} < 0.35 \cdot \left(\frac{m_S}{\SI{1}{\giga\electronvolt}}\right)^\frac{9}{8}  \cdot  \left(\frac{ 10^6\;\text{GeV}}{N f_a}\right)^\frac{3}{8}
      \cdot \left(\frac{0.1\;M_\text{Pl.}}{S_i}\right)^\frac{9}{8}.
 \end{align}
Additionally we need to ensure that the Saxion does not receive a large thermal correction from its coupling to the abundant Higgses $m_S(S_i)> \sqrt{\lambda_{\sigma H}} T_\text{osc.}$ which can be cast as
\begin{align}
    \theta_{S H} < \SI{2.6e-2}{}\cdot \left(\frac{m_S}{\SI{1}{\giga\electronvolt}}\right) \cdot \left(\frac{S_i}{0.1\;M_\text{Pl.}}\right).
\end{align}
The bounds in \eqref{eq:therm1} and \eqref{eq:therm2} are the strongest thermalization constraints and we depict them in figure \ref{fig:therm} by fixing $f_a/S_i$ via \eqref{eq:fa1}, which explains the dependence on $\sum m_\nu^2$ and $\Delta N_\text{eff.}$. We find that all the available parameter space would be excluded by the constraints from meson decays unless we take $S_i \lesssim 0.1 M_\text{Pl.}$. Furthermore for this initial field value we have to require that  $\Delta N_\text{eff.}>2.8\times 10^{-3}$, or else thermalization before Saxion domination would be excluded by LHCb. The original Lepto-Axiogenesis parameter space for a quartic potential involves Saxion masses that typically lie below the GeV-scale, whereas we will show in \eqref{sec:Diss} that our Saxion can be  heavier.

\subsubsection{Thermalization during reheating}
For the case where the oscillations begin before the completion of reheating our analysis finds that typically $T_S < T_\text{RH}$ which together with $T_\text{th}< T_S$ leads to the conclusion that thermalization will again proceed during radiation domination. Stil we need to demand that  $m_S(S_i)> \sqrt{\lambda_{\sigma H}} T_\text{osc.}^\text{RH}$ 
\begin{align}
    \theta_{S H} < \SI{0.25}{}\cdot \left(\frac{m_S}{\SI{1}{\giga\electronvolt}}\right)^\frac{3}{2} \cdot \sqrt{\frac{10^6\;\text{GeV}}{N f_a}}\cdot
    \left(\frac{S_i}{0.1\;M_\text{Pl.}}\right)^\frac{3}{2}\cdot \left(\frac{10^{14}\;\text{GeV}}{T_\text{RH}}\right).
\end{align}
For our parameter space this constraint is subdominant compared to \eqref{eq:therm1} and \eqref{eq:therm2}.

\subsection{Thermalized Saxion decays}\label{sec:late} 
In section \ref{sec:DRdecay} we showed that a significant part of the parameter space would be excluded by dark radiation constraints if the thermalized Saxions decayed long after their decoupling from the bath. Hence we require that the Saxions are in thermal equilibrium when they become non-relativistic. The mixing with the SM like Higgs allows for the following decay modes to the light SM leptons
\begin{align}
    \Gamma(S\rightarrow \overline{l}l) \simeq \frac{\theta_{SH}^2}{8 \pi} \left(\frac{m_l}{v_H}\right)^2 m_S \left(1-\frac{4 m_l^2}{m_S^2}\right)^\frac{3}{2} \quad \text{with} \quad  l=e,\; \mu,
\end{align}
where we neglect the phase space suppression for our first estimate and the corresponding decay temperature reads
\begin{align}
    T_\text{dec} &= \frac{0.15}{g_*(T_\text{dec})^\frac{1}{4}} \lambda_{\sigma H} \frac{\sqrt{m_S M_\text{Pl.}}\; N f_a\; m_l}{m_h^2}\\
    &\simeq \left(\frac{\theta_{SH}}{10^{-6}}\right)\cdot
    \left(\frac{10}{g_*(T_\text{dec.})}\right)^\frac{1}{4}\cdot
     \begin{cases}
        \SI{0.2}{\mega\electronvolt}\cdot \sqrt{\frac{m_S}{\SI{100}{\mega\electronvolt}}},&\; m_S<2 m_\mu \simeq \SI{200}{\mega\electronvolt}\\
        \SI{97}{\mega\electronvolt}\cdot\sqrt{\frac{m_S}{\SI{250}{\mega\electronvolt}}},&\; m_S\geq 2 m_\mu \simeq \SI{200}{\mega\electronvolt}\\
     \end{cases}.
\end{align} 
BBN will not be affected if the Saxions decay before the neutrino decoupling at $T=\SI{2}{\mega\electronvolt}$  implying
\begin{align}\label{eq:BBNbound}
     \theta_{SH} > 
        \SI{2e-5}{}\cdot \sqrt{\frac{\SI{100}{\mega\electronvolt}}{m_S}},&\quad m_S<2 m_\mu \simeq \SI{200}{\mega\electronvolt}.
\end{align}
This limit was depicted in figure \ref{fig:therm}.  Additionally the Saxions need to be Boltzmann-suppressed when the SM neutrinos decouple, because else at $T<T_\text{dec}$ the Saxions are still kept in equilibrium via decays and inverse decays to electrons and inject entropy into the SM plasma. This process would heat only the SM bath so the already decoupled neutrinos are cooled leading to a reduction in $\Delta N_\text{eff}$ \cite{Fradette:2017sdd}. Reference  \cite{Co:2020dya}  obtained that the lower bound $\Delta N_\text{eff}<-0.44$ \cite{Fields:2019pfx} is only compatible with (see also   \cite{Dunsky:2022uoq} for a similar analysis involving heavy QCD axions)
\begin{align}\label{eq:BBnlimit}
    m_S > \SI{4}{\mega\electronvolt}.
\end{align}
As it turns out our scenario typically requires Saxions with masses around the GeV-scale (see sections \ref{sec:kinmis} and \ref{sec:parRes}), which means that on the one hand decays to mesons are possible and on the other that the Saxion would be Boltzmann-suppressed at BBN.
The decay to SM fermions $\psi$ is the dominant mode compared to Diraxion final states as long as 
\begin{align}
    \theta_{SH} \frac{m_\psi}{v_H}> \frac{m_S}{2 N f_a}.
\end{align}
From the isocurvature limits \eqref{eq:bound11} and \eqref{eq:bound22} one can deduce that the right hand side in the above is of  $\mathcal{O}(10^{-5}-10^{-11})$, which is much smaller than the coupling to SM states for GeV-scale Saxions.

\subsection{Maximum Yield}
 The thermalization temperature allows us to determine an upper limit on the charge yield \cite{Co:2019jts,Co:2020jtv,Co:2021qgl,Barnes:2022ren}
 \begin{align}\label{eq:maxyield}
     Y_\theta <  Y_\theta^\text{max}\equiv \frac{3}{4}\varepsilon\frac{T_\text{th}}{m_S} 
 \end{align}
 where we used $T_\text{th.}<T_S$ so that the result only depends on $m_S$. This relation can be understood as follows: During thermalization the energy density of the oscillations $\rho_S = (1-\varepsilon) m_S n_S$ is dumped into the bath \cite{Gouttenoire:2021jhk} and only the pure rotation with $\rho_\theta = \varepsilon m_S n_S$  will remain. The entropy density released during thermalization is found from the first law of thermodynamics to be $s_f \sim \rho_S /T_\text{th}$ and consequently we obtain for the dimensionless dilution factor \footnote{The above estimate only holds for $\Delta \gg 1$.}
 \begin{align}
     \Delta \equiv \frac{s_f\; a^3(T_f)}{s\; a^3(T)} \sim \frac{1-\varepsilon}{\varepsilon} \frac{m_S}{T_\text{th}}Y_\theta
 \end{align}
 that leads to $Y_\theta / \Delta \sim T_\text{th}/m_S$. The entropy generation is maximal for the case of Saxion domination, for which an equal sign in \eqref{eq:maxyield} would apply. We express the maximum charge yield in terms of the mixing angle $\theta_{SH}< 3\times 10^{-3}$ by using  \eqref{eq:Ttherm} together with \eqref{eq:maxyield} and $\varepsilon\simeq 1$ (see the discussion in the beginning of \ref{sec:kinmis} and the next section). On the upper axis of figure \ref{fig:SaxionOverview} one can read off the value of $ Y_\theta^\text{max}$ for a given $m_S$.

\section{Discussion}\label{sec:Diss}

\begin{figure} 
    \centering
    \includegraphics[width=0.6\textwidth]{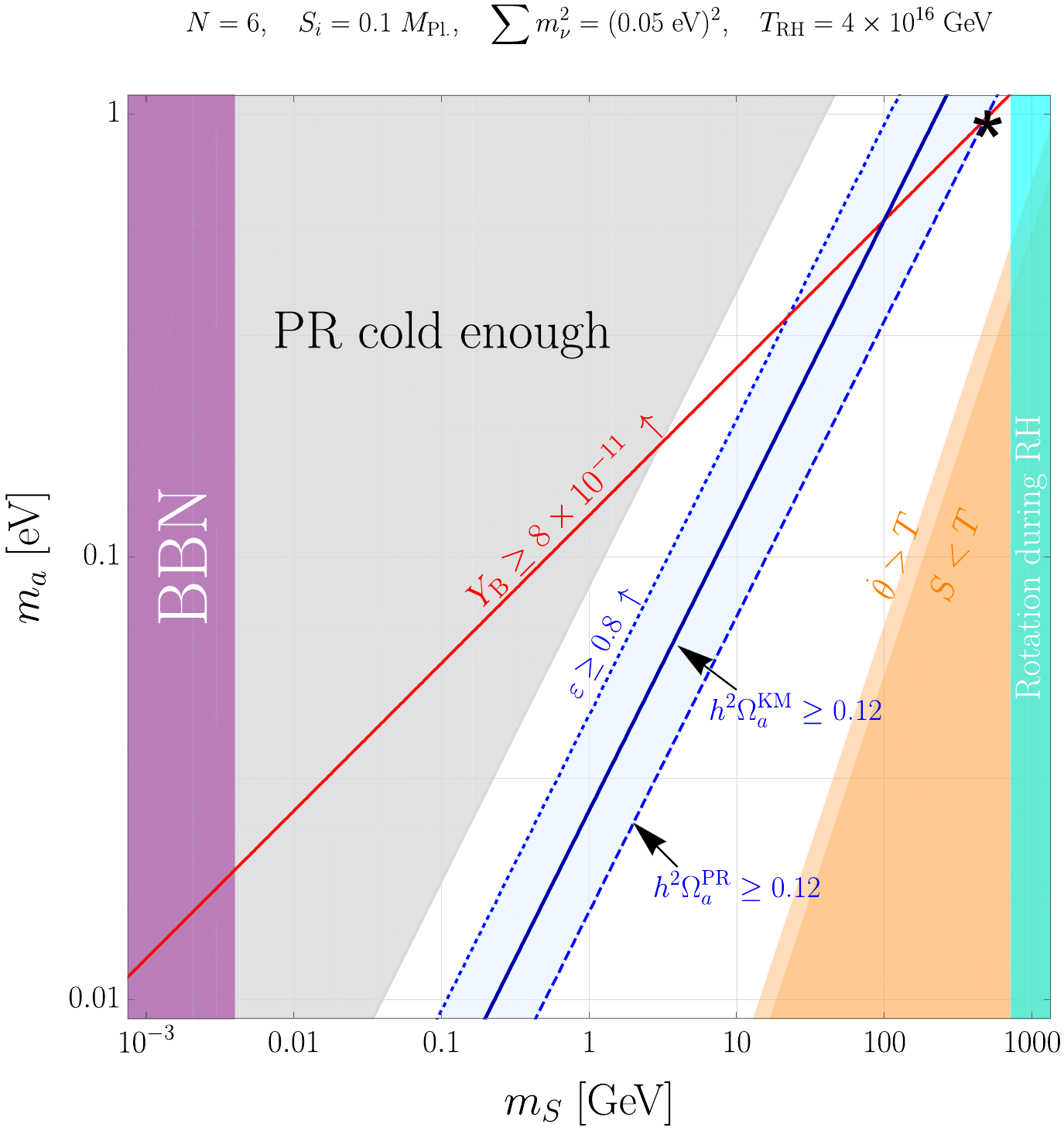}
    \includegraphics[width=0.6\textwidth]{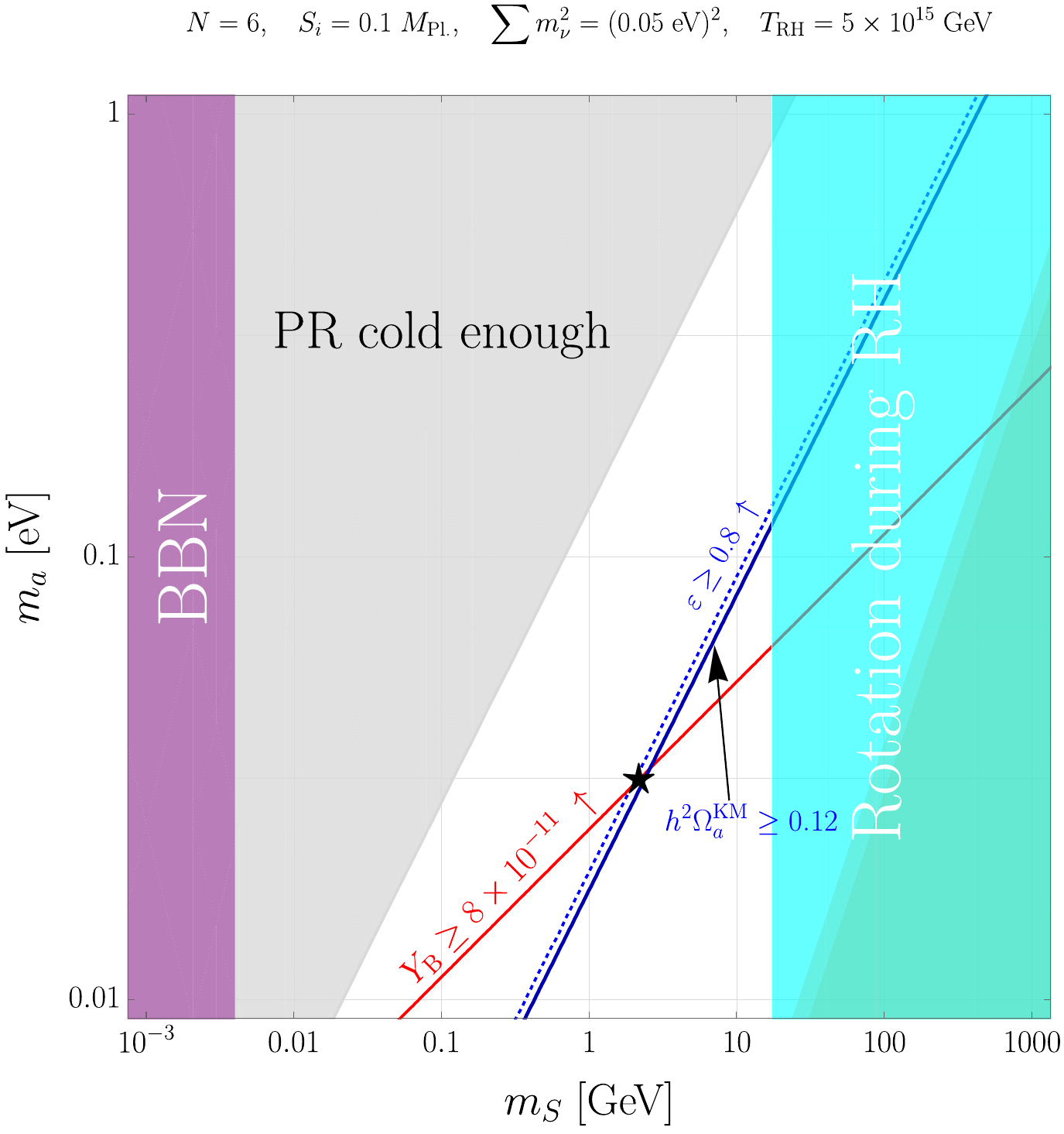}
    \caption{
    Parameter space for the cogenesis of the baryon asymmetry and dark matter relic abundance via kinetic misalignment or parametric resonance spanned by the Diraxion mass $m_a$ and the Saxion mass $m_S$. The decay constant $f_a$ was determined via \eqref{eq:fa1} in terms of $\Delta N_\text{eff.}$ and $\sum m_\nu^2$.   In the \textit{upper} plot we fixed $\Delta N_\text{eff.} = 1.4\times 10^{-3}$ and the point marked with a star corresponds to $(m_S,m_a,f_a)=(\SI{500}{\giga\electronvolt},\;\SI{1}{\electronvolt},\;\SI{1.1e6}{\giga\electronvolt})$. Here the Diraxion has a lifetime of $4\times 10^4$ in units of the age of our universe. In the \textit{lower} plot we fixed $\Delta N_\text{eff.} = 5\times 10^{-3}$ and the point marked with a star corresponds to $(m_S,m_a,f_a)=(\SI{2.2}{\giga\electronvolt},\;\SI{31}{\milli\electronvolt},\;\SI{2.8e6}{\giga\electronvolt})$.}
    \label{fig:ResRD1}
\end{figure}

\begin{figure} 
    \centering
    \includegraphics[width=0.6\textwidth]{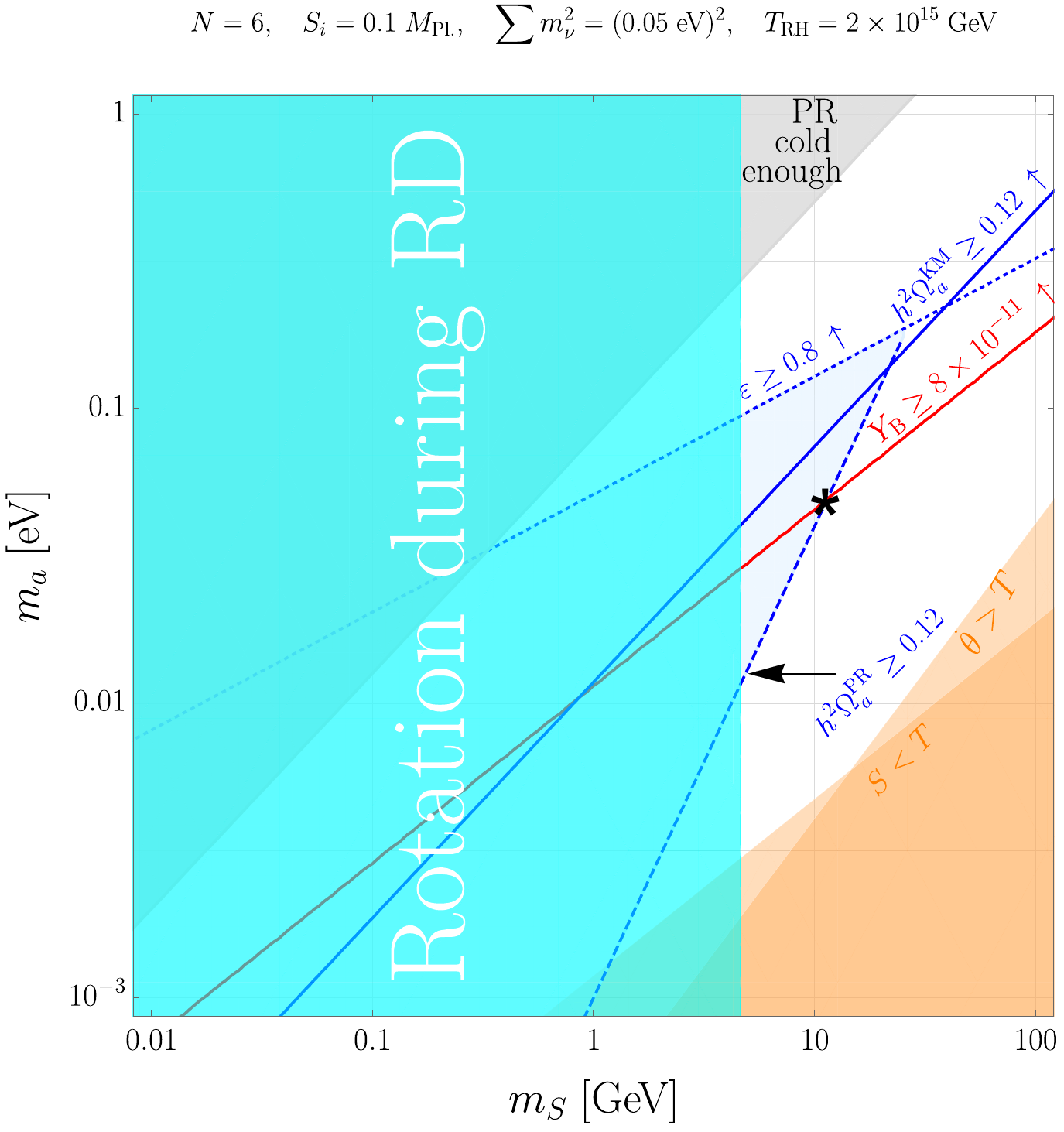}
    \includegraphics[width=0.6\textwidth]{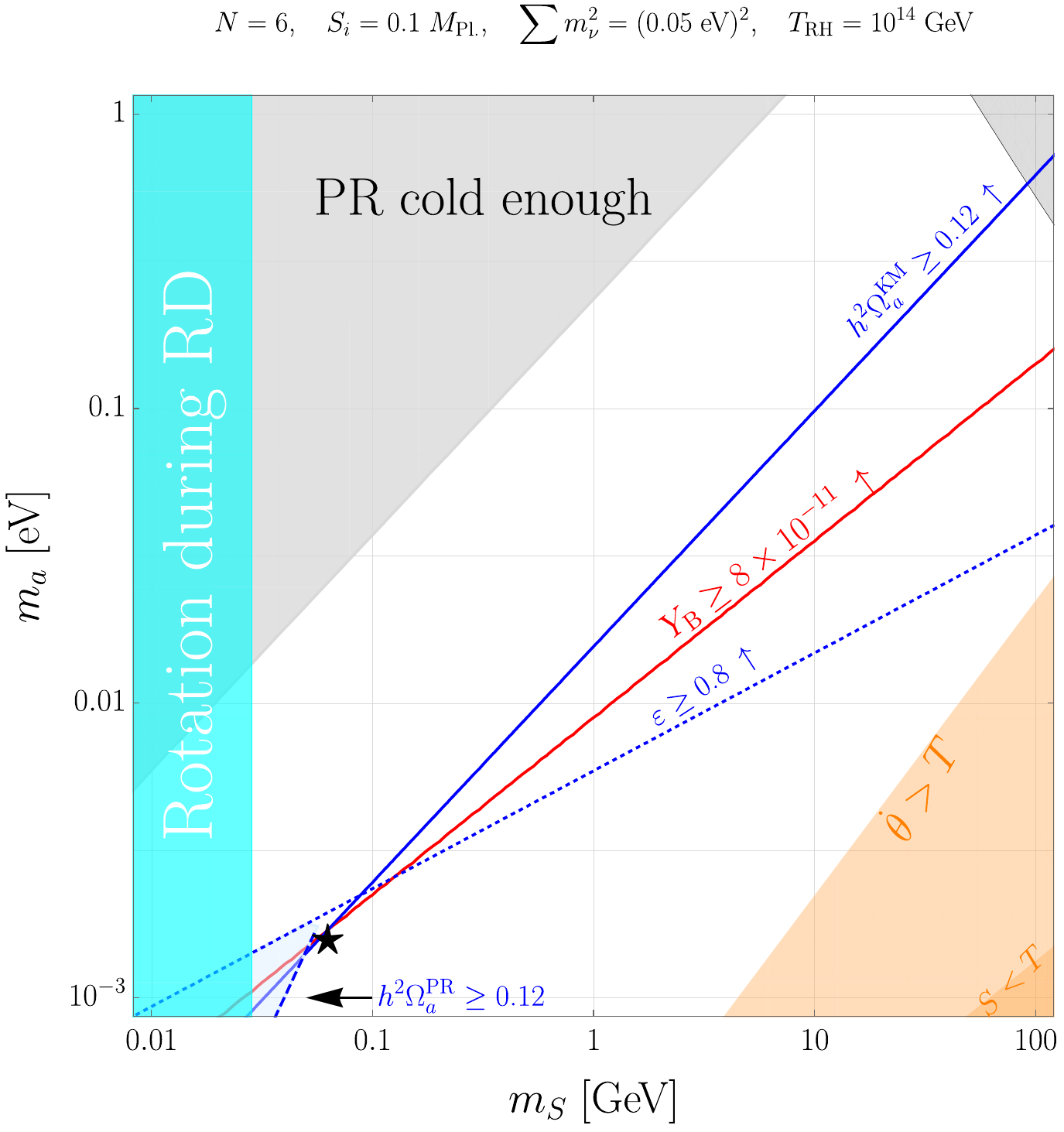}
   \caption{
   Parameter space for the cogenesis of the baryon asymmetry and dark matter relic abundance via kinetic misalignment or parametric resonance spanned by the Diraxion mass $m_a$ and the Saxion mass $m_S$. The decay constant $f_a$ was determined via \eqref{eq:fa2} in terms of   $\Delta N_\text{eff.}$ (the upper limit during radiation domination defined in \eqref{eq:neffmax})  and $\sum m_\nu^2$.
   In the \textit{upper} plot we fixed the upper limit $\Delta N_\text{eff.} = 4\times 10^{-3}$ and the point marked with a star corresponds to $(m_S,m_a,f_a)=(\SI{11.2}{\giga\electronvolt},\;\SI{50}{\milli\electronvolt},\;\SI{2.2e6}{\giga\electronvolt})$ with $\varepsilon=0.1$. In the \textit{lower} plot we fixed the upper limit $\Delta N_\text{eff.} = 0.014$ and the point marked with a star corresponds to $(m_S,m_a,f_a)=(\SI{63}{\mega\electronvolt},\;\SI{1.5}{\milli\electronvolt},\;\SI{6e7}{\giga\electronvolt})$ with $\varepsilon=0.4$. The dark grey region in the upper right corner is excluded by the Diraxion life-time.}
    \label{fig:ResRH}
\end{figure}

Late Saxion thermalization via the Higgs portal before Saxion domination requires $S_i \lesssim 0.1 M_\text{Pl.}$ (see the plot \ref{fig:therm}) and for concreteness we saturate this value. This leads to a high oscillation temperature for the Saxion in  \eqref{eq:TOSC} and \eqref{eq:TOSCRH}, which is why we typically need reheating temperatures above around $10^{14-15}\;\text{GeV}$, which is close to the limit $T_\text{RH}<\SI{6.5e15}{\giga\electronvolt}$ inferred from $H_I<\SI{6e13}{\giga\electronvolt}$ \cite{Planck:2018jri} and could be seen as a disadvantage of our scenario compared to Leptogenesis or Lepto-Axiogenesis \cite{Co:2020jtv}, that both work for reheating temperatures as low as  $10^{9}\;\text{GeV}$~\cite{Davidson:2002qv}. Thermalization with $S_i = 0.1 M_\text{Pl.}$ is only viable as long as $\Delta N_\text{eff.}>2.8\times 10^{-3}$ for Saxions around the GeV-scale.
To ensure that the processes encoded in $\Gamma_S$ never thermalize, we fix $f_a/S_i$ via equations \eqref{eq:fa1} (radiation domination) and \eqref{eq:fa2} (during reheating), which is equivalent to fixing the amount of $\nu_R$ dark radiation produced in \eqref{eq:neffmax}.
We find that  we typically need $\Gamma_S(S_i)/H(T_\text{osc})<0.1$ to avoid $\varepsilon>1$ (see  \eqref{eq:effdesc} and \eqref{eq:masses2}), so the upper limit on $\Delta N_\text{eff.}$ is never saturated and our cogenesis setup does not lead to observable amounts of dark radiation. For $N=5$ the bounds from the self-consistency criteria in \eqref{eq:cond1} and \eqref{eq:cond2} would be the most relaxed, however we find that cogenesis of the baryon asymmetry and dark matter would take place in regions with $m_a=\mathcal{O}(\text{keV})$. While we can generate the correct relic abundance in this regime, dark matter will not be long-lived enough due to its decay to neutrinos (see \eqref{eq:tauDM} for the lifetime). Hence we fix $N=6$ for the rest of our analysis, which leads to stable enough Diraxions with $m_a=\mathcal{O}(\text{meV}-\text{eV})$.
The downside of this parameter range of $N$ and $S_i$ is that we find ourselves in a region, where the warmness constraint \eqref{eq:warmness} on parametric resonance dark matter will always be violated. Thus we either have to invoke a scenario for early thermalization of the coupled Saxion-Diraxion system as sketched for the Type III Dirac Seesaw in \ref{app:earlysketch} or tune the eccentricity parameter $\varepsilon$ defined in \eqref{eq:varepsilon} close to  $\varepsilon\simeq 0.8$ (consult section \ref{sec:parRes} for details).\\
\\
In figures \ref{fig:ResRD1} and \ref{fig:ResRH}  we plot the baryon asymmetry as red line and the dark matter relic abundance from kinetic misalignment as a blue line.
Both quantities are overproduced above their respective lines. Parametric resonance dark matter is possible in the blue shaded region surrounded by the blue dashed line indicating $\Omega_a^\text{PR}\;h^2 \gtrsim  0.12$ and the dotted line for $\varepsilon= 0.8$. Inbetween these lines dark matter from parametric resonance is overproduced. 
Our parameter space for oscillations during radiation domination (reheating) can be seen in figure \ref{fig:ResRD1} (\ref{fig:ResRH}). 
In the gray shaded area parametric resonance dark matter would be cold enough and in the orange regions we begin to violate the self-consistency criteria $\dot{\theta}<T$  from \eqref{eq:cond1} and $S>T$ from \eqref{eq:cond2}. The region in which the upper limit on the yield from thermalization in \eqref{eq:maxyield} would not be able to explain the dark matter relic abundance via kinetic misalignment is completely within the orange area. The purple region is excluded by neutrino cooling from too light Saxions during BBN, see \eqref{eq:BBnlimit}. In the cyan area of figure \ref{fig:ResRD1} (\ref{fig:ResRH})  the Saxion oscillation would take place during reheating (radiation domination) as described by equation \eqref{eq:condRH}.
The first panel in \ref{fig:ResRD1} shows a point $(m_S,m_a,f_a)=(\SI{500}{\giga\electronvolt},\;\SI{1}{\electronvolt},\;\SI{1.1e6}{\giga\electronvolt})$, which would work with early thermalized parametric resonance (here the reheating temperature would actually be to large compared to the bound inferred from $H_I$). While $m_a$ is of the right order of magnitude to be detected by PTOLEMY \cite{Betts:2013uya,PTOLEMY:2018jst}  via decays to neutrinos \cite{McKeen:2018xyz,Chacko:2018uke} its lifetime of  about $4\times 10^4$ times the age of the universe is far too large, so there would be no appreciable number of produced neutrinos. Note that here we can have  $\Delta N_\text{eff.}<2.8\times 10^{-3}$ because the Saxion mass is around 100 GeV and thus not constrained by B-meson decays.
The second panel in the same figure shows a point  $(m_S,m_a,f_a)=(\SI{2.2}{\giga\electronvolt},\;\SI{31}{\milli\electronvolt},\;\SI{2.8e6}{\giga\electronvolt})$, where cogenesis with kinetic misalignment is possible for $\varepsilon\simeq 0.8$. 
If we try to generate a signal from kinetic misalignment  Diraxions decaying to neutrinos detectable with PTOLEMY, we have to choose the same value  $\varepsilon$ (see the discussion in section \ref{sec:kinmis}). While $\varepsilon=0.8$ is allowed by the limits in \eqref{eq:effdesc} and \eqref{eq:masses2}, here the $S_i$-dependent Diraxion mass $m_a(S_i)$ starts to become comparable in magnitude to the initial Saxion mass $m_SS(S_i)$, which might potentially hinder the onset of the coherent rotation. Reference \cite{Gouttenoire:2021jhk} solved the equations of motion for $N=6$ and $\varepsilon=0.8$ numerically and found that a rotation can be initiated, however their analysis only covered  an approximately quadratic potential.\\
\\
This conclusion can be somewhat ameliorated for rotations during reheating as depicted in \ref{fig:ResRH}: Again the upper panel showcases a point $(m_S,m_a,f_a)=(\SI{11.2}{\giga\electronvolt},\;\SI{50}{\milli\electronvolt},\;\SI{2.2e6}{\giga\electronvolt})$ that would work with early thermalization for parametric resonance dark matter. This panel also demonstrates that kinetic misalignment and parametric resonance scale differently during reheating, which can also be deduced by comparing the second line of \eqref{eq:yieldrelic} with the second line of \eqref{eq:yieldPR}. One finds that the relic abundance for parametric resonance scales with $f_a/S_i$ and becomes independent of $S_i$ after one uses \eqref{eq:fa2}. The different scaling allows us to find the point $(m_S,m_a,f_a)=(\SI{63}{\mega\electronvolt},\;\SI{1.5}{\milli\electronvolt},\;\SI{6e7}{\giga\electronvolt})$ depicted in the lower panel of \ref{fig:ResRH}, where the lines for the baryon asymmetry and kinetic misalignment meet for $\varepsilon\simeq 0.4$. The required Saxion mass of $\SI{63}{\mega\electronvolt}$ is however excluded, if thermalization is supposed to occur via the Higgs portal, see figure \ref{fig:therm}. Note that the values of $\Delta N_\text{eff.}$ mentioned in the caption of \ref{fig:ResRH} where computed using the estimate for the maximum amount of dark radiation during radiation domination in \eqref{eq:neffmax}.\\
\\
For our parameter space of interest we find that $f_a\simeq (10^5-10^6)\;\text{GeV}$, which means that regular misalignment and topological defect decay can at most only contribute a tiny fraction of the DM relic abundance. Inspection of \ref{fig:DMoverview} reveals that for these values of $f_a$ and kinetic misalignment one expects a completely fragmented Diraxion, so in principle a lattice study is needed. Consequently the Diraxion will not be a zero mode condensate for either production from parametric resonance or kinetic misalignment. As argued in the beginning of this section, we find $\Delta N_\text{eff.}< 0.028$ in the regions, where we can explain both the observed baryon asymmetry and dark matter. A detection of a lager amount of dark radiation by next generation experiments would exclude  only  the cogenesis scenario. Thus such a detection would imply that this construction can be responsible for  \textit{either} baryogenesis or the origin of dark matter.

\section{Conclusion}\label{sec:Conclusion}
\begin{itemize}
    \item \textbf{The Diraxion  from the Dirac Seesaw:}\\
    We showed that all three versions of the Dirac Seesaw mechanism automatically provide us with a PNGB that we call the Diraxion, whose cosmological implications have previously  not been analyzed.  This Goldstone mode  originates from a global $\text{U}(1)_\text{D}$ symmetry that forbids couplings of $\nu_R$ to the SM Higgs and we assume that this symmetry is explicitly broken by quantum gravity making the Diraxion massive.
    Furthermore we impose a separate (gauged or residual) symmetry like e.g. $\text{U}(1)_\text{B-L}$ to forbid all Majorana masses.
    The Diraxion has no direct coupling to SM fields and its connection to charged fermions or photons arises only at one or two loops respectively, which is why our set-up is unconstrained by stellar and supernova cooling arguments.

    \item \textbf{Diraxion rotation:}\\
    The Diraxion is accompanied by a radial mode called the Saxion, whose vev can undergo a large excursion during inflation and which oscillates in the early universe.
    We discuss two ways of inducing this large vev with a Hubble-dependent mass during inflation being the least excluded (see figure \ref{fig:SaxionOverview}). If the Diraxion mass comes from a higher dimensional operator, it can convert a part of the Saxion's oscillatory motion into a coherent angular rotation around the bottom of the scalar potential. For our parameter space we find that we need a dimension six operator to do so. If this operator involves additional insertions of the $\text{U}(1)_\text{B-L}$-breaking scalar, we can also generate the cosmologically required mass scale $m_a=\mathcal{O}(\text{meV}-\text{eV})$ without assuming a small Wilson coefficient. Conventionally we express the abundance of Diraxions $\varepsilon n_S$ in terms of the Saxion number density $n_S$ and a dimensionless eccentricity parameter $0<\varepsilon<1$. We have to keep $\varepsilon$ below unity, because else the Diraxion would start oscillating before the Saxion leading to it being trapped in a false vacuum, which hinders the onset of the coherent rotation.

    \item \textbf{Dirac-Leptogenesis:}\\
    We assume that all of the Seesaw messenger fields are too heavy to be present in the plasma.  The Diraxion rotation is thermodynamically stable and can act as a background field enabling spontaneous baryogenesis via Dirac-Leptogenesis \cite{Dick:1999je} from the Dirac-Weinberg operator. Scattering processes like $H S \rightarrow \overline{L} \nu_R, \quad L S\rightarrow H^\dagger \nu_R$ conserve the total B-L, but not the individual SM and right handed neutrino lepton numbers. This effectively leads to equal and opposite chemical potentials between the SM leptons and the right handed neutrinos $\mu_\text{B-L}^\text{SM} = \mu_{\nu_R}$.  Since the electroweak sphaleron process only couples to the left-chiral lepton doublet one can convert the asymmetry stored in the lepton doublet into a baryon asymmetry. The leptonic asymmetry is not washed out, because we ensure that the only interaction producing $\nu_R$ is always out of equilibrium. As elucidated in appendix \ref{sec:CPI} we do not encounter an overproduction of baryon asymmetry from the chiral hypermagnnetic instability.

    \item \textbf{Dark Matter:}\\
    The Diraxion rotation can be responsible for dark matter via the kinetic misalignment mechanism. Additionally fluctuations of the Diraxion can be produced via parametric resonance from the Saxion oscillation.  There is no domain wall problem in this construction, because domain walls can immediately decay to Diraxions either via a string-wall network with a single domain wall or for more domain walls via the well known bias-term from the Diraxion mass. Both contributions from topological defect decay are comparable to the relic abundance from standard misalignment, which is negligible for our range of decay constants $f_a\simeq (10^5-10^6)\;\text{GeV}$.
    Kinetic misalignment can lead to Diraxion dark matter with $\SI{0.1}{\electronvolt}\lesssim m_a \lesssim \mathcal{O}(\SI{1}{\electronvolt})$ and $f_a\simeq 10^5\;\text{GeV}$ (see figure \ref{fig:DMoverview}), which can be potentially tested via its decay to neutrinos with future cosmic neutrino background experiments like PTOLEMY. For these parameters we expect that the Diraxion zero-mode condensate is completely fragmentated into higher momentum excitations and in principle a lattice study is needed to treat this regime. 
    Additionally the resulting Saxion mass is in tension with isocurvature constraints, which is why for this range of parameters we can only accomodate a fraction of the relic abundance. Dark matter from Saxion induced parametric resonance  is constrained by its warmness. This channel can not produce cold enough Diraxions detectable with PTOLEMY, because we require $f_a \gg 10^5\;\text{GeV}$ to ensure $\varepsilon<1$. In order to set the relic abundance via kinetic misalignment and to avoid too much and too warm parametric resonance dark matter we have to set $\varepsilon\simeq 0.8$, which shuts off Saxion induced parametric resonance \cite{Co:2020dya}. This implies that the initial (evaluated at $S_i$) Diraxion mass is comparable in size to the initial Saxion mass, which brings our scenario closer to the mechanisms relying on heavy axion oscillations like e.g. \cite{Kusenko:2014uta,Ibe:2015nfa}. For a quadratic potential with $N=6$ and  $\varepsilon =0.8$ it was shown numerically by the authors of \cite{Gouttenoire:2021jhk} that a rotation can be initiated.

    \item \textbf{Parameter space and Cogenesis:}\\
    Our parameter space is spanned by the Diraxion and Saxion masses $m_a, m_S$ together with the initital Saxion field value $S_i$ and the Saxion vev today $N f_a$, where $N$ is the domain wall number. Thermalization fixes $S_i\lesssim 0.1 M_\text{Pl.}$ and we can eliminate $f_a$ by demanding that the Saxion does not have fast interactions with the bath, which also fixes the amount of right handed neutrino dark radiation. Throughout this work we use $N=6$ to keep the Diraxion light and long-lived enough. These inputs allow us to determine the values of $m_a, m_S$ for generating the correct dark matter abundance together with the right baryon asymmetry.
    Our parameter space for cogenesis with $m_a=\mathcal{O}(\text{meV}-\text{eV})$ and     $m_S= \mathcal{O}(100\;\text{MeV}-100\;\text{GeV})$ 
    suffers from too warm Diraxions from parametric resonance, which is why we either need thermalization from additional degrees of freedom in the bath, present e.g. in the Type III Dirac Seesaw (see appendix \ref{app:earlysketch}), or a region where  $\varepsilon\simeq 0.4\; (0.8)$ for oscillations during reheating (radiation domination) depicted in figure \ref{fig:ResRH} (\ref{fig:ResRD1}). 
    Our Saxion is typically predicted to be heavier than for the standard Lepto-Axiognesis scenario. It could be produced by (heavy) meson decays as depicted in figure \ref{fig:therm}.   Due to the size of $S_i$ and $m_S$ the Saxion oscillations occur at temperatures around $10^{15}\;\text{GeV}$, which implies comparable reheating temperatures close to the limit $T_\text{RH}<\SI{6.5e15}{\giga\electronvolt}$ inferred from $H_I<\SI{6e13}{\giga\electronvolt}$ \cite{Planck:2018jri}. This is a drawback compared to regular Leptogenesis or Lepto-Axiogenesis which work for reheating temperatures as low as $10^{9}\;\text{GeV}$ \cite{Davidson:2002qv,Co:2020jtv}.

    \item \textbf{Dark Radiation:}\\
    This setup can produce $\nu_R$ dark radiation with $2.8\times 10^{-3}\leq \Delta N_\text{eff.}\leq 0.028$, where the lower limit applies for $S_i = 0.1\;M_\text{Pl.}$ and Saxion masses below around 5 GeV due to thermalization via the Higgs portal.    While explaining both the baryon asymmetry and dark matter relic abundance involves smaller values of  $\Delta N_\text{eff.}$, fixing only one of these observables allows us to generate more dark radiation and to saturate the previous upper limit (see equations \eqref{eq:case1KM}-\eqref{eq:case2KM} and \eqref{eq:case1PR}-\eqref{eq:case2PR}). There is no observable dark radiation for cogenesis, because this would require $\varepsilon>1$.  Consequently we can use next generation measurements of  $\Delta N_\text{eff.}$ to test which cosmological history is realized in this framework.
    
    \item \textbf{Isocurvature constraints:}\\
    Dark matter and baryon isocurvature constraints enforce a value of $m_S$ that is very small compared to the $\text{U}(1)_\text{D}$ breaking scale $N f_a$ (see figure \ref{fig:SaxionOverview}). 
    For $S_i$ generated from quantum fluctuations we need $m_S/(N \;f_a) < \mathcal{O}(10^{-10})$ and for a Hubble-dependent mass we find $m_S/(N \;f_a) < \mathcal{O}(10^{-5})$.   We showed that one-loop corrections in the various Dirac Seesaw models do not upset this tuning. This comes at the price that the heavy messenger fields responsible for the Dirac-Weinberg-operator could be potentially light enough to be present in the plasma, so the description in terms of the Dirac-Weinberg-operator breaks down. To avoid this we assume an accidental cancellation between the one loop corrections to $m_S^2$ from the heavy Dirac fermion $N$ for the Type I Seesaw and the heavy doublet $\eta$ for the Type II Seesaw (see \eqref{eq:tuned}). Thus our scenario is a Hybrid-Seesaw \cite{Wetterich:1998vh,Chen:2005jm} and each contribution could be responsible for sourcing one of the two mass splittings observed in neutrino oscillation experiments implying that only one generation of $N$ would be needed. The messenger fields for the two versions of the Type III Dirac Seesaw are usually much lighter than $N,\;\eta$ so a separate study is needed to investigate their cosmological evolution and impact on the presented mechanism.  Our work predicts dark radiation isocurvature correlated with the dark matter and baryon isocurvature fluctuations, potentially detectable as neutrino isocurvature modes in the CMB.
    The dark radiation isocurvature signal is subleading if we saturate the bound from dark matter isocurvature.

    \item \textbf{Extensions and Outlook:}\\
    It would be worthwile to consider \textbf{(a)} the case of thermalized, lighter messenger fields for all Dirac Seesaws, \textbf{(b)} a supersymmetric set-up to ameliorate the strong isocurvature bounds on the non-supersymmetric quartic potential and \textbf{(c)} whether the Saxion could play the role of the inflaton, so Saxion thermalization is automatically obtained from successful reheating. Since each of these aspects constitutes a substantial modification of our analysis, we leave them for future investigation.
\end{itemize}

\section*{Acknowledgments}
\noindent  We would like to thank Andreas Trautner for valuable feedback about the manuscript.

\appendix 
\section*{\Huge{Appendix}}
\section{Equations of motion}\label{app:eom}
The coupled equations of motion for the Saxion $S$ and the Diraxion $\theta$ for the case of $N=1$ are given by \cite{Gouttenoire:2021jhk}
\begin{align}
    \dot{S}+ 3 H \dot{S} +\frac{\partial V_\sigma}{\partial S}+\frac{\partial V_\slashed{D}}{\partial S} - S \dot{\theta}^2=0\label{eq:eom1},\\
    S \ddot{\theta} + 3 H S \dot{\theta} + \frac{1}{S}\frac{\partial V_\slashed{D}}{\partial \theta} + 2 \dot{S} \dot{\theta}=0.\label{eq:eom2}
\end{align}

\section{Saxion Couplings}\label{sec:couplings}
The relevant couplings of the Saxion to the SM fields can be parameterized as 
\begin{align}
    \mathcal{L}_\text{ferm.} &= Y_{S\nu}\;  S\; \overline{\nu}\nu +  \sum_{\psi=l,q} Y_{S\psi}\;  S\;\overline{\psi} \psi \\
    \mathcal{L}_\text{bos.} &=\frac{\lambda_{\sigma}}{4} S^4 + \frac{\lambda_{\sigma H}}{4} S^2 h^2 + g_{S\gamma\gamma}\; S\; F_{\mu\nu} F^{\mu\nu}.
\end{align}
Apart from the interaction with the neutrinos given by
\begin{align}
    Y_{S\nu} = \frac{m_\nu}{N f_a}
\end{align}
all other interactions originate from Saxion-Higgs mixing or loop diagrams.

\subsection{Couplings from Saxion-Higgs mixing}
The Saxion-Higgs mixing angle reads
\begin{align}\label{eq:mixangle}
    \tan(2\theta_{SH}) \equiv 2 \lambda_{\sigma H} \frac{v_H N f_a}{m_h^2-m_S^2},
\end{align}
where $m_h=\SI{125}{\giga\electronvolt}$ is the mass of the SM like Higgs. The coupling to charged leptons and quarks $\psi$ is given by
\begin{align}
    Y_{S\psi} \equiv \sin(\theta_{SH}) \frac{m_\psi}{v_H}.
\end{align}
The on-loop Higgs di-photon coupling also leads to a Saxion-photon coupling in terms of the fine structure constant $\alpha$ \cite{Ellis:1975ap,Shifman:1979eb}
\begin{align}
    g_{S\gamma\gamma} \equiv \frac{\sin(\theta_{SH}) \alpha}{2\pi v_H} \left(\frac{N_c Q_t^2}{3}-\frac{7}{4}\right),
\end{align}
where the first term in parentheses is the contribution from the top-quark with $N_c=3$ colors and an electric charge $Q_t = 2/3$ and the second term is due to the $W$-boson.
Here the Saxion only couples to the strongly interacting quarks and gluons, or eqivalently the hadronic sector, via mass mixing with the Higgs. This is the reason why strong limits \cite{Ishizuka:1989ts} from light Saxion emission via couplings to nucleons inside SN1987A are abesent.

\subsection{Loop induced couplings}
The Saxion can also obtain one- and two-loop couplings to SM fermions and photons respectively via similar diagrams as the Diraxion in \ref{sec:DirLoop}. However since these couplings are suppressed by at least $m_\nu^2/v_H^2$, we take them to be negligible compared to the couplings from Saxion-Higgs mixing. Note that for Saxion masses below the QCD confinement scale $\mathcal{O}(100\;\text{MeV})$ we would need to replace the quarks running in the photonic loops with hadrons \cite{Garcia-Cely:2017oco,Heeck:2019guh}, which is beyond the scope of this work.

\section{Higher dimensional operator from a second scalar field}\label{app:dirOp}
The domain wall problem can be remedied by considering another effective operator constructed   from $\sigma$ and its conjugate   that essentially reduces to a tadpole, where the operator dimension reads $d=5+2n$ with $n>0$ 
\begin{align}\label{eq:Dmass}
     N^{(ii)}=1: \quad  V_\slashed{D}^{(ii)} =  c_d^{(ii)} \frac{|\sigma|^{2(2+n)} \sigma}{M_\text{Pl.}^{2n+1}}  + \text{h.c.}\;,\quad m_a^{2\;(ii)} \equiv  \frac{2}{\sqrt{2}^{2n+5}}  \left|c_d^{(ii)}\right| \left(\frac{f_a}{M_\text{Pl.}}\right)^{2n+1} f_a^2 
\end{align}
and the domain wall number is unity for all choices of $n$.
Alternatively one can switch on multiple effective operators with different dimensions $d$ of lowest common denominator one so that no residual $\mathcal{Z}_N$ symmetry remains intact \cite{Reig:2019sok}. However this comes at the price of having at least two sources for the Diraxion mass. 
One can actually engineer a scenario where the desired operator dimension appears due to an accidental symmetry after the spontaneous breaking of an additional  gauge symmetry such as e.g. $\text{U(1)}_\text{B-L}$, hypercharge or a linear combination of them. We assume a particle spectrum and charge assignment to cancel all gauge anomalies. Let us concentrate on the case of $\text{U(1)}_\text{B-L}$ since this symmetry can have the additional benefit of prohibiting Majorana masses. A second scalar singlet $\varphi$ with a vev $v_\phi \gg v_\sigma$ dominantly breaks the local symmetry, under which the scalars have charges $Q_\text{B-L}[\varphi]$ and $Q_\text{B-L}[\sigma]$   with lowest common denominator one.
The only gauge invariant operators of dimension $d= n_\phi +n_\sigma>n_\phi +5$ then read \cite{PhysRevD.46.539,Rothstein:1992rh}
\begin{align}\label{eq:cosine}
  N^{(iii)}:\quad    V_\slashed{D}^{(iii)} = c_d^{(iii)} \frac{\varphi^{*\; n_\varphi}\; \sigma^{n_\sigma} }{M_\text{Pl.}^{n_\sigma+n_\varphi-4}}  + \text{h.c.}\;,\quad m_a^{2\;(iii)} \equiv \frac{2 N_\text{glob.}^{n_\sigma}  }{\sqrt{2}^{n_\sigma+n_\varphi}}  \left|c_d^{(iii)}\right|  \frac{v_\sigma^{n_\sigma-2}\;v_\varphi^{n_\varphi}}{M_\text{Pl.}^{n_\sigma+n_\varphi-4}} 
\end{align}
and the gauge charges must satisfy
\begin{align}
    n_\sigma  Q_\text{B-L}[\sigma] - n_\varphi  Q_\text{B-L}[\varphi]=0.
\end{align}
After $\varphi$ condenses we see that the potential has an accidental $\mathcal{Z}_{n_\sigma}$ symmetry, that can be understood as a remnant of the original gauge symmetry. The Diraxion mass then depends on its decay constant and the second larger vev $v_\varphi$. 
When it comes to the domain wall issue the situation is more complicated due to the presence of two symmetries: Once $\varphi$ gets a vev local strings from the $\text{U(1)}_\text{B-L}$ breaking are formed. Afterwards when $\sigma$ condenses global strings from the breaking of $\text{U(1)}_\text{D}$ are formed. However $\sigma$ also carries a gauge charge, so the local $\text{U(1)}_\text{B-L}$ string obtains a defect in $v_\sigma$, and the winding number $\omega_\sigma^{(\varphi)}$ of  $\sigma$ around the local string is determined from the minimization of the system's kinetic energy leading to\footnote{Here we assumed that $v_\sigma$ wraps once around the global string and $v_\varphi$ winds once around the local string} \cite{PhysRevD.46.539,Rothstein:1992rh}
\begin{align}
     N^{(iii)}= \begin{cases} \left|n_\sigma\right|\;&\text{global}\\\text{Min}\left|n_\varphi-\omega_\sigma^{(\varphi)} n_\sigma\right|\;&\text{local} \end{cases}.
\end{align}
The important point is that the system of domain walls and two types of strings will be unstable if \textit{either} of the domain wall numbers is one \cite{PhysRevD.46.539,Rothstein:1992rh}. Since we need $n_\sigma>5$ for Axiogenesis \cite{Co:2020jtv} this leaves the two choices $n_\varphi = 1, \;\omega_\sigma^{(\varphi)} = 0$ or $n_\varphi \neq  1, \;\omega_\sigma^{(\varphi)} = (n_\varphi -1)/n_\sigma$. A non-trivial winding number $\omega_\sigma^{(\varphi)}$ implies mixed anomalies between $\text{U(1)}_\text{D}$ and $\text{U(1)}_\text{B-L}$. Since the gauge symmetry is assumed to be abelian, this does not lead to non-perturbative contributions from instantons to the Diraxion mass. On the other hand, if $\text{U(1)}_\text{B-L}$ is embedded into a larger non-abelian group such as the Pati-Salam hypercolor $\text{SU(4)}_\text{c}$ \cite{PhysRevD.10.275}, these unwanted  effects reappear and might be even compounded by  small size UV-instantons  \cite{Agrawal:2017evu,Agrawal:2017ksf}, which could  limit  the possible UV-completions of our setup. The dynamics of $\varphi$ in the early universe can be neglected as long as
\begin{align}
    \sqrt{|\lambda_{\sigma \varphi}|} \; S_i > \text{Max}\left[T_\text{RH},T_\text{max}, \frac{H_I}{2\pi}\right].
\end{align}
As long as $\lambda_{\sigma \varphi}<0$ there is no danger of restoring the B-L symmetry via the large vev $S_i$.

\section{Initial field value from a coupling to the Inflaton}\label{app:initial}
A second way to generate a Hubble dependent mass contains one (or more)  coupling(s) to the inflaton field $\chi$ that could take the following forms
\begin{align}\label{eq:operators}
    V(\sigma) \supset c_1\frac{ |\chi|^{2m} |\sigma|^{2n}}{M_\text{Pl}^{2(m+n)-4}}, \quad c_2 \frac{V(\chi) |\sigma|^{2} }{M_\text{Pl}^2}, \quad c_3 \frac{\left(\partial_\mu \chi \partial^\mu \chi\right) |\sigma|^{2}}{M_\text{Pl}^2},
\end{align}
that were suggested by references  \cite{Kusenko:2014lra,Yang:2015ida,Kawasaki:2017ycl}, \cite{Kearney:2016vqw} and \cite{Bao:2022hsg} respectively.
One can understand the effect of these couplings by focusing e.g. on the second term and using the Friedmann equation to eliminate $V(\chi)$ for slow roll inflation, which  implies an effective mass squared $3 c_2 H_I^2$. In general the idea is, that the large initial field value $\chi_i$ of the slowly rolling inflaton field sources an effective  mass term that is initially larger than the Hubble rate $m_S(\chi_i)>H_I$. As the inflaton field decreases the effective mass will become smaller than the Hubble rate after $N_\text{last}$ e-folds of inflation (counted from the end of inflation) so that quantum fluctuations can push $S$ to the value (we assume $m_S < H_I$) \cite{Kusenko:2014lra,Yang:2015ida,Kawasaki:2017ycl}
\begin{align}
    S_i \simeq \sqrt{\braket{S^2}} = \frac{\sqrt{N_\text{last}}}{2\pi} H_I.
\end{align}
That implies that isocurvature fluctuations will  not be produced before $N_\text{last}$. Observations by the Planck and WMAP collaborations only constrain isocurvature modes whose momenta are below the pivot scale $k_{*} = 0.1\;\text{Mpc}^{-1}$
\cite{Planck:2018jri}. The matter power spectrum extracted from Lyman-$\alpha$ data is only sensitive to perturbations at scales of $0.2\;\text{MPc}^{-1}\lesssim k \lesssim 10\;\text{Mpc}^{-1}$ \cite{Beltran:2005gr}. If the Saxion fluctuation has comoving momenta $k \geq 10\;\text{Mpc}^{-1}$, current experiments do not set a limit on the isocurvature power spectrum. Reference \cite{Kawasaki:2017ycl} determined the required value of $N_\text{last}$ to be
\begin{align}
    N_\text{last} \lesssim 46.4 - \text{Log}\left(\frac{k}{10\;\text{Mpc}^{-1}}\right) + \frac{1}{3}\text{Log}\left(\frac{H_I}{10^{13}\;\text{GeV}}\right) + \frac{1}{3}\text{Log}\left(\frac{T_\text{RH}}{10^{12}\;\text{GeV}}\right),
\end{align}
where we suppressed subleading terms depending on the temperature today and number of relativistic degrees of freedom. Determining the evolution of the effective mass from the evolution of $\chi$ and checking whether $ N_\text{last} $ is realizable, requires specifying an inflationary model and a dedicated analysis beyond this work.
It is important to note that we chose only effective couplings between the inflaton and the singlet scalar in \eqref{eq:operators} in order to not destabilize the flat inflaton potential too much \cite{Kearney:2016vqw} and to avoid additional, potentially large, contributions to the energy density of the universe \cite{Kearney:2016vqw}. The second and third operator in the above  are motivated by scenarios where the inflaton is protected by a shift symmetry \cite{Freese:1990rb}, that is only explicitly broken by  $V(\chi)$. Consequently these operators do not introduce a new   source of shift symmetry breaking.

\section{Absence of chiral hypermagnetic instability}\label{sec:CPI}
The network of Yukawa and gauge interactions in the Diraxion background used in section \ref{sec:Boltz}  will in general also source a chiral chemical potential for all the fermionic species $\psi_i$ carrying hypercharge $Q_\text{Y}[\psi_i]$ \cite{Co:2022kul} 
\begin{align}
    \mu_{\text{Y}5}= \sum_i s_i g_i Q_\text{Y}[\psi_i]^2 \mu_i \equiv c_5 \dot{\theta}, 
\end{align}
where $g_i$ is the number of degrees of freedom for each multiplet and $s_i$ equals $+1\;(-1)$ for left-(right-)chiral fermions. 
The presence of such a chiral chemical potential can convert fermionic asymmetries into hypermagnetic gauge fields via the chiral plasma instability (CPI) \cite{Joyce:1997uy} from the chiral magnetic effect \cite{Fukushima:2008xe}, which can also be understood as a tachyonic instability in one of the two  helicities of the hypercharge gauge fields \cite{Boyarsky:2011uy}. The change in hypermagnetic helicity leads to a violation of B+L, which is not washed-out by the weak sphalerons and can act as a separate source for producing the baryon asymmetry \cite{Giovannini:1997eg,Giovannini:1997gp,Kamada:2016eeb,Kamada:2016cnb,Kamada:2018tcs} compared to our intended mechanism. Reference \cite{Co:2022kul} analyzed these dynamics for the case of a rotating axion background and found that CPI is only relevant when the rate of growth for the fastest growing helicity mode
\begin{align}
    \Gamma_\text{CPI} = \frac{\alpha_\text{Y}^2 \mu_{\text{Y}5}^2}{2\pi^2 \sigma_\text{Y}}
\end{align}
is much larger than the Hubble rate before the Saxion settles at its minimum at $T_S$, where in the above we denote the hypercharge finestructure constant as $\alpha_\text{Y}\simeq 0.01$ and $\sigma_\text{Y}\simeq 54 T$ denotes the thermal conductivity in the SM plasma before the electroweak phase transition \cite{Arnold:2000dr}. CPI is not relevant as long as \cite{Co:2022kul} 
\begin{align}
    \frac{ \Gamma_\text{CPI}}{H}\Big|_{T_S} < c_\text{CPI} = 10,
\end{align}
because for $T<T_S\;(T>T_S)$ one can see from the scaling relations in \eqref{eq:sclingtheta} that $\Gamma_\text{CPI}/H\sim \dot{\theta}^2/T^3\sim T^3 \;(1/T)$ during radiation domination. If the Dirac Weinberg operator were to thermalize and all the SM Yukawas were in thermal equilibrium, which is only true below $10^6\;\text{GeV}$  \cite{Nardi:2005hs,Bodeker:2019ajh} due to the electron Yukawa, then one would find that
\begin{align}
      c_5^\text{eq.}(T_S<10^6\;\text{GeV})= 0.
\end{align}
This temperature regime is the relevant one for our scenario as 
\eqref{eq:TS} implies that $T_S= \mathcal{O}(1\;\text{TeV})$ for GeV-scale Saxions and hence CPI is absent.
If $T_S$ was larger than $10^6\;\text{GeV}$ then in general we expect a non-zero $c_5^\text{eq.}$.
However since we work in the Freeze-In regime with an $S$-dependent rate for the Dirac Weinberg operator, we have to repeat the same steps that lead to \eqref{eq:asym}.
Furthermore since Freeze-In occurs predominantly at $T_\text{osc.}$ and chemical potentials redshift as $\sim 1/T$ during radiation domination we would find that 
\begin{align}
      \mu_{\text{Y}5}(T_S>10^6\;\text{GeV}) = \frac{T_S}{T_\text{osc.}} \mu_{\text{Y}5}(T_\text{osc.}) = \sqrt{\frac{3}{2}} c_5^\text{eq.}(T_S>10^6\;\text{GeV})\cdot  \frac{\Gamma_S(S_i)}{H(T_\text{osc.})} \cdot \varepsilon m_S.
\end{align}
This implies that $\Gamma_\text{CPI}$ is suppressed by two powers of the small parameter $\Gamma_S(S_i)/H(T_\text{osc.})\leq 0.1$ and hence we would not expect strong constraints on our parameter space even for a non-zero $c_5^\text{eq.}$.

\section{Early Thermalization from Type III Dirac Seesaw}\label{app:earlysketch}
To realize early thermalization, that occurs some time after the initiation of oscillations, but still early enough to thermalize the parametric resonance Diraxions, we need a particle with a thermal abundance and a renormalizable coupling to $S$, such as the iso-triplet $\Delta$ from the Type III scenario \ref{sec:typeIII}, which may be much lighter than the other new degrees of freedom (see \eqref{eq:delta}). Early thermalization could proceed as follows: $\Delta$ is relativistic and rapidly scatters with the bath due to its gauge interactions. For this to stay true we need to ensure that the corrections to the triplet mass squared  $\lambda_{H\Delta} S_i^2$ (tree level) and $\lambda_4^2 / (16 \pi^2) S_i^2$ (one-loop) stay below $T_\text{osc.}^2$, which has typical values of $\mathcal{O}(10^{15}\;\text{GeV})$ for successful cogenesis. This implies that
\begin{align}
    \lambda_{H\Delta} &< 10^{-6} \cdot \left(\frac{0.1 M_\text{Pl}}{S_i}\right)^2 \cdot \left(\frac{T_\text{osc.}}{10^{15}\;\text{GeV}}\right)^2 \label{eq:lambda1},\\
    \lambda_4 &< 10^{-2} \cdot \left(\frac{0.1 M_\text{Pl}}{S_i}\right) \cdot \left(\frac{T_\text{osc.}}{10^{15}\;\text{GeV}}\right).\label{eq:lambda2}
\end{align}
In general the Saxion may also receive a thermal mass from its coupling to $\Delta$. The thermal mass is smaller than  the tree level mass as long as
\begin{align}
  \text{Max}(\lambda_{H\Delta},\lambda_4) <    \left(\frac{S_i}{0.1 M_\text{Pl}}\right) \cdot \left(\frac{10^{15}\;\text{GeV} }{T_\text{osc.}}\right) \cdot 
 \begin{cases}
     10^{-6}\quad&\text{quant. fluct.}\\
     0.1 \quad&\text{$H$-mass from $R$}
    \end{cases}.
\end{align}
We expect interaction rate between triplets and Saxions to scale as\footnote{Here for simplicity we ignore the factor of $\alpha_2(T)^2$ that would appear with the coupling $\lambda_{H\Delta}$, see \eqref{eq:thermrate}.} e.g. \\
$\Gamma\sim \text{Max}[\lambda_{H\Delta}^2,\lambda_4^2]\; S^2/T$ \cite{Mukaida:2012qn,Mukaida:2012bz}  so that during radiation domination $\Gamma/H \sim S^2/T^3\sim 1/T$, which is IR dominated. Hence the scattering with $\Delta$ can be taken to be slow when the oscillations start implying $T_\text{th.}<T_\text{osc.}$  and further
\begin{align}
    \text{Max}(\lambda_{H\Delta},\lambda_4) <1.4\times 10^{-3} \cdot \left(\frac{0.1 M_\text{Pl}}{S_i}\right) \cdot \left(\frac{T_\text{osc.}}{10^{15}\;\text{GeV}}\right)^2.
\end{align}
The thermalization rate peaks while the $\Delta$ are relativistic and then rapidly decreases once they are Boltzmann-suppressed. 
In order to avoid warmness bounds on the Diraxion abundance from parametric resonance, thermalization during radiation domination needs to occur before the Saxion field value has reached $S\simeq 10^{-2}  S_i$ \cite{Kasuya:1998td,Kasuya:1999hy,Kawasaki:2013iha}, hence we require $T_\text{th.}>0.01\; T_\text{osc.}$, from which 
\begin{align}
    \text{Max}(\lambda_{H\Delta},\lambda_4) >1.4\times 10^{-4} \cdot \left(\frac{0.1 M_\text{Pl}}{S_i}\right) \cdot \left(\frac{T_\text{osc.}}{10^{15}\;\text{GeV}}\right)^2
\end{align}
follows. Comparing this to \eqref{eq:lambda1} and \eqref{eq:lambda2}
reveals that only $\lambda_4$ is large enough for thermalization, which is not in conflict  the isocurvature bounds in \eqref{eq:cond41} and \eqref{eq:cond4} (at least for the Hubble induced Saxion field value). Since $\Delta$ is charged under the $\text{U}(1)_\text{D}$ symmetry, its interactions with the condensate can transmit a fraction of the Noether charge into an asymmetry of $\Delta$. Reference \cite{Domcke:2022wpb} considers the charge transfer between rotating condensates in detail.
The asymmetry in $\Delta$ is then converted via its Yukawa couplings  (essentially the same reactions as in \ref{sec:dissp} but with $S$ replaced by $\Delta$) into an asymmetry of the left chiral leptons that gets reprocessed into a baryon asymmetry via the sphalerons.
Once $\Delta$ becomes Boltzmann suppressed we can integrate it out and recover the Weinberg-operator.

\bibliographystyle{JHEP}
\bibliography{references}

\end{document}